\documentclass[a4paper,11pt]{article} 
\usepackage{graphicx} 
\usepackage[  backend=bibtex,   style=alphabetic,maxalphanames=4,minalphanames=4 ,maxbibnames=6,refsegment=section]{biblatex}

\bibliography{bibliography}
\usepackage{tocloft}

\cftsetindents{subsection}{1em}{4em}
\cftsetindents{section}{0em}{2em}

\usepackage{amsmath,amssymb, amsthm}

\usepackage{enumerate}
\usepackage{xcolor}
\usepackage{bm}
\usepackage{amsbsy}
\usepackage[unicode]{hyperref}
\usepackage{cleveref}
\usepackage{textgreek}
\usepackage{fontawesome5}
\usepackage{dsfont}
\usepackage{mathtools}

\newtheorem{theorem}{Theorem}[subsection]

\hypersetup{
  colorlinks   = true, 
  urlcolor     = blue, 
  linkcolor    = {blue!75!black}, 
  citecolor   = {blue!50!black} 
}
\newtheorem{proposition}[theorem]{Proposition}

\newtheorem{corollary}[theorem]{Corollary}
\newtheorem{lemma}[theorem]{Lemma}

\newtheorem{remark}{Remark}

\newtheorem{definition}{Definition}[section]
\numberwithin{equation}{section}

\newcommand{\grouppi}{\texorpdfstring{$\boldsymbol{\pi}$}{\textpi}}
\newcommand{\grouplambda}{\texorpdfstring{$\boldsymbol{\lambda}$}{\textlambda}}
\newcommand{\groupmu}{\texorpdfstring{$\boldsymbol{\mu}$}{\textmu}}
\newcommand{\groupgamma}{\texorpdfstring{$\boldsymbol{\gamma}$}{\textgamma}}

\newcommand{\bb}{\mathbb}
\newcommand{\R}{\mathbb R}
\newcommand{\E}{\mathbb E}
\newcommand{\abs}[1]{\lvert #1 \rvert}
\newcommand{\ip}[1]{\left\langle #1\right\rangle}

\newcommand{\lt}{\left}
\newcommand{\rt}{\right}
\DeclarePairedDelimiter\ceil{\lceil}{\rceil}

\DeclareMathOperator{\Var}{Var}         
\newcommand{\norm}[1]{\left\| #1 \right\|}

\usepackage{environ}
\usepackage[framemethod=tikz]{mdframed}
\NewEnviron{hint}{%
  \begin{tikzpicture}
    \node[rotate=180, inner sep=10pt] {%
      \parbox{0.8\linewidth}{\emph{Hint: \BODY}}
    };
  \end{tikzpicture}%
}

\newif\ifnotes
\notestrue

\usepackage{pifont}
\newcommand*\colourcheck[1]{%
  \expandafter\newcommand\csname #1check\endcsname{\textcolor{#1}{\ding{52}}}%
}

\colourcheck{green}
\DeclareMathOperator*{\sign}{sign}

\newcommand{\NNN}{\mathcal{N}}
\newcommand{\PPP}{\mathcal{P}}
\newcommand{\EE}{\mathbb{E}}
\newcommand{\SK}{\mathrm{SK}}

\newcommand{\cert}{\mathrm{cert}}
\newcommand{\srch}{\mathrm{srch}}
\newcommand{\defeq}{\mathrel{\mathop:}=}

\DeclareMathOperator{\GOE}{GOE}

\usepackage{subcaption}

\title{Average-case complexity in statistical inference: \\  A puzzle-driven research seminar}
\author{Anastasia Kireeva \thanks{Department of Mathematics, ETH Z\"{u}rich. Email: \texttt{anastasia.kireeva@math.ethz.ch}.}
\and
Afonso S. Bandeira \thanks{Department of Mathematics, ETH Z\"{u}rich. Email: \texttt{bandeira@math.ethz.ch}.}}

\date{}

\begin{document}

\maketitle
\begin{abstract}
    These notes describe our experience with running a student seminar on average-case complexity in statistical inference using the jigsaw learning format at ETH Zurich in Fall of 2024. The jigsaw learning technique is an active learning technique where students work in groups on independent parts of the task and then reassemble the groups to combine all the parts together. We implemented this technique for the proofs of various recent research developments, combined with a presentation by one of the students in the beginning of the session. We describe our experience and thoughts on such a format applied in a student research seminar: including, but not limited to, higher engagement, more accessible talks  by the students, and increased student participation in discussions. In the Appendix, we include all the exercises sheets for the topic, which may be of independent interest for courses on statistical-to-computational gaps and average-case complexity. 
\end{abstract}

\section{Introduction}
In several universities, student research seminars for undergraduate students are designed to acquaint students with advanced or current research topics through reading and preparing a presentation to their peers on a certain topic, with each student presenting different content. It also enhances  the students' speaking and presentation skills, through the exercise of introducing the topic to their fellow seminar participants.

The seminar content is chosen beforehand by the seminar organizers. They prepare a list of publications or chapters of a book related to the seminar topic and assign the topics among the students at the beginning of the semester. A student prepares the assigned topic and presents it to their peers using slides or blackboard in one of the seminar sessions. The other participants have the opportunity to ask the speaker questions during or after the presentation. 

A common problem encountered in this format is the relatively low engagement level of the audience during presentations and discussion times. It is challenging to successfully encourage students to ask more questions during their colleagues presentations, especially considering that some of them may feel reluctant to speak in front of the group. In some situations, the opportunity to ask questions first in a small group, can help some students to be more participative. 
With this in mind, and aiming for increased engagement, we introduced a modified student seminar structure that blends in a classical seminar format with the jigsaw learning method, also sometimes referred to as the puzzle learning method \cite{aronson1978jigsaw,slavin1978using}. 

\begin{figure}[htbp]
  \centering
  \begin{subfigure}[b]{0.45\textwidth}
    \includegraphics[width=\textwidth]{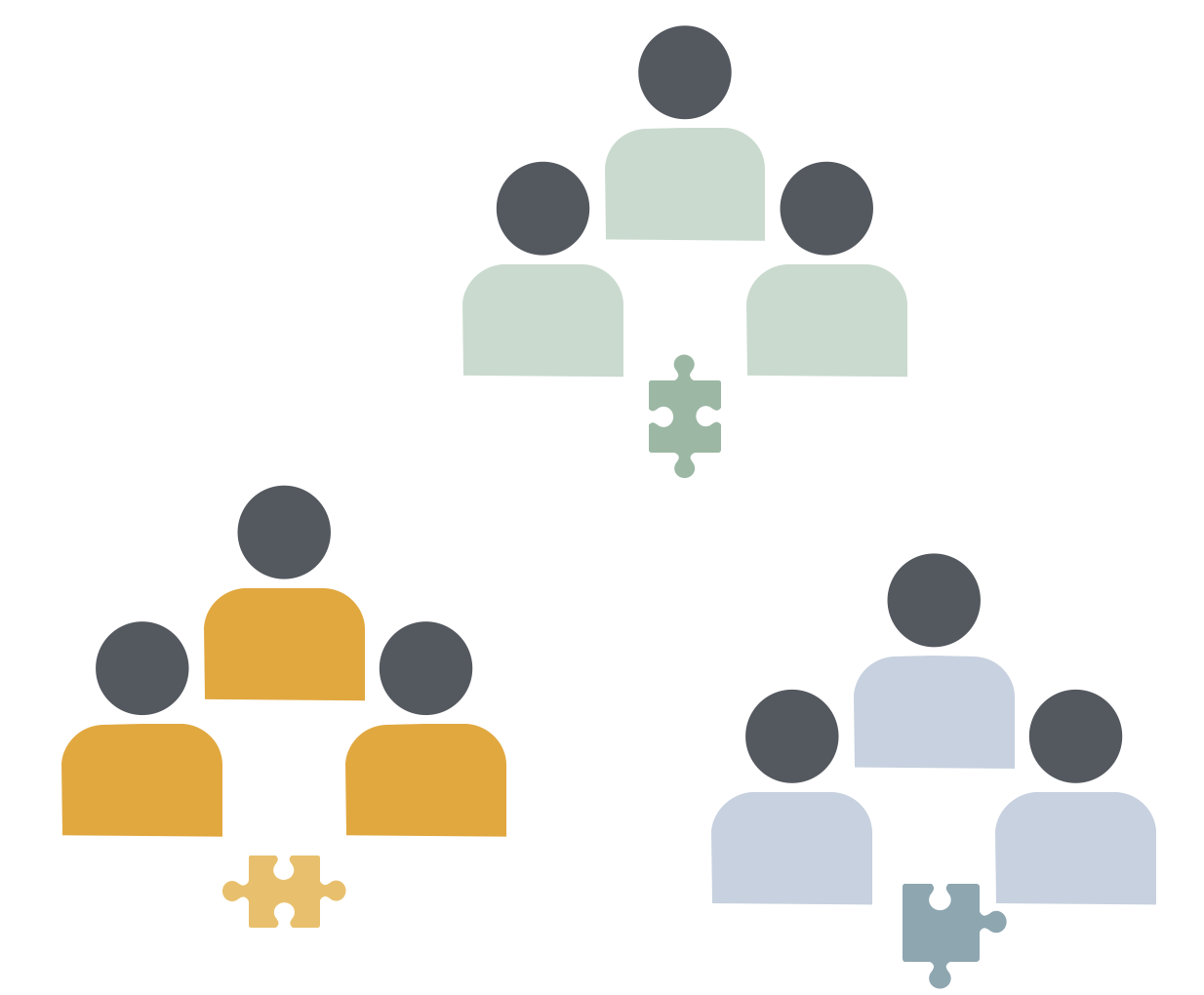}
    \caption{Students study and become experts on their part of the topic in groups.}
    \label{fig:jigsaw1}
  \end{subfigure}
  \hfill
  \begin{subfigure}[b]{0.45\textwidth}
    \includegraphics[width=\textwidth]{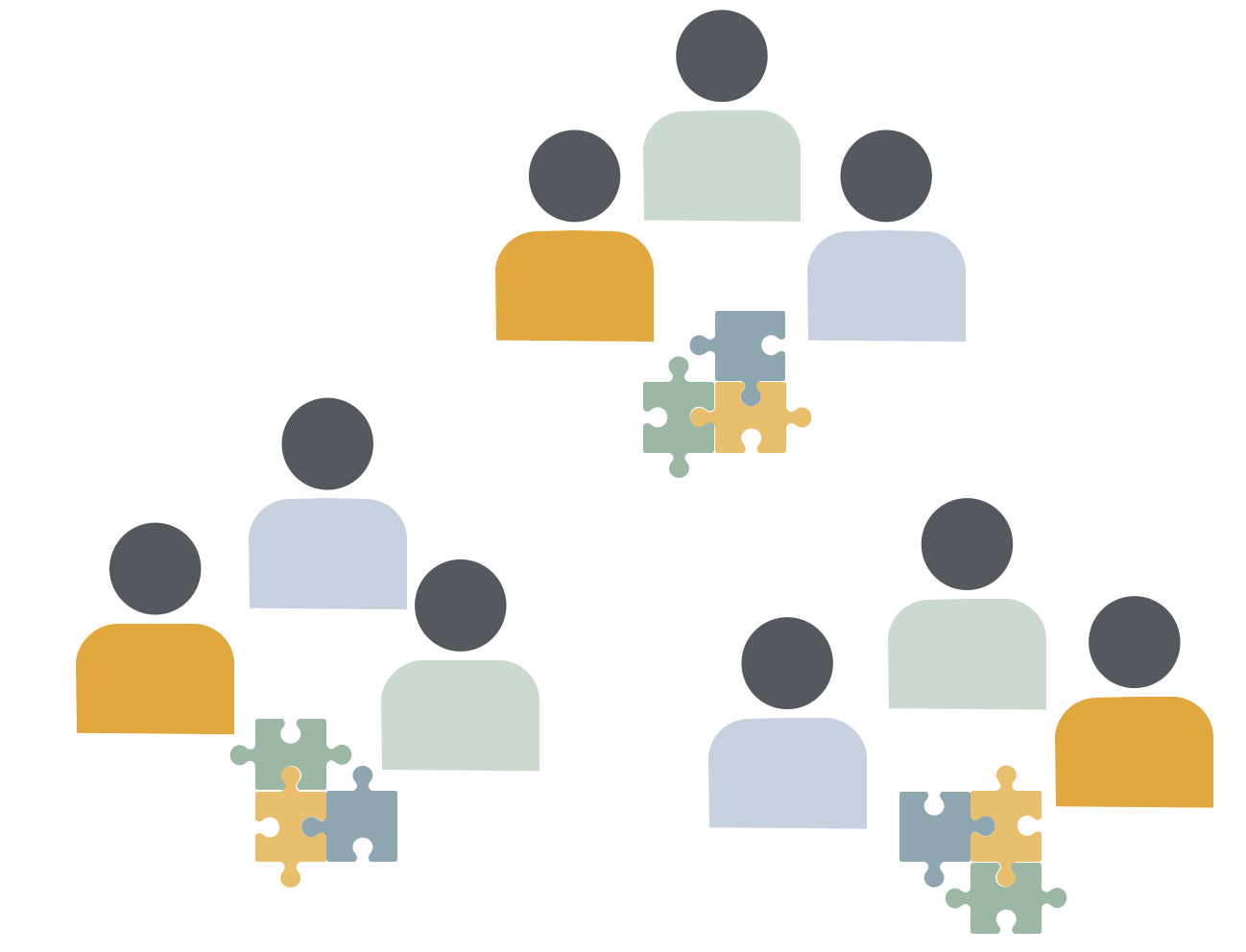}
    \caption{Students are reassembled into jigsaw groups to share their findings and work on the final task together.}
    \label{fig:jigsaw2}
  \end{subfigure}
  \caption{Schematic illustration of the jigsaw technique.}
  \label{fig:jigsaw}
\end{figure}

The jigsaw technique, developed by \cite{aronson1978jigsaw}, is a cooperative learning strategy designed to promote teamwork among students and increase active learning. It consists of two phases. In the first phase, students work on their group assignment in small groups, and each member of a group becomes an ``expert'' on that part of the topic (see \Cref{fig:jigsaw1}). 
In the second phase, the groups are reassembled so that each new group has a representative of each part, see \Cref{fig:jigsaw2}. During this phase, the students share their acquired knowledge with their peers and work on a task that requires expertise of each original part of the topic. 
Since each student is the most knowledgeable in their topic part in their new group, this tends to promote interdependency and a sense of responsibility among the students.

In our seminar, each session began with a student presenting the topic to provide necessary background for other participants and to give the presenter an opportunity to practice their presentation and public speaking skills. For the jigsaw part of the seminar, we divided a research result related to the week topic into several parts so that each part can be proven independently of others. These intermediate results were distributed among groups of students in the form of the assignment sheets. After 20 to 35 minutes, the groups are reassembled, and students work on combining all the parts together to finish the proof. We describe the format in more detail in \Cref{sec:seminar-structure}. 

This format gives students an opportunity to discuss their questions with their peers first. This seemed to us to have many benefits, including giving an opportunity for students to work out questions with fellow students, resembling the the process of real collaborative research work. 

In this note, we describe the process of setting up the jigsaw seminar format and our impressions after the first edition of the seminar on the topic of average-case complexity in statistical inference at ETH Zurich in Fall 2024. We hope that our experience will motivate other educators to try engaging methods for teaching mathematics to bachelor and master-level students. 

\paragraph{Acknowledgments} AK expresses gratitude to organizers of EPT summer camp 2024, where she learned a lot about teaching methods in higher education. She is grateful to Siara Isaac for introducing the jigsaw learning method to her and inspiring her to try it in a research seminar. AK and ASB thank Norbert Hungerb\"{u}hler for encouraging us to write this note.

\section{Seminar structure}\label{sec:seminar-structure}

Each seminar session lasts ninety minutes, divided into three parts -- a 20-minute presentation by a student, first part of the jigsaw work in groups (35 to 50 minutes in groups), and second part after reassembling the groups (20 to 35 minutes).

In the first session, the topics for the entire semester were distributed among the students. Each week, one student (acknowledged as \emph{the expert of the week}), is responsible for studying the assigned topic to explain it to their peers at the session. This student prepares a twenty-minute presentation and becomes familiar with the proofs assigned for the session. Before the session, the presenting student has a possibility to meet one of the seminar organizers to go through the remaining questions and discuss the presentation. 

\begin{figure}[!b]
    \centering
    \includegraphics[width=0.5\linewidth]{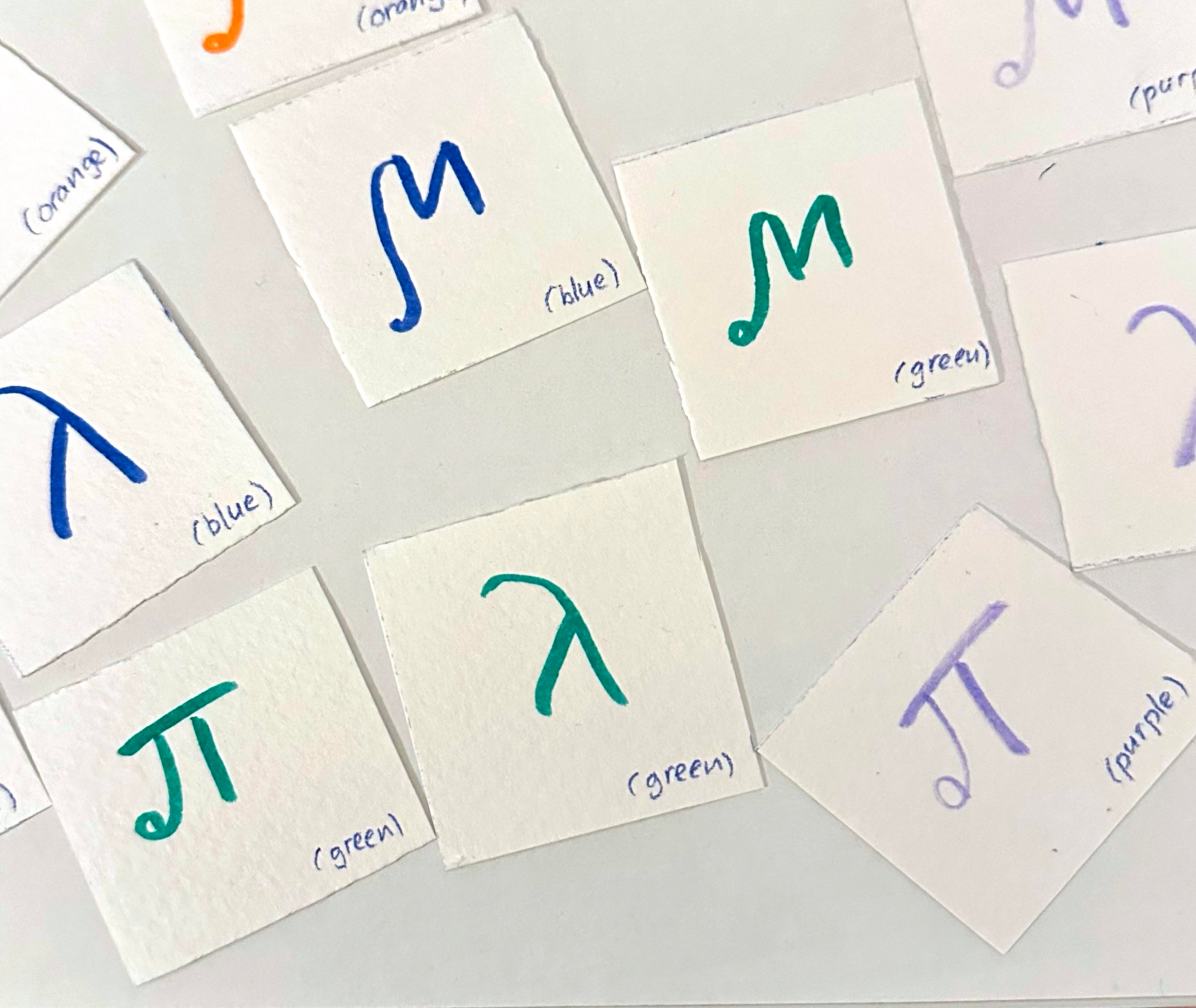}
    \caption{Cards for assigning the groups.}
    \label{fig:cards}
\end{figure}

After the presentation, followed by a five-minute discussion session, students are divided into three or four groups by distributing colored cards with a group symbol, see \Cref{fig:cards}. In the first phase, students are split into groups according to the symbol on the card ($\bm \lambda$, $\bm \mu$, $\bm \pi$) and work on their corresponding part of the result for 35 to 50 minutes. The students are also encouraged to ask questions or request hints from the expert of the week, or, if necessary, from the seminar organizers. Each group receives several copies of their assignment sheet marked with their group symbol. Besides the tasks, the assignment sheets contained the main concepts and intermediate technical lemmas to assist the students in the proving more substantial results.

During the second phase, students are reassembled into new groups according to the color of their cards. Each new group is composed of students who share the same card color but represent different symbols. For instance, one group might consist of students holding purple $\bm \lambda$, purple $\bm \mu$, and purple $\bm \pi$ cards. In this part, the students share with each other their findings and combine their intermediate results to complete the proof. For this part, a separate assignment sheet was available with light guidance towards finishing the proof.  

The seminar had 13 student participants, with one student presenting the topic, and the remaining 12 students  divided into three groups (or, for some sessions, four groups) and later reshuffled into four groups (or three, respectively). To split students into random groups every session, we used 12 cardboard cards, each labeled with $\bm \lambda$, $\bm \pi$, or $\bm \mu$ (and, if necessary, $\bm \gamma$) and colored purple, blue, green, or orange. The color was also indicated in text at the bottom of the card to ensure accessibility.

In most sessions, PhD students joined the seminar. In such cases, additional star-marked cards were used to balance groups with fewer members or to assist groups with more difficult tasks.

\subsection{Icebreaker session: SBM game}
At ETH, the students taking the same course may not know each other as they may come from different programs and years. Since our seminar involved a significant amount of teamwork during the semester, we organized an ice-breaker session in the first week. 

The game is based on a theoretical concept, namely, the stochastic block model \cite{abbe2018community}. Informally, the (assortative) stochastic block model describes random graphs with planted communities, where nodes in the same community are more likely to be connected than nodes in different communities. The inter- and intra-community edge probabilities are fixed, and all edges are drawn independently of each other. A classical problem is to determine the regime of these probabilities for which it is possible to recover the underlying communities by observing only the edges of the graph. 

Inspired by this model, we developed a game, in which each participant represents a vertex and gest a list of the names of students representing their adjacent neighbors. The underlying graph is drawn according to the assortative stochastic block model. The goal of the students is to recover the underlying partition in communities. Given that each student only has local information, it is crucial that students communicate and share graph information with one another. Since the problem is highly non-trivial even with the full graph information, a solution also requires significant brainstorming. The successful strategies tend to correspond to algorithms that do most of their operations locally, so that each student can mentally ``implement'' their part of the algorithm. 

The game was aimed at fostering friendlier and more comfortable atmosphere among the students. In the beginning of the game, students learned each others names to find their neighbors; moreover, we encouraged them to properly introduce themselves to other participants regardless of their 'adjacency' to each other. Not only the students got to know each other, they collaborated on the winning strategy by sharing their thoughts and discussing it together. 

\begin{remark}
    We repeated this game several times at larger scale courses at ETH. As there were significantly more participants  (around 45-50 students), we replaced the names by the numbers and shared with the students the parameters of the model, namely, intra- and inter-community edge probabilities. Despite the less intimate setting, students remained very collaborative, suggesting and actively discussing their solutions with their peers. Our experience is that the game gets mathematically more interesting as the number of students grow. In larger setting,  it also shows the students the phenomenon of parallelization in algorithms.
\end{remark}

\section{Thoughts and Impressions}

A particularly interesting phenomenon that we observed was that the presentations of the expert of the week were, in our opinion, of significantly higher quality than in a more classical student seminar setting. The presentations were more ``high-level'' and focusing on the main concepts and ideas, rather than on technical details. We believe that since the expert of the week knew that the students were going to do the technical part of the topic in the assignments, they did not feel a need to cover technical details and thus focused on the broader concepts and ideas of the week. This, in our view, led to a significant improvement of presentation accessibility to other students. We remark that, in other years when running classical student seminars, we have encouraged students to skip technical content and focus on big-picture concepts and ideas. However, this encouragement did not, in our experience, necessarily result in truly big-picture presentations, while the jigsaw format appears to have naturally fostered such presentations. We were very positively surprised with this (a priori) unexpected feature. 

Additionally, in our view, the role of the expert of the week promoted a sense of responsibility in the presenters, which was also visible during the group work. 
During the presentations, we felt that the audience was considerably more active at asking questions, as the students anticipated working on this topic and were motivated to understand it better. This increased engagement reflects one of the outcomes we aimed to achieve by introducing the new format.

During the session, students were mostly independent, requiring only occasional support from the seminar organizers, as the expert of the week was available and generally well-prepared for the questions. With only minimal guidance from the organizers to signal the transitions between seminar phases, the seminar involved high peer-to-peer engagement and student interdependence, which, in our view, promoted a positive and collaborative atmosphere. 

We remark that, on the other hand, preparing such sessions required significantly more effort than for seminars in a traditional format. It was important for us to balance introduction to new concepts, relatively low technicality of the arguments in the proofs, and interpretability of the proven result. Finding such a proof and dividing it into relatively equal and independent parts required careful preparation and creativity to adjust some of the parts to simpler cases.

During the semester, the students had opportunities to provide (anonymous) feedback on the seminar format and content. 
Overall, the students gave positive feedback on the collaborative and more autonomous atmosphere and felt that hands-on problem-solving improved their understanding, while teamwork increased motivation. Some suggested improvements included ending each session with a brief summary by an organizer to clarify key concepts --- something done informally in some sessions but not consistently as it was not planned initially. Others noted time constraints and recommended providing reading materials in advance, which we implemented mid-semester while still aiming to keep sessions self-contained due to students' busy schedules.

\section{Introduction to average-case complexity}
The main topic of the seminar was studying the computational-to-statistical gaps of the statistical inference problems. This term refers to regimes where a statistical inference problem is information-theoretically solvable, but we are unaware of any time-efficient algorithm for doing so. 

\begin{figure}
    \centering
    \includegraphics[width=0.9\linewidth]{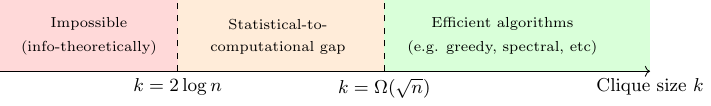}
    \caption{Statistical-to-computational gap in Erd\"os-R\'enyi graph}
    \label{fig:clique}
\end{figure}

To illustrate this concept, consider a problem of finding a planted clique in Erd\"os-R\'enyi graph. This problem is a canonical example of a problem exhibiting a statistical-to-computational gap. While it is possible to detect and recover the hidden clique by brute-force search when its size exceeds $(2+\varepsilon) \log n$, where $n$ is the number of vertices and $\varepsilon$ is an arbitrary positive constant, we are unaware of any efficient algorithm to do so unless the clique size exceeds $\Omega(\sqrt{n})$, see \Cref{fig:clique}. 

Apart from the planted clique, there are many more examples of statistical inference problems, with applications ranging from electronic microscopy~\cite{Singer_CryoEM} to building a recommendation system~\cite{koren2009matrix}. 
Two distinctive features of modern statistical problems are high-dimensionality and the randomness of the measurement process. 

In many modern applications, typical instances do not resemble worst-case scenarios. Consequently, worst-case complexity predictions may often be too pessimistic for typical instances of the problem. For this reason, we aim to understand average-case complexity of the problem, for example, how much noise we can tolerate to solve, with high probability, a \emph{typical} instance of the problem of interest.

To characterize the information-theoretic threshold, that is, the noise level beyond which reliable inference is fundamentally impossible, we can rely on tools from statistical theory. However, identifying the computational threshold, which separates the regimes where efficient algorithms succeed or fail, requires a separate toolbox. 
Although ruling out all time-efficient algorithms is currently out of reach, we can exclude instead certain classes of algorithms or rely on heuristic arguments to predict the location of the computational threshold.

For a more comprehensive overview of this topic, we point the reader to the following (non-exhaustive) list of surveys~\cite{wein2025computational,Kunisky2019NotesLowDeg,bandeira2018notes,gamarnik2022disordered}.

In the seminar, the main focus was to study various theoretical frameworks for establishing information-theoretical lower bounds and predicting computational thresholds. We briefly outline the main topics of the seminar in the next section.

\subsection{Seminar topics}
In the last section, \Cref{s1:contiguity} to \Cref{s16:Sk-model}, we provide the assignment sheets for the seminar topics. Below is the high-level description of the covered topics.  

\paragraph{Statistical distinguishability} The statistical decision theory concerns with making an optimal choice between statistical conclusions based on some qualitative criteria. We consider the simple hypothesis testing to explore the tools from Le Cam's contiguity theory to provide information-theoretical lower bounds for a spiked Wigner model and explore different notions of statistical distinguishability. This topic is covered in sessions \ref{s1:contiguity} and \ref{sec:s2-stat-decition-theory}.

\paragraph{Low-degree polynomials}
Low-degree polynomials framework allows us to study computational hardness of the statistical inference tasks in average case~\cite{Hopkins_lowdeg}. Originally, this method arose in the study of Sum-of-Squares hierarchy~\cite{barak2019nearly}, and subsequently was developed in numerous works as an independent tool for analyzing computational hardness for hypothesis testing and other statistical inference problems \cite{Hopkins2017EfficientBE,hopkins2017power,Hopkins_lowdeg}. Sessions \ref{s3:low-degree-spike-model} to \ref{s7:two-planted} cover this framework on various examples ranging from the hypothesis testing of a spiked model to counting communities in the stochastic block model.

\paragraph{Statistical physics}
Statistical physics tools are now routinely used for analysis of both statistical and computation hardness of Bayesian inference problems (see, e.g., \cite{dominguez2024Statistical}). The connection lies in the fact that in many inference problems posterior distribution of the signal given the observation follows Gibbs distribution that is widely studied in statistical physics in analysis of the disordered systems. Sessions \ref{s8:free-energy-I} to \ref{sec:franz-parisi-I} cover this framework. In particular, session~\ref{sec:franz-parisi-I} connects low-degree polynomials framework with the statistical physics tools, and sessions~\ref{sec:MCMC} and \ref{s13:franz-parisi-II} provide a connection to Monte Carlo Markov chains.

\paragraph{Overlap Gap Property}
Overlap Gap Property (OGP) is a tool for analyzing complexity of the optimization statistical problems based on studying the structural properties of the objective landscape. This property originated in the study of spin glasses in \cite{talagrand2010mean}, and its first formal algorithmic implications were explored by Gamarnik and Sudan \cite{gamarnik2014limits}. The idea of this approach is that in some cases computational hardness of an optimization problem translates into non-trivial geometry of the objective function in the neighborhood of optimal and near-optimal solutions. Sessions~\ref{s14:ogp-I} and \ref{s15:e-OGP} cover two notions of OGP, namely, a classical notion of OGP and the ensemble version.

\bigskip
The last session~\ref{s16:Sk-model} presents various ideas for the Sherrington-Kirkpatrick model, covering quiet planting, certifying the optimization upper bounds and computational lower bounds.

\begin{remark}
    Some of the exercise sheets were edited based on the feedback of the students and thus slightly differ from the original exercise sheets. We split several exercise sheets into separate sessions, removed some of the tasks and added more hints to reduce the load of each individual session. Additionally, we fixed some of typos and rearranged the sheets structure in this report. In particular, in the seminar each student would get a separate sheet containing an introduction to their group tasks, however, in this report, we compiled the introduction into the beginning of each section to avoid repetition.
\end{remark}

\paragraph{Disclaimer} These notes have not been subjected to the scrutiny of a formal publication and may contain typos or inaccuracies. If you notice any errors, please contact the authors at anastasia.kireeva@math.ethz.ch, or bandeira@math.ethz.ch.

\newpage

\printbibliography
\newpage

\tableofcontents

\appendix
\renewcommand\thesection{S\arabic{section}}

\renewcommand\thesubsection{S\arabic{section}.\sectionletter}

\section{Contiguity and distinguishability}\label{s1:contiguity}

We start by studying various criteria for the distributions to be distinguishable in a hypothesis testing setting. We will cover the notions of weak and strong detection, separability, and contiguity, and show connections between them. In later sessions, we will see that some of these connections remain true in the computationally bounded setting. The main references are \cite[Section 1]{Kunisky2019NotesLowDeg} and \cite[Sections 1.1-1.2]{bandeira2022franz}.

\newcommand{\sectionletter}{1}
\subsection{Background}
Suppose ${\mathbb{P}}=\left(\mathbb{P}_n\right)_{n \in \mathbb{N}}$ and $\mathbb{Q}=\left(\mathbb{Q}_n\right)_{n \in \mathbb{N}}$ are two sequences of probability distributions over a common sequence of measurable spaces $\mathcal{S}=$ $\left(\left(\mathcal{S}_n, \mathcal{F}_n\right)\right)_{n \in \mathbb{N}}$.  Suppose we observe $\boldsymbol{Y} \in \mathcal{S}_n$ which is drawn from one of $\mathbb{P}_n$ or $\mathbb{Q}_n$. We hope to recover this choice of distribution in the following sense.
\begin{definition}
    A sequence of measurable functions $f_n: \mathcal{S}_n \rightarrow\{p, q\}$ is said to \emph{strongly distinguish} ${\mathbb{P}}$ and ${\bb Q}$ if $\bb P_n(f_n(\boldsymbol{Y})=p) = 1 - o(1)$ and $\bb Q_n(f_n(\boldsymbol{Y})=q) = 1 - o(1)$ . If such $f_n$ exist, we say that ${\mathbb{P}}$ and $\mathbb{Q}$ are \emph{statistically distinguishable}.
\end{definition}

\begin{definition}[Strong/Weak Separation]
Let $\bb P_n$, $\bb Q_n$ be as above and $f_n : \mathcal{S}_n \to \mathbb{R}$ be a measurable function such that $\Var_{\bb P_n} f_n, \Var_{\bb Q_n} f_n$ are well-defined. 

\begin{itemize}
    \item we say $f$ \textbf{strongly separates} $\mathbb{P}_n$ and $\mathbb{Q}_n$ if, as $n \to \infty$,
    \[
    \frac{\sqrt{\max \left\{ \text{Var}_{\mathbb{P}_n}[f_n], \text{Var}_{\mathbb{Q}_n}[f_n] \right\}}}{\left| \mathbb{E}_{\mathbb{P}_n}[f_n] - \mathbb{E}_{\mathbb{Q}_n}[f_n] \right|} = o(1),
    \]
    
    \item we say $f$ \textbf{weakly separates} $\mathbb{P}$ and $\mathbb{Q}$ if, as $n \to \infty$,
    \[
    \frac{\sqrt{\max \left\{ \text{Var}_{\mathbb{P}_n}[f_n], \text{Var}_{\mathbb{Q}_n}[f_n] \right\}}}{\left| \mathbb{E}_{\mathbb{P}_n}[f_n] - \mathbb{E}_{\mathbb{Q}_n}[f_n] \right|}  = O(1).
    \]
\end{itemize}
\end{definition}

\begin{definition}
    We say that $\mathbb P_n$ is \emph{contiguous} with respect to $\mathbb Q_n$, written $\mathbb P_{n} \triangleleft \mathbb Q_n$ if for every event $A_n$ such that $\mathbb Q_n(A_{n})\to 0$ it holds $\mathbb P_n(A_{n})\to 0$. 
\end{definition} 

We will explore connections between these different notions, and in particular, we will consider different parameter regimes of a toy model for which they come into play. Define the univariate Gaussian model as 
\begin{align*}
&\mathbb{P}_{n}: Y = \mu_{n} + w, \quad \text{where } w\sim \mathcal{N}(0, 1) \\
&\mathbb{Q}_{n}: Y = w, \quad \text{where } w \sim \mathcal{N}(0, 1)
\end{align*}
for some $\mu_n \in \bb R$ that may depend on $n$.

\renewcommand{\sectionletter}{$\bm \lambda$} 
\subsection{Group \grouplambda}
For tasks (a)–(c), assume the general case of distributions $\bb P_n$ and $\bb Q_n$ for which all necessary notions (e.g., expectation, variance) are well defined.
\begin{enumerate}[(a)]
    \item If $\|L_{n}\|^2 = O(1)$ then there is no strong separation.
    \item If $\|L_n\|^2 - 1 = o(1)$ then there is no weak separation.
    \item Prove that when $\|L_n\|^2 = O(1)$, then $\mathbb P_{n} \triangleleft \mathbb Q_n$.
    \item For the univariate Gaussian model, show that $\|L_{n}\| = O(1)$ if $\mu_{n} = O(1)$ and $\|L_{n}\| = 1 + o(1)$ if $\mu_{n} = o(1)$.
\end{enumerate}

\renewcommand{\sectionletter}{$\bm \mu$} 
\subsection{Group \groupmu}
For tasks (a)–(b), assume the general case of distributions $\bb P_n$ and $\bb Q_n$ for which all necessary notions (e.g., expectation, variance) are well defined.
\begin{enumerate}[(a)]
    \item If $\mathbb P_{n}=\mathbb P$ and $\mathbb Q_{n} = \mathbb Q$ then contiguity implies absolute continuity, i.e., $\mathbb P\triangleleft \mathbb Q \Rightarrow \mathbb P \ll \mathbb Q$.
    \item Prove that if $\mathbb P_{n}\triangleleft \mathbb Q_n$ or $\mathbb Q_{n}\triangleleft \mathbb P_n$ then they are statistically indistinguishable, i.e., there is no test that  strongly distinguishes them.
    \item For the univariate Gaussian model, show that $\mathbb P_n \ll \mathbb Q_n$ for any $\mu_n$.
    \item For the same model, show that $\mathbb P_{n} \triangleleft \mathbb Q_{n}$ if and only if $\mu_{n} = O(1)$.
\end{enumerate}

\renewcommand{\sectionletter}{$\bm \pi$} 
\subsection{Group \grouppi}
For tasks (b)–(c), assume the general case of distributions $\bb P_n$ and $\bb Q_n$ for which all necessary notions (e.g., expectation, variance) are well defined.
\begin{enumerate}[(a)]
    \item For the univariate Gaussian model, show that $\mathbb P_{n}$ and $\mathbb Q_{n}$ are strongly separable by $f(x) = x$ if and only if $\mu_{n} = \omega(1)$. Similarly, show that $\mathbb P_{n}$ and $\mathbb Q_{n}$ are weakly separable by $f(x) = x$ if and only if $\mu_{n} = \Omega(1)$.
    \item Prove that strong separation implies that strong detection is achievable by thresholding $f$, but the converse is not true in general.
    \item (optional) Show that weak separation implies that weak detection is achievable. 
    
    \begin{hint}
        Show that $P_{\text{error}} \ge 1 - d_{TV}(\bb P, \bb Q)$, where $P_{\text{error}}$ is the sum of the type I and type II error, and $d_{TV}$ is the total variation distance. Show that TV distance is not vanishing as $n\to\infty$ if $\bb P_n$ and $\bb Q_n$ are weakly separable. 
    \end{hint} 
\end{enumerate}

\renewcommand{\sectionletter}{\faPuzzlePiece}
\subsection{Jigsaw}
Discuss your results and show that 
\begin{enumerate}[(a)]
    \item If $\|L_{n}\|^2 = O(1)$ then there is no strong detection between $\bb P_n$ and $\bb Q_n$. 
    \item If $\|L_n\|^2 - 1 = o(1)$ then there is no weak detection between $\bb P_n$ and $\bb Q_n$. 
    \item Discuss your results on the univariate Gaussian model for different asymptotics on $\mu_n$.

\end{enumerate}

\printbibliography[segment=\therefsegment]
\section{Statistical Decision Theory}\label{sec:s2-stat-decition-theory}
In this session, we consider the spiked Wigner model to study the statistical decision theory concepts we analyzed in the previous session. We will derive the information-theoretical threshold for detection the spike in this model and review some of the basic probability concepts.

The main reference for this session is \cite[Section 1]{Kunisky2019NotesLowDeg}. 
\renewcommand{\sectionletter}{1}
\subsection{Background}

\begin{definition}
    For $\lambda \geq 0$ and a spike prior $\mathcal{X}$, we define the spiked Gaussian Wigner model  as follows. We first draw a spike $x \in \mathbb{R}^n$ from the prior $\mathcal{X}_n$. Then we reveal

$$
Y=\lambda x x^{\top}+\frac{1}{\sqrt{n}} W,
$$

where $W$ is drawn from the $n \times n$ GOE (Gaussian orthogonal ensemble), i.e. $W$ is a random symmetric matrix with off-diagonal entries $\mathcal{N}(0,1)$, diagonal entries $\mathcal{N}(0,2)$, and all entries independent (except for symmetry $\left.W_{i j}=W_{j i}\right)$. 
\end{definition}

In what follows, denote the spiked Gaussian Wigner model with the signal-to-noise ratio $\lambda > 0$ by $\bb P_{n, \lambda}$, where $n$ denotes the dimension, and denote the null model by $\bb Q_n$ (i.e., the signal-to-noise ratio $\lambda=0$, and the observation contains the pure noise).

Finally, we define the likelihood ratio as 
$$
L_n(Y) := \frac{d \bb P_{n, \lambda}}{d \bb Q_n}
$$
and its norm induced by $\bb Q_n$ as
$$
\| L_n(Y) \|^2 = \bb E_{\bb Q_n}[ L_n(Y)^2]
$$

\begin{definition}
We say that a real-valued random variable $X$ is \emph{sub-Gaussian with variance proxy} $\sigma^2$ if
\[
\mathbb{E}[X] = 0 \quad \text{and} \quad \mathbb{E}[\exp(tX)] \leq \exp\left( \frac{1}{2} \sigma^2 t^2 \right)
\]
for all $t \in \mathbb{R}$.
\end{definition}
\renewcommand{\sectionletter}{$\bm \lambda$} 
\subsection{Group \grouplambda}
Our goal is to show the following sub-Gaussian tail bound. For any $a > 0$ and sub-Gaussian random variable $X$ with variance proxy $\sigma^2$,
\begin{equation}\label{eq:tail}
\bb P(|X| \ge a) \le 2 \exp\lt(-\frac{a^2}{2\sigma^2}\rt).
\end{equation}
\begin{enumerate}[(a)]
    \item We start with bounding the probability $\bb P(X \ge a)$. Use Chernoff bound followed by the definition of sub-Gaussian random variable to show that 
    $$
    \bb P(X \ge a) \le \exp\lt(\frac{1}{2}\sigma^2 t^2 - ta\rt),
    $$
    for any $t \in \bb R$.
    \item Since the inequality above holds for any $t$, choose $t$ to minimize the right-hand side and substitute it.
    \item To conclude the proof, use union bound to get a two-sided variant of the inequality.
\end{enumerate}

\renewcommand{\sectionletter}{$\bm \mu$} 
\subsection{Group \groupmu}
\begin{enumerate}[(a)]
    
\item  Let $\xi$ be a Rademacher random variable ($\xi = \pm 1$ with  with probability $1/2$). Show that $\xi$ is a sub-Gaussian with variance proxy $1$.

\begin{hint}
    Use Taylor's theorem.
\end{hint}

\item Let $\xi$ be a sub-Gaussian variable with variance proxy $\sigma^2$. Let $a \in \bb R$ be a fixed scalar. Show that $\xi / a$ is also sub-Gaussian, and find the variance proxy parameter.

\item  Let $X_1, \dots, X_n$ be independent sub-Gaussian variables with variance proxy $\sigma^2$. Show that $\sum_{i=1}^n X_i$ is sub-Gaussian with variance proxy $n \sigma^2$.

\item  Suppose that $x \in \bb R^n$ is a random vector with independent Rademacher coordinates ($x_i = \pm 1$ with probability $1/2$ for each $i$). Show that $\lt \langle \frac{x}{\sqrt{n}}, \frac{x^\prime}{\sqrt n} \rt \rangle$ is sub-Gaussian with proxy $1$. 

\end{enumerate}

\renewcommand{\sectionletter}{$\bm \pi$} 
\subsection{Group \grouppi}
Our goal is to prove the following proposition. 
\begin{proposition}
    Let $\mathcal{X}$ be any spike prior. With $x$ and $x^{\prime}$ drawn independently from $\mathcal{X}_n$, suppose $\frac{n \lambda^2}{2} \left\langle x, x^{\prime}\right\rangle$ is $\left(\lambda^2 \sigma^2 / 2\right)$-sub-Gaussian for some constant $\sigma$. If $\lambda<1 / \sigma$, then $\mathbb{E}_{x, x^{\prime}} \exp \left(\frac{n \lambda^2}{2}\left\langle x, x^{\prime}\right\rangle^2\right)$ is bounded.
\end{proposition}

\begin{enumerate}[(a)]
    \item Express the expectation as an integral using the tail probability expectation formula (for a non-negative random variable $\xi \ge 0$, it holds $\bb E \xi = \int_0^\infty \bb P(\xi > t) dt$).
    \item To bound the integral, you may use the following sub-Gaussian tail bound without proof. For a sub-Gaussian random variable $X$ with proxy $\sigma^2$, it holds
$$
\bb P(|X| \ge t) \le 2 \exp\lt(-\frac{t^2}{2\sigma^2}\rt).
$$
\begin{hint}
    You may encounter a problem where the resulting integral is convergent at infinity when $\lambda < 1 / \sigma$ but divergent at zero. How can you bound the integral near zero before applying the sub-Gaussian tail bound to circumvent this issue?
\end{hint}
\end{enumerate}

\renewcommand{\sectionletter}{$\bm \gamma$} 
\subsection{Group \groupgamma}
\begin{enumerate}[(a)]
    \item Read Propositions 2.4, 2.5 of \cite{Kunisky2019NotesLowDeg} and the respective proofs.
    \item Prove the following proposition (you may use the results of Propositions 2.4, 2.5 without reproving them). See also \Cref{rmk:symm-asymm-model} below.
\begin{proposition}
    Let $\lambda \geq 0$ and let $\mathcal{X}_n$ be a spike prior. 
    Let $x$ and $x^{\prime}$ be independently drawn from $\mathcal{X}_n$. Then

$$
\| L_n(Y)\|^2=\underset{x, x^{\prime}}{\mathbb{E}} \exp \left(\frac{n \lambda^2}{2}\left\langle x, x^{\prime}\right\rangle^2\right).
$$
\end{proposition}
\end{enumerate}
\begin{remark}[Symmetric and asymmetric variants of the model]\label{rmk:symm-asymm-model}
    Note that the signal counterpart is symmetric, which is why we constrain the noise matrix to be symmetric. 
    
    To see why, suppose the noise matrix is asymmetric and contains i.i.d. entries. By taking the average $\frac{1}{2} \left( (xx^\top + W) + (xx^\top + W)^\top \right)$, we can reduce the variance of each entry $W_{ij}$ by a factor of $\sqrt{2}$. We can still define an equivalent asymmetric model by adjusting the signal-to-noise ratio $\lambda$ by a factor of $\sqrt{2}$.

Define
$$
Y_{\text{asym}}=\frac{\lambda}{\sqrt 2} x x^{\top}+\frac{1}{\sqrt{n}} W_{\text{asym}},
$$
where $W_{\text{asym}}$ is a random matrix with i.i.d. Gaussian entries $\mathcal{N}(0, 1)$. To see that these models are equivalent, notice that
$$
\frac{1}{\sqrt 2}(Y_{\text{asym}} + Y_{\text{asym}}^\top) = \lambda x x^{\top}+\frac{1}{\sqrt{n}} W,
$$ 
and $Y$ are equally distributed. For more details, see also Appendix A.2 of \cite{Kunisky2019NotesLowDeg}.

For the proof, it may be more convenient to use the asymmetric version with the adjusted SNR.
\end{remark}

\renewcommand{\sectionletter}{\faPuzzlePiece}
\subsection{Jigsaw}
Combine your results to show that when $\lambda \le 1$, then the spike model with signal strength $\lambda$ is contiguous to the null model, $\bb P_{n, \lambda } \triangleleft \bb Q_n$.
\printbibliography[segment=\therefsegment]
\section{Low-degree hardness of fixed-rank spike detection}
\label{s3:low-degree-spike-model}
\renewcommand{\sectionletter}{1}
In this session, we explore how to apply the low-degree polynomial framework to find the parameter regime where detection a low-rank spike in the Wigner model is low-degree hard. This session is closely connected to the next one, where we will use the derived results to show low-degree hardness of detecting communities in the stochastic block model in the certain parameter regime. 

The main references are \cite[Sections 1-2]{Kunisky2019NotesLowDeg} and \cite[Section 6]{bandeira2021spectral}. 

\subsection{Background}

\begin{definition}
    We say that $W$ is drawn from the $n \times n$ GOE$(n)$ (Gaussian orthogonal ensemble), i.e. $W$ is a random symmetric matrix with off-diagonal entries $\mathcal{N}(0,1/n)$, diagonal entries $\mathcal{N}(0,2/n)$, and all entries independent (except for symmetry $W_{i j}=W_{j i}$). 
\end{definition}

\begin{definition}[Wigner model with low-rank spike, or Gaussian additive model]\label{def:GAM}
Let $\mathcal X$ be a probability measure over symmetric matrices of size $n$, $\R^{sym}_{n\times n}$.
\begin{itemize}
    \item  Under $\mathbb Q$ we observe $Y \sim GOE(n)$;
    \item Under $\mathbb P$ we draw $X\in \mathcal X$ and observe $Y = \lambda X + W$ where $W\sim GOE(n)$.
\end{itemize}

\end{definition}
We define a class of priors of PSD matrices with fixed rank in the following way.

\begin{definition}\label{def:prior-lowrank}
    Let $\pi$ be a probability measure supported on a bounded subset of $\R^k$ such that $\E \pi = 0$ and $\|Cov(\pi)\| = 1$. Draw $U \in \R^{n\times k}$ so that each row is sampled independently according to $\pi$. Let $\mathcal X(\pi)$ denote the distribution of matrices $X = \frac{1}{n} U U^\top$. 
\end{definition}

\begin{definition}
A random vector ${x}$ is $\varepsilon$-local $c$-subgaussian if for any fixed vector $v$ with $\|v\| \leq \varepsilon$,
$$
\mathbb{E} \exp (\langle v, {x}\rangle) \leq \exp \left(\frac{c}{2}\|v\|^2\right)
$$
\end{definition}

\begin{proposition}\label{prop:small-deviations-local-subgaussian}
    Let $\delta>0$. If a random vector $\boldsymbol{x} \in \mathbb{R}^k$ is $\varepsilon$-local $c$-subgaussian with $c<(1-\delta)^2 / 2$ then

$$
\mathbb{E}\left[\mathbf{1}\lt[\|\boldsymbol{x}\| \leq \frac{\varepsilon c}{ (1-\delta)}\rt] \exp \left(\|\boldsymbol{x}\|^2\right)\right] \leq 1+\frac{C(\delta, k)}{(1-\delta)^2 /(2 c)-1},
$$

where $C(\delta, k)$ is a constant depending only on $\delta$ and $k$.

\end{proposition}

\begin{proposition}\label{prop:large-deviations-local-subgaussian}
    
    If a random vector $\boldsymbol{x} \in \mathbb{R}^k$ is $\varepsilon$-local $c$-subgaussian then for any $\delta>0$ and any $0 \leq t \leq \varepsilon c /(1-\delta)$,

$$
\operatorname{Pr}\{\|x\| \geq t\} \leq C(\delta, k) \exp \left(-\frac{1}{2 c}(1-\delta)^2 t^2\right)
$$

where $C(\delta, k)$ is a constant depending only on $\delta$ and $k$.
\end{proposition}

\renewcommand{\sectionletter}{\grouplambda}
\subsection{Group \grouplambda}
\begin{enumerate}[(a)]
    \item 
    Let $u_i$ be a row of $U$ from \Cref{def:prior-lowrank}, i.e., $u_i \sim \pi$, and let $u_i^\prime$ be its independent copy. 

Denote $R_i := u_i^\top u_i^\prime$ and $R := \sum_{i=1}^n R_i$. 
    \item Our goal is to prove that for any $\eta > 0$ there exists $\varepsilon> 0$ such that $R_i$ is $\varepsilon$-local $(1 + \eta)$-subgaussian for some $\varepsilon>0$. To this end, we will show that the MGF $M(T) := \bb E (\ip{T, R_i})$ is bounded in a neighborhood of $T=0$ (its existence follows from the boundedness of $\bm \pi$). Use Taylor's theorem to bound $M(T)$ for $T$ such that $\| T\|_F \le \varepsilon$.

    \begin{hint}
        Observe that $\nabla M(0) = \bb E R_i$ and $(\operatorname{Hess} M)(0) = \operatorname{Cov}(R_i) = \operatorname{Cov}(\bm \pi)^{\otimes 2}$.
    \end{hint}
    \item Suppose $\boldsymbol{x}$ is $\varepsilon$-local $c$-subgaussian. Show that, for a (non-random) scalar $\alpha \neq 0, \alpha \boldsymbol{x}$ is $\varepsilon /|\alpha|$-local $c \alpha^2$-subgaussian, and the sum of $n$ independent copies of $\boldsymbol{x}$ is $\varepsilon$-local $cn$-subgaussian.
    \item Using (c), show that $\lambda R / \sqrt{2n}$ is $\varepsilon\sqrt{2n}/\lambda$-local $(1+ \eta)\lambda^2 /2$-subgaussian.
\end{enumerate}

\renewcommand{\sectionletter}{\groupmu}
\subsection{Group \groupmu}

Recall that for the Gaussian additive model as in \Cref{def:GAM}, the low-degree likelihood ratio can be expressed as following:
$$
\|L^{\le D}_n\|^2 = \bb E_{X,X^{\prime}}  \exp^{\le D}\left( \frac{\lambda^2 n}{2} \langle X,X^{\prime}  \rangle  \right) = \bb E_{X,X^{\prime}}  \sum_{d=0}^D  \frac{\lambda^{2d} n^d}{2^d} \langle X,X^{\prime}  \rangle^d.
$$

Our goal is to bound the LDLR norm. To this end, we will split it into large and small deviations parts,
\begin{equation*}
\begin{split}
\| L^{\le D}\|^2 &= \E_{X,X^{\prime}} \mathbb I [\langle X,X^{\prime} \rangle \le\Delta ]\exp^{\le D}\left( \frac{\lambda^2n}{2} \langle X,X^{\prime} \rangle  \right) +  \E_{X,X^{\prime}} \mathbb I [\langle X,X^{\prime} \rangle >\Delta ]\exp^{\le D}\left( \frac{\lambda^2n}{2} \langle X,X^{\prime} \rangle  \right) \\
& =: L_1 + L_2,
\end{split}
\end{equation*}
where $\Delta$ is to be chosen later. 

Group $\bm \lambda$ will show that there exists a random matrix $R \in \bb R^{k\times k}$ such that $ \langle X,X^{\prime}  \rangle =\frac{1}{n^2} \|R\|^2_F$, and for any $\eta > 0$ there exists $\varepsilon$ such that $R$ is $\varepsilon$-local $(1+\eta)n$-subgaussian. We will focus on bounding $L_1$ using properties of local subgaussian variables.
\begin{enumerate}[(a)]
 \item  Show that 
$$
L_{1} =  \bb E\lt[ \mathbb I [\lVert R \rVert^2_{F}  \le\Delta n^2 ]\exp\left( \frac{\lambda^2}{2n} \lt\lVert R\rt \rVert_{F}^2  \right)\rt].
$$
    
\item Using  \Cref{prop:small-deviations-local-subgaussian}, show that when $\lambda < 1$, we can choose $\eta$ and $\delta$ so that $L_1 \le C(\delta, \eta) = O(1)$ for $\Delta = \lt ( \varepsilon (1 + \eta)/ (1 - \delta)\rt)^2$.

\begin{hint}
    By assumptions of \Cref{prop:small-deviations-local-subgaussian}, $c < (1 - \delta)^2/2$, where $c = (1+ \eta)\lambda^2 /2$. Since we can choose $\delta$ and $\eta$ freely, we can pick appropriate values to satisfy this assumption. 
\end{hint}

\end{enumerate}

\renewcommand{\sectionletter}{\grouppi}
\subsection{Group \grouppi}
The introduction for this group is the same as for group $\bm \mu$. We omit this introduction here to avoid repetition; however, in the actual problem sheets, each group sheet includes the necessary details for their tasks.
\begin{enumerate}[(a)]
 \item  Show that 
$$
L_{2} =  \bb E\lt[ \mathbb I [\lVert R \rVert^2_{F}  \ge\Delta n^2 ]\exp\left( \frac{\lambda^2}{2n} \lt\lVert R\rt \rVert_{F}^2  \right)\rt].
$$
\item Show  that $\ip{X, X^\prime}\le Cn^2 $ for some constant $C > 0$ due the boundedness of prior $\pi$.
\item Using \Cref{prop:large-deviations-local-subgaussian}, show that $\bb P(\|R\|_F > \sqrt{\Delta}n) = \exp(-\Omega(n))$.
\item Substituting previous results into the expression for $L_2$, conclude that $L_2 = o(1)$ provided that $D = o(n/\log n)$.
\end{enumerate}

\renewcommand{\sectionletter}{\faPuzzlePiece}
\subsection{Jigsaw}
\begin{enumerate}[(a)]
    \item Discuss your findings with your peers.
    \item Using your results, deduce the following theorem. 
    \begin{theorem}[Low-degree hardness of the Wigner model with low-rank spike]\label{thm:gam-ldlr-bound}
Fix constants $k\ge 1$ and $\lambda \in\R$. Consider $\mathcal X$ as in \Cref{def:prior-lowrank}. If $|\lambda|<1$ then $\|L^{\le D}\|= O(1)$ for all $D = o(n / \log n)$.
\end{theorem}
\end{enumerate}

\printbibliography[segment=\therefsegment] 
\section{Low-degree hardness of community detection}
\label{sec4:sbm-low-degree}
\renewcommand{\sectionletter}{1}

In this session, we study the problem of community detection in the stochastic block model. We will use the results from the previous session to predict the parameter regime in which the problem is low-degree hard. To do so, we will express the community detection problem as a binary model and reduce the binary model to the Gaussian additive model in the low-degree sense. The low-degree hardness then follows by comparing the respective parameters of the models.

The main reference for this session is \cite[Appendix B]{bandeira2021spectral}. For foundations on the stochastic block model, refer to the survey \cite{abbe2018community}.

\subsection{Background}

\begin{definition}[Definition 2.19 in \cite{bandeira2021spectral}]\label{def:sbm}
    Let $k \geq 2$ (number of communities), $d > 0$, (average degree), $\eta \in [-1/(k-1), 1]$ (signal-to-noise ratio). Suppose that $k, d, \eta$ are constants that do not depend on $n$. We define the stochastic block model with parameters $k, d, \eta$ by the following null and planted distributions over the $n$-vertex graphs.
\begin{itemize}
    \item Under $\mathbb{Q}$, for every $u < v$ the vertices $u$ and $v$ are connected with probability $d /n$ independently of other pairs of vertices.
    \item Under $\mathbb{P}$, each vertex is independently assigned a community label drawn uniformly from $[k]$. Conditioned on these labels, the edges occur independently. If $u, v$ belong to the same community, then edge $(u,v)$ occurs with probability $(1+(k-1)\eta)d/n$; otherwise $(u,v)$ occurs with probability $(1-\eta)d/n$.
\end{itemize}
\end{definition}

Known polynomial-time algorithms only succeed in distinguishing $\mathbb{P}$ from $\mathbb{Q}$ above the so-called Kesten–Stigum (KS) threshold, i.e., when $d\eta^2 > 1$. We will show the evidence supporting this conjecture by showing that the problem is low-degree hard. 

\begin{theorem}
    Consider the stochastic block model as in \Cref{def:sbm} with parameters $k, d, \eta$ fixed. If $d\eta^2 < 1$ then $\| L^{\leq D} \| = O(1)$ for any $D = o(n/\log n)$.
\end{theorem}

\begin{remark}
    You may have seen another definition of the stochastic block model, where the model is parametrized by the (scaled) average degree of nodes inside the same community and nodes across different communities. Under certain reparametrization, these definitions are equivalent.
\end{remark}

\begin{definition}\label{def:binary-model} We define the binary model as follows.
\begin{itemize}
    \item Under $\bb Q$, we observe a random variable $Y\in\bb R^N$ takes two values $\{a_{i}, b_{i}\}$ where $b_{i}> a_{i}$ so that $\bb E Y_{i}= 0$ and $\bb E Y_{i}^2 = 1$ (for this we need $a_{i} b_{i} = -1$). 
    \item Under $\mathbb P$, we sample $X \in \R^N$ according to some prior $X \sim \pi$ such that $X_{i}\in[a_{i}, b_{i}]$. We observe a random variable $Y\in \{a_i, b_i\}^N$ such that, conditioned on ${X}$, $Y_i$ are drawn independently and $\mathbb{E}[Y_i \mid X_i] = X_i$.
\end{itemize}
\end{definition}

\renewcommand{\sectionletter}{\grouplambda}
\subsection{Group \grouplambda}

Define the Fourier character as $\chi_{S}(Y)=\prod_{i\in S}Y_{i}$ with $|S|\le D$. To show that Fourier characters form an orthonormal basis for polynomials of degree at most $D$ under $\bb Q$, we need to show orthogonality and completeness. Then we can decompose the LDLR using Fourier characters.

\begin{enumerate}[(a)]
    \item Prove $\bb E_{Y \sim \bb Q} \chi_S(Y) \chi_T(Y) = \bb I(S=T)$.
    \item Show that for any $r\in \bb N$, $Y_i^r$ can be written as degree-1 polynomial in $Y_i$, and thus $\{\chi_S\}_{|S|\le D}$ span the subspace of degree $\le D$ polynomials.
    \item Prove that
$$
\lVert L^{\le D} \rVert ^2 = \sum_{|S|\le D} \langle  L^{\le D}, \chi_{S} \rangle ^2 = \sum_{|S|\le D} [\mathbb E_{\mathbb P} \chi_{S}(Y)]^2.
$$
\end{enumerate}

\renewcommand{\sectionletter}{\groupmu}
\subsection{Group \groupmu}
Our goal is to show the following theorem.
\begin{theorem}[Low-degree likelihood ratio for the binary model]\label{thm:ldlr-binary}
For the binary model as in \Cref{def:binary-model}, we have $$
\lVert L^{\le D} \rVert^2 \le \bb E_{X,X^{\prime}} \sum_{d=0}^{D} \frac{1}{d!} \langle X, X^\prime \rangle ^d,
$$
where $X, X^\prime$ are independent copies.
\end{theorem}

Group $\bm \gamma$ will show the following fact, which you may use as given.
$$
\lVert L^{\le D} \rVert ^2 = \sum_{|S|\le D} [\mathbb E_{\mathbb P} \chi_{S}(Y)]^2,
$$
where $\chi_{S}(Y)=\prod_{i\in S}Y_{i}$ with $|S|\le D$ are Fourier characters.
\begin{enumerate}[(a)]

\item Let $\alpha$ denote an ordered multi-set of $[N]$ (i.e., it is ordered and allows for multiple instances for each of its elements), and $S$ is a set. Show that for an integer $d$,
$$
|\{S: |S| = d\}| \le \frac{1}{d!}\abs{\{\alpha: \abs{\alpha} = d\}},
$$
where $\abs{M}$ denotes cardinality of a set $M$.
\item Show that for vectors $u, v \in \bb R^N$,
$$
\ip{u, v^\prime}^d = \sum_{\alpha: \abs{\alpha}=d} \chi_{\alpha} (u)\chi_{\alpha}(v),
$$
where the sum ranges over all ordered multi-sets $\alpha$ of size $d$.
\item Conclude the proof of \Cref{thm:ldlr-binary}.
\end{enumerate}

\renewcommand{\sectionletter}{\grouppi}
\subsection{Group \grouppi}

Consider SBM as in \Cref{def:sbm}. 
For simplicity of the proof, we will modify the model so that we allow self-loops. 
\begin{itemize}
    \item Under $\bb P$, let the edge $(i,i)$ occur with probability $d /n$.
    \item Under $\bb Q$, let the edge $(i,i)$ occur with probability $(1 + \frac{\eta}{\sqrt{2}}(k-1))\frac{d}{n}$.
\end{itemize}

Denote $p:=\frac{d}{n}$.
\begin{enumerate}[(a)]
    \item Check that revealing the extra information can only increase $\|L^{\le D}_n\|^2$. 

    \begin{hint}
        Recall the variational formulation of the low-degree likelihood ratio, Proposition~1.15 in \cite{Kunisky2019NotesLowDeg}.
    \end{hint}

    \item Prove that SBM model can be written as a binary model with $N := n(n+1)/2$, and find parameters $a$ and $b$.

    \begin{hint}
        To ensure all assumptions of \Cref{def:binary-model}, we take $a := - \sqrt{p/(1-p)}$ and $b := \sqrt{(1-p)/p}$.
    \end{hint}
    \item  Note that conditioned on community structure, the edge $(i,j)$ occurs with probability $p(\Delta_{ij}+1)$, where
    $$
    \Delta_{ij} = \begin{cases}
        \frac{\eta(k-1)}{\sqrt{2}}&\quad \text{if } i=j,\\
        \eta(k-1) &\quad\text{if } i, j \text{ belong to the same community},\\
        -\eta  &\quad\text{if } i, j \text{ are from the different communities}.
    \end{cases}
    $$
    Using that $pa + (1-p) b = 1$ and $\bb E[Y_{ij}|X_{ij}]=X_{ij}$, find $X_{ij}$ in terms of $\Delta_{ij}$ and $p$. 
\end{enumerate} 

\renewcommand{\sectionletter}{\groupgamma}
\subsection{Group \groupgamma}
You will show the essential connection between SBM and the corresponding Gaussian model. Group $\bm \pi$ will find the parameters of the corresponding binary model for SBM, and your goal is to prove that a certain choice of parameters of Gaussian model corresponds to the parameters of SBM.
\begin{enumerate}[(a)]
    \item Define $U\in \bb R^{n\times k}$ whose $i$-th row is $\sqrt{k} e_{k_i} - \mathds 1_k / \sqrt{k}$, where $k_i \in [k]$ is the community assignment of vertex $i$. Check that $(UU^\top)_{ij} = k \bb I[k_i=k_j] - 1$.
    \item Let $X_{ij} = \Delta_{ij}\sqrt{\frac{p}{1-p}}$, where 
    $$
    \Delta_{ij} = \begin{cases}
        \frac{\eta(k-1)}{\sqrt{2}}&\quad \text{if } i=j,\\
        \eta(k-1) &\quad\text{if } i, j \text{ belong to the same community},\\
        -\eta  &\quad\text{if } i, j \text{ are from the different communities}.
    \end{cases}
    $$
    Check that $X_{ij} = \eta\sqrt{\frac{p}{1-p}}(UU^\top)_{ij}$ if $i\ne j$, $X_{ii} = \frac{\eta}{\sqrt{2}}\sqrt{\frac{p}{1-p}}.(UU^\top)_{ii}$, and so
    $$
    \left\langle{X}, {X}^{\prime}\right\rangle=\frac{\eta^2}{2} \frac{p}{1-p}\left\langle{U} {U}^{\top}, {U}^{\prime}\left({U}^{\prime}\right)^{\top}\right\rangle.
    $$
\end{enumerate}

\renewcommand{\sectionletter}{\faPuzzlePiece}
\subsection{Jigsaw}
\begin{enumerate}[(a)]
    \item Discuss your findings with your peers.
    \item Using \Cref{thm:ldlr-binary}, show that detection in SBM is at least as low-degree hard as detection in a fixed-rank spike model as in \Cref{def:GAM}. Combine your results to find the corresponding parameters. In particular, show that 
    $$
\left\|L^{\leq D}\right\|^2 \leq \sum_{d=0}^D \frac{1}{d!} \mathbb{E}\left(\frac{\eta^2}{2} \frac{p}{1-p}\left\langle{U} {U}^{\top}, {U}^{\prime}\left({U}^{\prime}\right)^{\top}\right\rangle\right)^d,
$$
where $U$ is defined in group $\bm \gamma$. 
\item By \Cref{thm:gam-ldlr-bound}, conclude that detection in SBM is low-degree hard when $d\eta^2 < 1$. 

\begin{hint}
    Note that $\eta^2 d / (1-p) = (1 + o(1)) \eta^2 d$.
\end{hint}
\end{enumerate}

\printbibliography[segment=\therefsegment] 
\section{Low-degree hardness of estimation in planted submatrix model}
In this session, we will study the application of the low-degree polynomial framework for statistical estimation problems. Our running example will be the planted submatrix model. Similarly as before, we will show low-degree hardness by reducing this model to the Gaussian additive model and matching the parameters to the low-degree hardness regime of the latter model.

The main reference is \cite{Schramm_Lowdeg_recovery}.

\renewcommand{\sectionletter}{1}
\subsection{Literature and background}

\begin{definition}
In \emph{general additive Gaussian noise model} we observe $Y = X + Z$ where $X \in \mathbb{R}^N$ is drawn from an arbitrary (but known) prior, and $Z$ is i.i.d. $\mathcal{N}(0,1)$, independent from $X$. The goal is to estimate a scalar quantity $x \in \mathbb{R}$, which is a function of $X$.
\end{definition}

Let $\mathbb{R}[Y]_{\leq D}$ denote the space of polynomials $f : \mathbb{R}^N \to \mathbb{R}$ of degree at most $D$. The degree-$D$ maximum correlation is defined as follows.
$$
\text{Corr}_{\leq D} := \sup_{f \in \mathbb{R}[Y]_{\leq D} \atop \mathbb{E}_{\mathbb{P}}[f^2] = 1} \, \mathbb{E}_{(x,Y) \sim \mathbb{P}} \left[ f(Y) \cdot x \right] 
= \sup_{f \in \mathbb{R}[Y]_{\leq D} \atop \mathbb{E}_{\mathbb{P}}[f^2] \neq 0} \, \frac{\mathbb{E}_{(x,Y) \sim \mathbb{P}}[f(Y) \cdot x]}{\sqrt{\mathbb{E}_{Y \sim \mathbb{P}}[f(Y)^2]}}.
$$

We wish to study the hardness of estimation in the planted submatrix model. 
\begin{definition}[Planted submatrix model]
    In the \emph{planted submatrix problem}, we observe the $n \times n$ matrix $Y=\lambda v v^{\top}+W$ where $\lambda \geq 0$, $v \in\{0,1\}^n$ is i.i.d. Bernoulli $(\rho)$ for some $\rho \in(0,1)$, and $W$ has entries $W_{i j}=W_{j i} \sim \mathcal{N}(0,1)$ for $i<j$ and $W_{i i} \sim \mathcal{N}(0,2)$, where $\left\{W_{i j}: i \leq j\right\}$ are independent. We assume the parameters $\lambda$ and $\rho$ are known. The goal is to estimate $x=v_1$.
\end{definition}

\paragraph{Notation}
In this session, we will use the following notation. 
    Suppose $v\in \mathbb{R}^N$ is a random variable (under $\bb P$ or $\bb Q$). For $\alpha, \beta \in \mathbb{N}^N$, define

$$
|\alpha|:=\sum_i \alpha_i, \quad \alpha!:=\prod_i \alpha_{i}!, \quad\binom{\alpha}{\beta}:=\prod_i\binom{\alpha_i}{\beta_i}, \quad \text { and } \quad X^\alpha:=\prod_i X_i^{\alpha_i}
$$

Also, define $\beta \leqslant \alpha$ to mean " $\beta_i \leqslant \alpha_i$ for all $i$ " and $\beta \leqslant \alpha$ to mean " $\beta_i \leqslant \alpha_i$ for all $i$ and for some $i$ the inequality is strict: $\beta_i<\alpha_i$."

\renewcommand{\sectionletter}{$\bm \lambda$} 
\subsection{Group \grouplambda}
Our goal is to prove the following theorem.
\begin{theorem}[Theorem~2.2 in \cite{Schramm_Lowdeg_recovery}]\label{thm:corr-bound}
In the general additive Gaussian model, 
$$
\text{Corr}^2_{\leq D} \leq \sum_{\alpha \in \mathbb{N}^N, \, 0 \leq |\alpha| \leq D} \frac{\kappa_{\alpha}^2}{\alpha!}, 
$$
where $\kappa_{\alpha}$ for $\alpha \in \mathbb{N}^N$ is defined recursively by
$$
\kappa_{\alpha} = \mathbb{E}[x X^{\alpha}] - \sum_{0 \leq \beta \lneq \alpha} \kappa_{\beta} \binom{\alpha}{\beta} \mathbb{E}[X^{\alpha - \beta}].
$$
\end{theorem}

\begin{enumerate}[(a)]
    \item Let $\hat f_{\alpha}$ be coefficients of $f$ in the Hermite expansion, i.e., $f (Y) = \sum_{|\alpha| \le D} \hat f_\alpha h_\alpha(Y)$. Prove that $\bb E x f(Y) = \ip{c, \hat f}$, where 
    $$
    c_\alpha = \bb E_X \frac{x X^\alpha}{\sqrt{\alpha!}}.
    $$
    
    \begin{hint}
        Write the Hermite expansion and use the proposition above. 
    \end{hint}
    \item Use Jensen's inequality to show that
    $$
    \bb E [f(Y)^2] \ge \bb E_Z g(Z)^2,
    $$
    where $g(Z) = \bb E_X f(X+ Z)$. Show that $\E_Z g(Z)^2 = \sum_{|\alpha| \le D}\hat g_\alpha^2 = \| \hat g\|^2$.
    \item We want to show that $\hat g = M \hat f$ for some upper-triangular matrix $M$. Find coefficients $\hat g_{\alpha}$ by writing the expansion of $g(Z)$. 

    You should obtain that 
    $$
    M_{\beta,\alpha} = \bb I[\beta \le \alpha] \bb E[X^{\alpha-\beta}]\sqrt{\frac{\beta!}{\alpha!}}\binom{\alpha}{\beta}
    $$
    for all $\alpha, \beta \in \bb N^N$.
    \item Note that $M$ is invertible. Denote by $w = (M^{-1})^\top c $ and show that 
    $$
    \text{Corr}^2_{\leq D} \le \| w\|.
    $$
    \item Find the recursion for coefficients $w_\alpha$. 
    
    \begin{hint}
        Write $c_\alpha$ in terms of $w_\alpha$ and express $w_\alpha$ recursively. 
    \end{hint}

    Denote $\kappa_\alpha =w_\alpha \sqrt{\alpha!}$ and conclude the proof.
    
\end{enumerate}

\renewcommand{\sectionletter}{$\bm \mu$} 
\subsection{Group \groupmu}

Group $\bm \lambda$ will show that $\operatorname{Corr}^{\le D}$ can be bounded using cumulants $\kappa_\alpha$ for $\alpha \in \mathbb{N}^N$, which are defined recursively by
$$
\kappa_{\alpha} = \mathbb{E}[x X^{\alpha}] - \sum_{0 \leq \beta \lneq \alpha} \kappa_{\beta} \binom{\alpha}{\beta} \mathbb{E}[X^{\alpha - \beta}].
$$
In order to get a closed-form bound on the correlation, we need to prove some properties of $\kappa_\alpha$. 

To cast planted submatrix as a special case of the additive Gaussian noise model, we take $X = (X_{ij})_{i\le j}$ defined by $X_{ij} = \lambda v_i v_j$. Note that we have removed the redundant lower-triangular part of the matrix, and decreased the noise on the diagonal from $\mathcal N (0, 2)$ to $\mathcal N (0, 1)$. This modification on the diagonal can only increase $\text{Corr}_{\le D}$ so an upper bound on $\text{Corr}_{\le D}$ in this new model implies the same upper bound on $\text{Corr}_{\le D}$ in the original model.

\begin{lemma}[Lemma~3.2 in \cite{Schramm_Lowdeg_recovery}]\label{prop:kappa-zero}
    Consider $\alpha = (\alpha_{ij})_{i\le j} \in \bb N^{n(n+1)/2}$ as a multi-graph (with self-loops allowed) on vertex set $[n]$, where $\alpha_{ij}$ represents the number of edges between $i$ and $j$. 

    If $\alpha$ has a non-empty connected component that does not contain vertex $1$, then $\kappa_\alpha=0$. (In particular, $\kappa_\alpha=0$ whenever $\alpha$ is disconnected.)

    Here, non-empty means that the connected component contains at least one edge.
\end{lemma}
\begin{enumerate}[(a)]
    \item  One way to prove this statement is by induction in $\abs{\alpha}$ with the base case $\abs{\alpha} = 1$. Show the base case.
    \item Consider $\gamma$ to be the connected component of $\alpha$ that contains node $1$ (note that if $1$ is an isolated node, $\gamma = \bf 0$). Deduce that by induction hypothesis for any $\beta$ such that $\beta \lneq \alpha$ and $\beta \nleq \gamma$, $\kappa_\beta = 0$.
    \item Show that $\binom{\alpha}{\gamma} = 1$ and $\bb E [x X^\gamma] \bb E[X^{\alpha-\gamma}] = \bb E [x X^\alpha ]$
    \item Show that for $\beta$ satisfying the same assumptions as in (b) and additionally $\beta \le \gamma$, 
    $$
    \bb E[X^{\gamma-\beta}]\bb E[X^{\alpha-\gamma}] = \bb E[X^{\alpha-\beta}]
    $$
    and $\binom{\gamma}{\beta} = \binom{\alpha}{\beta}$
    \item Show that $\kappa_{\alpha} = 0$.
    
    \begin{hint}
        Write definition of $\kappa_\alpha$ and $\kappa_\gamma$, and use that $\kappa_\beta=0$ under assumptions of (b).
    \end{hint}
\end{enumerate}

\renewcommand{\sectionletter}{$\bm \pi$} 
\subsection{Group \grouppi}
Same introduction as in group $\bm \mu$.

\begin{lemma}[Lemma~3.4 in \cite{Schramm_Lowdeg_recovery}]\label{prop:kappa-connected}
    Consider $\alpha = (\alpha_{ij})_{i\le j} \in \bb N^{n(n+1)/2}$ as a multi-graph (with self-loops allowed) on vertex set $[n]$, where $\alpha_{ij}$ represents the number of edges between $i$ and $j$. 

    Show that $\kappa_0 = \rho$. Suppose that $\alpha$ has one connected component that includes node $1$ (it may also have isolated nodes). Then for $\abs{\alpha} \ge 1$, we have
    $$
    \abs{\kappa_\alpha}\le (\abs{\alpha}+1)^{\abs{\alpha}} \lambda^{\abs{\alpha}} \rho^{\abs{V(\alpha)}}.
    $$
\end{lemma}
\begin{enumerate}[(a)]
    \item One way to prove this statement is by induction in $\abs{\alpha}$ with the base case $\abs{\alpha} = 0$. Show the base case.
    \item Show that for any multigraph $\gamma$,
    $$
    \bb E X^\gamma = \lambda^{\abs{\gamma}} \rho^{\abs{V(\gamma)}}
    $$
    and 
    $$
    \bb E x X^\gamma = \lambda^{\abs{\gamma}} \rho^{\abs{V(\gamma)\cup \{1\}}}
    $$
    \item Use triangle inequality to bound $\kappa_\alpha$ and use results from (b) and induction hypothesis.
    \item Note that $\abs{V(\alpha)} \le \abs{V(\beta)} + \abs{V(\alpha - \beta)} $ and use it to simplify the expression.
    \item At this point of proof you should get the following bound:
    $$
\kappa_\alpha \leq \lambda^{|\alpha|} \rho^{|V(\alpha)|}\left[2+\sum_{0 \neq \beta \leq \alpha}(|\beta|+1)^{|\beta|}\binom{\alpha}{\beta}\right],
$$
    
    To bound the parenthesized quantity, you may use Vandermonde's identity without proof. For integers $n_1, \dots, n_p, m$, it holds
    $$
    \binom{n_1+\cdots+n_p}{m}=\sum_{k_1+\cdots+k_p=m}\binom{n_1}{k_1}\binom{n_2}{k_2} \cdots\binom{n_p}{k_p}
    $$

    Finish the proof and conclude.
    
\end{enumerate}

\renewcommand{\sectionletter}{$\bm \gamma$} 
\subsection{Group \groupgamma}

\begin{enumerate}[(a)]
    \item Show that $\text{MMSE}_{\leq D} = \bb E [x^2 ] - \text{Corr}^2_{\leq D}$.
    \item  What $\text{MSE}$ achieves the trivial estimator $f(Y) = \bb E[v_1]$?
    \item Prove the following lemma.
    \begin{lemma}[Lemma~3.5 in \cite{Schramm_Lowdeg_recovery}]\label{lm:number-connected-multigraphs}
    For integers $d\ge 1$ and $0 \le h \le d$, the number of connected multigraphs $\alpha$ on vertex set $[n]$ such that 
    \begin{enumerate}[(i)]
        \item $\abs{\alpha} = d$,
        \item $1 \in V(\alpha)$,
        \item $\abs{V(\alpha)} = d+1-h$,
    \end{enumerate}
    is at most $(dn)^d (\frac{d}{n})^h$.
\end{lemma}

\begin{hint}
     What is the procedure for spanning the $d+1-h$ vertices and how many choices do we have to place the edges? 
   Once all vertices are spanned, how many choices do we have for placing edges among the chosen vertices?
\end{hint}

\end{enumerate}

\renewcommand{\sectionletter}{\faPuzzlePiece} 
\subsection{Jigsaw}
Our goal is to show the following result on the low-degree harndess of estimation in the planted submatrix model.
\begin{theorem}[Theorem~1.2 in \cite{Schramm_Lowdeg_recovery}]\label{thm:planted-submatrix-low-degree-hard}
    Consider the planted submatrix problem with $n\to\infty$ with parameters $\lambda = n^{-a}$ and $\rho = n^{-b}$ for fixed constants $a > 0, 0 < b < 1/2$. If $a > 1/2 - b$ then there exists some constant $\varepsilon = \varepsilon(a, b) > 0$ such that
then 
$$
\text{MMSE}_{\leq D} = \rho - (1 + o(1)) \rho^2,
$$
where $D \le n^\varepsilon$.

In particular, it implies that no degree-$n^\varepsilon$ polynomial outperforms the trivial estimator $f(Y) = \bb E[v_1]=\rho.$

\end{theorem}

\begin{enumerate}[(a)]
\item Discuss your results with your peers.
    \item Use \Cref{thm:corr-bound} to bound $\text{Corr}^2_{\leq D}$. In the first step, you may use a crude bound $1/d! \le 1$. 
    \item Apply \Cref{prop:kappa-zero} to remove all zero terms from the sum and bound the remaining terms using \Cref{prop:kappa-connected} and \Cref{lm:number-connected-multigraphs}.
    \item Show that
    $$
    \text{Corr}^2_{\leq D} \le \rho^2 \sum_{h=0}^{D} \left[ D^2(D+1)^2 \lambda^2 \right]^h \sum_{d=h}^{D} \left[ D(D+1)^2 \lambda^2 \rho^2 n \right]^{d-h}.
    $$
    \item Substitute parameter regime in the formula above and bound the finite series with infinite series. Deduce that 
    $$
    \text{Corr}^2_{\leq D} = \rho^2 ( 1+ o(1)).
    $$
    Using the bound on the MMSE, conclude the proof of \Cref{thm:planted-submatrix-low-degree-hard}.
\end{enumerate}
\printbibliography[segment=\therefsegment] 
\section{Low-degree hardness of testing between two planted distributions}
\renewcommand{\sectionletter}{1}

Before, we mostly considered the setting of hypothesis testing between two distributions where one distribution contains a signal (planted distribution) and another distribution represents "pure" noise (null distribution). In this session, we will study hypothesis testing between two planted models using the low-degree polynomials framework.

Although interesting in its own right, this setting can be applied to show low-degree hardness of the estimation problems. In particular, in the next session, we will show the low-degree hardness of counting communities in the stochastic block model using the results of this session.

The main reference for this session is \cite{Rush_LowDegPlantedvsPlanted}.

\subsection{Background}

In this session, we assume that both distributions $\bb P$ and $\bb Q$ contain the planted signal. We adjust the definition of the binary model accordingly. 
\begin{definition}\label{def:general-binary-model}
    We define the general binary model as follows. To sample $Y \sim \mathbb{P}$ (respectively $Y \sim \bb Q$), we sample $X \in \mathbb{R}^N$ from an arbitrary prior $\mathbb{P}_X$ (resp. $X \sim \bb Q_X$ supported on $X \in\left[\tau_0, \tau_1\right]^N$ with $0<\tau_0 \leqslant \tau_1<1$. We sample $Y \in\{0,1\}^N$ with entries $Y_i$ conditionally independent given $X$ and $\mathbb{E}\left[Y_i \mid X_i\right]=X_i$. 
\end{definition}

To establish hardness lower bounds, we use the low-degree likelihood ratio as before, but this time, we refer to it as the degree-$D$ advantage to emphasize that it measures the ability of low-degree polynomials to outperform random guessing.
\begin{definition}
    Define the degree-$D$ "advantage" as

$$
\operatorname{Adv}_{\leqslant D}(\mathbb{P}, \mathbb{Q}):=\sup _{f \operatorname{deg} D} \frac{\mathbb{E}_{\mathbb{P}}[f]}{\sqrt{\mathbb{E}_{\mathrm{Q}}\left[f^2\right]}}
$$

where $f$ ranges over polynomials $\mathbb{R}^N \rightarrow \mathbb{R}$ of degree at most $D$. 

\end{definition}

\begin{proposition}[General binary observation model]\label{prop:binary}
Assume $\bb P$ and $\bb Q$ follow the general binary model. Then
$$
\operatorname{Adv}_{\leqslant D}(\mathbb{P}, \mathbb{Q}) \leqslant \sqrt{\sum_{\alpha \in\{0,1\}^N,|\alpha| \leqslant D} \frac{r_\alpha^2}{\left(\tau_0\left(1-\tau_1\right)\right)^{|\alpha|}}},
$$
where $r_\alpha=r_\alpha(X) \in \mathbb{R}$ for $\alpha \in \{0, 1\}^N$ are defined recursively by

\begin{equation}\label{eq:r}
r_\alpha=\mathbb{E}_{\mathbb{P}}\left[X^\alpha\right]-\sum_{0 \leqslant \beta \leqslant \alpha} r_\beta\mathbb{E}_{\bb{Q}}\left[X^{\alpha-\beta}\right]
\end{equation}
\end{proposition}

\begin{proposition}[Proposition~2.1 in \cite{Rush_LowDegPlantedvsPlanted}, Advantage upper bound for general additive Gaussian model]\label{prop:gaussian}
Suppose $\mathbb{P}$ and $\mathbb{Q}$ take the following form: to sample $Y \sim \mathbb{P}$ (or $Y \sim \mathbb{Q}$, respectively), first sample $X \in \mathbb{R}^N$ from an arbitrary prior $\mathbb{P}_X$ (or $\bb Q_X$, resp.), then sample $Z \sim \mathcal{N}\left(0, I_N\right)$, and set $Y=X+Z$. Define $r_\alpha$ as in \eqref{eq:r}. Then

$$
\operatorname{Adv}_{\leqslant D}(\mathbb{P}, \mathbb{Q}) \leqslant \sqrt{\sum_{\alpha \in \mathbb{N}^N,|\alpha| \leqslant D} \frac{r_\alpha^2}{\alpha!}}
$$
\end{proposition}

Recall additionally the definition of strong separability. 
\begin{definition}[Strong/Weak Separation]
For a polynomial $f_n : \mathcal{S}_n \to \mathbb{R}$

\begin{itemize}
    \item we say $f$ \textbf{strongly separates} $\mathbb{P}_n$ and $\mathbb{Q}_n$ if, as $n \to \infty$,
    \[
    \frac{\sqrt{\max \left\{ \Var_{\mathbb{P}_n}[f_n], \Var_{\mathbb{Q}_n}[f_n] \right\}}}{\left| \mathbb{E}_{\mathbb{P}_n}[f_n] - \mathbb{E}_{\mathbb{Q}_n}[f_n] \right|} = o(1),
    \]
    
    \item we say $f$ \textbf{weakly separates} $\mathbb{P}$ and $\mathbb{Q}$ if, as $n \to \infty$,
    \[
    \frac{\sqrt{\max \left\{ \Var_{\mathbb{P}_n}[f_n], \Var_{\mathbb{Q}_n}[f_n] \right\}}}{\left| \mathbb{E}_{\mathbb{P}_n}[f_n] - \mathbb{E}_{\mathbb{Q}_n}[f_n] \right|}  = O(1).
    \]
\end{itemize}
\end{definition}

\renewcommand{\sectionletter}{$\bm \lambda$} 
\subsection{Group \grouplambda}

Your task is to prove \Cref{prop:binary}. 
To this end, follow the proof of Theorem 2.7 in Section 3.3 of \cite{Schramm_Lowdeg_recovery}. Make adjustments where needed for the new setting.

\begin{hint}
    Redefine $c_\alpha := \bb E_{\bb P} \tilde{X}^\alpha$. Adjust $M_{\beta\alpha}$ accordingly.  
\end{hint} 

\renewcommand{\sectionletter}{$\bm \mu$} 
\subsection{Group \groupmu}\label{sec:s6mu}
Similarly as in the last session, we will bound the advantage using recursively defined quantities $r_\alpha$, which we define as follows.

Let $r_\alpha=r_\alpha(X) \in \mathbb{R}$ for $\alpha \in \mathbb{N}^N$ be recursively defined by

\begin{equation}\label{eq:r}
r_\alpha=\mathbb{E}_{\mathbb{P}}\left[X^\alpha\right]-\sum_{0 \leqslant \beta \lneq \alpha} r_\beta\binom{\alpha}{\beta} \mathbb{E}_{\mathrm{Q}}\left[X^{\alpha-\beta}\right]
\end{equation}

\begin{enumerate}[(a)]
    \item The first task is to show properties of $r_\alpha$ under the shifts and scaling. Namely, let $\tilde{X}$ be defined by $\tilde X_{ij} = c X_{ij} + y_{ij}$ for some $c, y_{ij}\in \bb R$, $c\ne 0$ deterministic constants for all $i, j$. For fixed probability spaces $\bb P$ and $\bb Q$, $$
    r_\gamma(\tilde{X}) = c^{|\gamma|} r_\gamma(X).
    $$
    As the proof is technical, we will focus only on one part of the proof. You may use that the invariance under shifts as given, i.e., $r_\gamma(\tilde{X}) = r_\gamma(X)$ for all $\tilde{X}_{ij} = X_{ij} + y_{ij}$. 
    
    Using induction, show that for $\tilde{X}$ defined as $\tilde X_{ij} = c X_{ij}$ with $c \ge 0$, and for any $\gamma \in \bb N^N$, 
    $$
    r_\gamma(\tilde{X}) = c^{|\gamma|} r_\gamma(X).
    $$
   
    \item The idea of the proof is to relate the general binary model and the general Gaussian model. Consider the Gaussian model with the signal defined by $X_{i}^{\rm{Gaussian}} = \frac{X_{i} - \tau_0}{\sqrt{\tau_0(\tau_1-\tau_0)}}$. Let $\bb P^{\rm{binary}}$, $\bb Q^{\rm{binary}}$ follow the general binary model induced by $X$ and $\bb P^{\rm{Gaussian}}$, $\bb Q^{\rm{Gaussian}}$ follow the general additive Gaussian model induced by $X^\lambda$. 

    Suppose that one has shown that
    $$
    \operatorname{Adv}_{\leqslant D}(\mathbb{P}^{\rm{Gaussian}}, \mathbb{Q}^{\rm{Gaussian}}) \le \sqrt{\sum_{\alpha \in \mathbb{N}^N,|\alpha| \leqslant D} \frac{r_\alpha^2}{\alpha!}} = o(1) \quad \text{or } O(1).
    $$ 
    Show that it implies that
    $$
    \operatorname{Adv}_{\leqslant D}(\mathbb{P}^{\rm{binary}}, \mathbb{Q}^{\rm{binary}}) = o(1) \quad \text{or } O(1) \text{ respectively.}
    $$
\end{enumerate}

\renewcommand{\sectionletter}{$\bm \pi$} 
\subsection{Group \grouppi}
Prove the following two statements. 

Fix a sequence $D=D_n$.
    \begin{enumerate}[(a)]
        \item If $\operatorname{Adv}_{\leqslant D}(\mathbb{P}, \mathbb{Q})=O(1)$ then no degree-D test strongly separates $\mathbb{P}$ and $\mathbb{Q}$.
        \item If $\operatorname{Adv}_{\leqslant D}(\mathbb{P}, \mathbb{Q})=1+o(1)$ then no degree-D test weakly separates $\mathbb{P}$ and $\mathbb{Q}$.
    \end{enumerate}

\begin{hint}
    One way to prove it is by contradiction. Without loss of generality, you may assume that polynomial $g$ that strongly/weakly separates $\bb P$ from $\bb Q$ is shifted and scaled so that $\bb E_{\bb Q} g = 0$ and $\bb E_{\bb P} g = 1$.
\end{hint}

\renewcommand{\sectionletter}{\faPuzzlePiece} 
\subsection{Jigsaw}

\begin{enumerate}[(a)]
\item Discuss your results and share your findings with your peers.
\item Let $ \bb P , \bb Q$ follow the general binary model with $\bb P_X$, $\bb Q_X$ be the corresponding priors on the signal. Let $\mathbb{P}^{\rm{Gaussian}}, \mathbb{Q}^{\rm{Gaussian}}$ be Gaussian additive model induced by the signal with coordinates $X_{i}^{\rm{Gaussian}} = \frac{X_{i} - \tau_0}{\sqrt{\tau_0(\tau_1-\tau_0)}}$, where $X \sim \bb P_X$ or $\bb Q_X$ respectively. Use your results to conclude that if 
    $$
    \operatorname{Adv}_{\leqslant D}(\mathbb{P}^{\rm{Gaussian}}, \mathbb{Q}^{\rm{Gaussian}}) \le \sqrt{\sum_{\alpha \in \mathbb{N}^N,|\alpha| \leqslant D} \frac{r_\alpha^2}{\alpha!}} = o(1),
    $$ 
    then $\bb P$ and $\bb Q$ are not strongly separable.
\end{enumerate}

\printbibliography[segment=\therefsegment] 
\section{Counting communities in the stochastic block model}\label{s7:two-planted}
\renewcommand{\sectionletter}{1}

In this session, we consider the problem of estimating number of communities in the stochastic block model. We consider testing between two models with different number of communities and show low-degree hardness of testing between them using techniques from the last session. 

The main reference for this session is \cite{Rush_LowDegPlantedvsPlanted}.

\subsection{Background}
\begin{definition}[Definition 2.3 in \cite{Rush_LowDegPlantedvsPlanted}, Binary observation model for community detection]\label{def:community-detection-binary} Given the number of vertices $n$, total community size $k$, edge probability parameters $q, s \geq 0$, number of communities $M$, define the \emph{binary observation model} $\mathbb{P} = \mathbb{P}_{\text{Binary}}(n, k, q, s, M)$ as follows. 

Under $\bb P$, independently for each $i\in [n]$, the community label $\sigma_i$ is sampled such that
$$
\sigma_i = \begin{cases}
    \ell \quad \text{with probability } \frac{k}{nM} \text{ for each } \ell\in[M]\\
    \star \quad\text{with probability } 1 - \frac{k}{n}.
\end{cases}
$$

For each pair of vertices $i, j \in [n]$, we say that the edge is present if $Y_{ij} = 1$ and absent if $Y_{ij} = 0$. Given the community labels, for $i, j \in [n]$  with $i < j$, $Y_{ij}$ is sampled from

    \[
    Y_{ij} \sim 
    \begin{cases} 
    \text{Bernoulli} \left( q + s M \right), & \sigma_i = \sigma_j = \ell \text{ for some } \ell \in [M], \\ 
    \text{Bernoulli}(q), & \text{otherwise}.
    \end{cases}
    \]
    
    For $i > j$, the edge weight $Y_{ij}$ is defined to be $Y_{ji}$ and the diagonal entries set to zero $Y_{ii} = 0$.
    
\end{definition}

\renewcommand{\sectionletter}{$\bm \lambda$} 
\subsection{Group \grouplambda}
Define distributions $\mathbb{P} = \mathbb{P}_{\text{Binary}}(n, k, q, s, M)$ and $\mathbb{Q} = \mathbb{P}_{\text{Binary}}(n, k, q, s, M^\prime)$ for some parameters $n, k, q, s, M, M^\prime$. Write $\widetilde{M} = |M - M'|$ and $\widehat{M} = \max\{M, M'\}$.

We will show that there exists a simple $3$-degree polynomial test that strongly separates $\bb P$ and $\bb Q$. 

Define the signed triangle count test statistic $\widehat{R}$ as follows.
\[
\widehat{R} = \sum_{i < j < k} R_{ij} R_{ik} R_{jk} \quad \text{where} \quad R_{ij} = Y_{ij} - q.
\]

\begin{proposition}
    If $\widetilde{M}^{2/3} (s^2/q) k^2 / n = \omega(1)$, $\widehat{M}^{-1/3} sk = \omega(1)$ and $\widetilde{M}^2 k / \widehat{M}^2 = \omega(1)$, then $\widehat{R}$ strongly separates $\mathbb{P}$ and $\mathbb{Q}$.
\end{proposition}
Let 
$$
\Delta = \{\{ij, ik, jk\}: i, j, k \in [n], i < j < k\} 
$$
be a set of all triangles on set $[n]$. For $S = \{ij, ik, jk\}\in \Delta $, denote $R_S = R_{ij}R_{jk}R_{ki}$ and observe that $\widehat{R} = \sum_{S\in \Delta} R_S$. 

In what follows, the expectations are taken with respect to $\bb P$ (for $\bb Q$ the computations are the same).

\begin{enumerate}[(a)]
    \item Find $\bb E[R_{ij}|\sigma_i \ne \sigma_j]$ and $\bb E[R_{ij}|\sigma_i = \sigma_j=c]$ for some $c \in [ M]$.
    \item Fix $S \in \Delta$ and w.l.o.g. assume $S = \{12, 13, 23\}$. Using the law of total expectation, find $\bb E R_S$.
    \item Show that $\abs{\Delta} = \frac{1}{6}n^3\lt(1 + O\lt(\frac{1}{n}\rt)\rt)$ and show that
    \[
\mathbb{E}_{\mathbb{P}} \left[ \widehat{R} \right] = \frac{1}{6} M s^3 k^3 \left( 1 + O(n^{-1}) \right).
\] 
    \item To conclude the proof you may use the following bound on the variance of $\widehat{R}$ without proof. 
\begin{align*}
    \mathrm{Var}\left[\hat{R}\right]&\le M^{2}k^{5}s^{6}+M k^{4}s^{4}q+M^{2}k^{4}s^{5}+\frac{1}{3}n^{3}q^{3}+n k^{2}s q^{2}\\
    &\qquad+k^{3}q^{2}s+k^{3}q s^{2}+\frac{1}{3}M k^{3}s^{3}.
\end{align*}
\end{enumerate}

\renewcommand{\sectionletter}{$\bm \mu$} 
\subsection{Group \groupmu}
Similarly as in the last session, we will bound the advantage using recursively defined quantities $r_\alpha$, which we define as follows.

Let $r_\alpha=r_\alpha(X) \in \mathbb{R}$ for $\alpha \in \mathbb{N}^N$ be recursively defined by

\begin{equation}\label{eq:r}
r_\alpha=\mathbb{E}_{\mathbb{P}}\left[X^\alpha\right]-\sum_{0 \leqslant \beta \lneq \alpha} r_\beta\binom{\alpha}{\beta} \mathbb{E}_{\mathrm{Q}}\left[X^{\alpha-\beta}\right]
\end{equation}
To relate the binary community detection model to its Gaussian additive counterpart, we will need the following definition.
    \begin{definition}[Definition 2.2 in \cite{Rush_LowDegPlantedvsPlanted}, Additive Gaussian model for community detection]\label{def:cd-gaussian}
        Given the number of vertices $n$, total community size $k$, edge probability parameters $\lambda \geq 0$, number of communities $M$, we define the \emph{additive Gaussian model} for community detection $\mathbb{P} = \mathbb{P}_{\text{Gaussian}}(n, k, \lambda, M)$ as follows. We sample community labels the same way as for the binary observation model. For each pair of vertices $i, j \in[n]$ with $i \leqslant j$, the \textbf{edge weight} $Y_{i j}$ is sampled from

$$
Y_{i j} \sim \begin{cases}\mathcal{N}\left(\lambda M , 1\right), & \sigma_i=\sigma_j=\ell \text { for some } \ell \in[M] \\ \mathcal{N}(0,1), & \text { otherwise. }\end{cases}
$$
For $i>j$, the edge weight $Y_{i j}$ is defined to be $Y_{j i}$.
    \end{definition}

\begin{enumerate}[(a)]
    \item Define $X^{(q, s)}$ so that the binary observation model for community detection (\Cref{def:community-detection-binary}) reduces to the general binary model from \Cref{def:general-binary-model}. Find $\tau_0$ and $\tau_1$.
    
    \begin{hint}
        You should get $X^{(q,s)} = q + sM$ if $\sigma_i=\sigma_j \ne *$ and $X^{(q,s)} = q $, otherwise, with, e.g., $\tau_0 = q$, $\tau_1 = q + s M < 1$. 
    \end{hint}
    \item Consider the Gaussian model with $\lambda = \frac{s}{\sqrt{q(1-\tau_1)}}$ and let $X_{ij}^\lambda = \lambda M \bb I(\sigma_{i} = \sigma_j \ne \star)$. Find the relation between $X^\lambda$ and $X^{(q,s)}$. 
    \item Recall the result from the last session \Cref{sec:s6mu}, task (b). Use this result and the following theorem to find the parameters of the binary community detection model for which the detection is low-degree hard. 

    \begin{proposition}
    Let distributions $\mathbb{P}=\mathbb{P}_{\text {Gaussian }}(n, k, \lambda, M)$ and $\mathbb{Q}=\mathbb{P}_{\text {Gaussian }}\left(n, k, \lambda, M^{\prime}\right)$. 
    If $D^5 \widehat{M}^2 \lambda^2\left(k^2 / n \vee 1\right)=o(1)$, then 
    $$
    \text{Adv}_{\le D} \leqslant \sqrt{\sum\limits_{\alpha \in \mathbb{N}^N,|\alpha| \leqslant D} \frac{r_\alpha^2}{\alpha!}} \le 1 + o(1).
    $$
    \end{proposition}
\end{enumerate}

\renewcommand{\sectionletter}{$\bm \pi$} 
\subsection{Group \grouppi}
Consider a toy problem of testing between two graph models: 
$\bb P = \mathbb{P}_{\text{Binary}}(n, k, q, q, 1)$, i.e., the graph contains one community of expected size $k$, and $\mathbb{Q} = \mathbb{Q}_{\text{Binary}}(n, k, q, q, 2)$, i.e., the graph contains two communities each of expected size $k / 2$. 
Any pair of vertices from the same community are connected independently with probability $2 q$ and $3 q$ under $\mathbb{P}$ and $\mathbb{Q}$, respectively, and all the other pairs of vertices are connected independently with probability $q$ under both models. 

\begin{enumerate}[(a)]
    \item Check that the expected degrees of the nodes under the two distributions matches. 
    \item Let 
$$
\Delta = \{\{ij, ik, jk\}: i, j, k \in [n], i < j < k\} 
$$
be a set of all triangles on set $[n]$. For $S = \{ij, ik, jk\}\in \Delta $, denote $R_S = R_{ij}R_{jk}R_{ki}$ and observe that $\widehat{R} = \sum_{S\in \Delta} R_S$. 

Under this model, show that $\bb E \hat R = O(k^3 q^3)$ and $\Var \hat R \le \Theta( n^3q^3)$ (both for $\bb P$ and $\bb Q$). 
\end{enumerate}
\renewcommand{\sectionletter}{\faPuzzlePiece} 
\subsection{Jigsaw}
\begin{theorem}[Theorem~2.2 in \cite{Rush_LowDegPlantedvsPlanted}]
    Given parameters $n, k, q, s, M, M^\prime$, define distributions $\mathbb{P} = \mathbb{P}_{\text{Binary}}(n, k, q, s, M)$ and $\mathbb{Q} = \mathbb{P}_{\text{Binary}}(n, k, q, s, M^\prime)$. Assume that $q + s M\leq \tau_1$ for some constant $\tau_1 < 1$. Write $\widetilde{M} = |M - M'|$ and $\widehat{M} = \max\{M, M'\}$. We have:
\begin{itemize}
    \item If $D^5 \widetilde{M}^2 (s^2/q) (k^2/n \vee 1) = o(1)$, then no degree-$D$ test weakly separates $\mathbb{P}$ and $\mathbb{Q}$.
    \item If $\widetilde{M}^{2/3} (s^2/q) k^2 / n = \omega(1)$, $\widehat{M}^{-1/3} sk = \omega(1)$ and $\widetilde{M}^2 k / \widehat{M}^2 = \omega(1)$, then there exists a degree-3 test that strongly separates $\mathbb{P}$ and $\mathbb{Q}$.
\end{itemize}
\end{theorem}

Note that if $k^2 \ge n$, $\widehat M = O(1)$, $D \le \operatorname{polylog}(n)$ the upper and lower bound match up to log factors (in this case, we also say that the bound is \emph{tight} up to log factors).

\begin{enumerate}[(a)]
    \item Discuss your results and put them together to conclude the theorem. Assuming the low-degree conjecture, what can you say about the hardness of counting communities?
    \item To illustrate the results, consider the toy model introduced in group $\bm \pi$. Check that in this special case, if $q(k^2/n \vee 1) \le 1 / \operatorname{polylog}(n)$, no degree-$D$ test weakly separates $\bb P$ and $\bb Q$.  
    \item For the same toy model, conclude that the triangle counting algorithm consistently distinguishes $\mathbb{P}$ and $\mathbb{Q}$ if $q k^2 / n \gg 1$.
\end{enumerate}

\printbibliography[segment=\therefsegment] 
\section{Free energy of scalar Gaussian channel}\label{s8:free-energy-I}
\renewcommand{\sectionletter}{1}

With this session, we start a new chapter about statistical physics tools for studying information-theoretical and computational lower bounds. In particular, in this session, we introduce the notions of the Hamiltonian and the free energy and explore their connections to the mean square error in statistical inference problems. In this session, we closely follow introduction of \cite{miolane2018phase}.

\subsection{Background}

We consider inference problems of the form:
\begin{equation}\label{eq:scalar-channel}
\mathbf{Y}=\sqrt{\lambda} \mathbf{X}+\mathbf{Z}
\end{equation}
where the signal $\mathbf{X} \sim P_X$, where $P_X$ is a probability distribution over $\mathbb{R}^n$, and the noise $\mathbf{Z}=\left(Z_1, \ldots, Z_n\right) \stackrel{\text { i.i.d. }}{\sim} \mathcal{N}(0,1)$ is independent from $\mathbf{X}$. The parameter $\lambda \geq 0$ is a signal-to-noise ratio. Assume that $\mathbb{E}\|\mathbf{X}\|^2<\infty$.

\begin{definition}
    Let $\hat{\theta}=\hat{\theta}(Y)$ be a measurable function of the observations~$\bf Y$, which we will refer to as an estimator. We define its \emph{Mean Squared Error} as:
$$
\operatorname{MSE}(\widehat{\theta})=\mathbb{E}\left[\|\mathbf{X}-\widehat{\theta}(\mathbf{Y})\|^2\right]
$$
We define also Minimum Mean Squared Error as
$$
\operatorname{MMSE}(\lambda)=\min _{\widehat{\theta}} \operatorname{MSE}(\widehat{\theta})=\mathbb{E}\left[\|\mathbf{X}-\mathbb{E}[\mathbf{X} \mid \mathbf{Y}]\|^2\right],
$$
where the minimum is taken over all measurable function $\hat{\theta}$ of the observations $\bf Y$. 
\end{definition}

By Bayes rule, the posterior distribution of $\mathbf{X}$ given $\mathbf{Y}$ is

\begin{equation}\label{eq:posterior}
d P(\mathbf{x} \mid \mathbf{Y})=\frac{1}{\mathcal{Z}(\lambda, \mathbf{Y})} e^{H_\lambda, \mathbf{Y}(\mathbf{x})} d P_X(\mathbf{x})
\end{equation}

where

\begin{equation}
H_{\lambda, \mathbf{Y}}(\mathbf{x})=\sqrt{\lambda} \mathbf{x}^{\top} \mathbf{Y}-\frac{\lambda}{2}\|\mathbf{x}\|^2=\sqrt{\lambda} \mathbf{x}^{\top} \mathbf{Z}+\lambda \mathbf{x}^{\top} \mathbf{X}-\frac{\lambda}{2}\|\mathbf{x}\|^2
\end{equation}

\begin{definition}
    $H_{\lambda, \mathbf{Y}}$ is called the \emph{Hamiltonian} and the normalizing constant
$$
\mathcal{Z}(\lambda, \mathbf{Y})=\int d P_X(\mathbf{x}) e^{H_{\lambda, \mathbf{Y}}(\mathbf{x})}
$$

is called the \emph{partition function}.
\end{definition}

We will denote the expectations with respect to the posterior distribution \eqref{eq:posterior} by the Gibbs brackets~$\langle\cdot\rangle_\lambda:$

$$
\langle f(\mathbf{x})\rangle_\lambda=\mathbb{E}[f(\mathbf{X}) \mid \mathbf{Y}]=\frac{1}{\mathcal{Z}(\lambda, \mathbf{Y})} \int d P_X(\mathbf{x}) f(\mathbf{x}) e^{H_{\lambda, \mathbf{Y}}(\mathbf{x})}
$$

for any measurable function $f$ such that $f(\mathbf{X})$ is integrable.

\begin{definition}
    $F(\lambda)=\mathbb{E} \log \mathcal{Z}(\lambda, \mathbf{Y})$ is called the \emph{free energy}. 
\end{definition}
    
We will see that the free energy is closely related to the mutual information between $\bf X$ and $\bf Y$, which is defined below.
\begin{definition}
    The mutual information $I(\mathbf{X} ; \mathbf{Y})$ is defined as the Kullback-Leibler divergence between $P_{(X, Y)}$, the joint distribution of $(\mathbf{X}, \mathbf{Y})$ and $P_X \otimes P_Y$ the product of the marginal distributions of $\mathbf{X}$ and $\mathbf{Y}$:
$$
I(\mathbf{X} ; \mathbf{Y})  =\mathbb{E} \log \left(\frac{d P_{(X, Y)}}{d P_X \otimes P_Y}(\mathbf{X}, \mathbf{Y})\right)
$$
\end{definition}
A useful property that we will prove next time is that we can reduce the computation of the MMSE to the computation of the free energy. 
\begin{equation}\label{eq:free-energy-mmse}
F^{\prime}(\lambda)=\frac{1}{2}\left(\mathbb{E}\|\mathbf{X}\|^2-\operatorname{MMSE}(\lambda)\right).
\end{equation}

\renewcommand{\sectionletter}{$\bm \lambda$} 
\subsection{Group \grouplambda}
We will prove several properties of MSE and MMSE. 

\begin{enumerate}[(a)]

    \item $\operatorname{MMSE}(0)=\mathbb{E}\|\mathbf{X}-\mathbb{E}[\mathbf{X}]\|^2$,
    \item $\operatorname{MMSE}(\lambda) \xrightarrow[\lambda \rightarrow+\infty]{ } 0$.

    \begin{hint}
        Consider estimator $\hat{\bf x} = \frac{1}{\sqrt \lambda} \bf Y$.
    \end{hint}
  \item $\lambda \mapsto \operatorname{MMSE}(\lambda)$ is non-increasing over $\mathbb{R}_{+}$. 

    To prove this statement, consider the following observations. Let $0<\lambda_2 \leq \lambda_1$. Define $\Delta_1=\lambda_1^{-1}, \Delta_2=\lambda_2^{-1}$ and

$$
\left\{\begin{array}{l}
\mathbf{Y}_1=\mathbf{X}+\sqrt{\Delta_1} \mathbf{Z}_1 \\
\mathbf{Y}_2=\mathbf{X}+\sqrt{\Delta_1} \mathbf{Z}_1+\sqrt{\Delta_2-\Delta_1} \mathbf{Z}_2
\end{array}\right.
$$
where $\mathbf{X} \sim P_X$ is independent from $\mathbf{Z}_1, \mathbf{Z}_2 \stackrel{\text { iid. }}{\sim} \mathcal{N}\left(0, \operatorname{Id}_n\right)$.

Show that $\operatorname{MMSE}(\lambda_1) = \bb E \norm{\bf X - \bb E [\bf X \mid \bf Y_1]}^2$ and $\operatorname{MMSE}(\lambda_2) = \bb E \norm{\bf X - \bb E [\bf X \mid \bf Y_2]}^2$ and use independence of $(\bf X, \bf Y_1)$ and $\bf Z_2$.
 
    \item[(d*)] Optional: Prove that the function $\theta_{\text{MMSE}} = \bb E [\bf X \mid \bf Y]$ achieves Minimum Mean Squared error. 
\end{enumerate}

\renewcommand{\sectionletter}{$\bm \mu$} 
\subsection{Group \groupmu}

\begin{enumerate}[(a)]
    \item Show that the free energy is related to the mutual information between $\mathbf{X}$ and $\mathbf{Y}$ by

$$
F(\lambda)=\frac{\lambda}{2} \mathbb{E}\|\mathbf{X}\|^2-I(\mathbf{X} ; \mathbf{Y})
$$
\item Using the Nishimori identity, show that $$
\operatorname{MMSE}(\lambda)=\mathbb{E}\|\mathbf{X}\|^2-\mathbb{E}\left\langle\mathbf{x}^{\top} \mathbf{X}\right\rangle_\lambda
$$
\begin{hint}
    Use the replica trick: $\bb E \|\ip{\bf x}_\lambda\|^2 = \bb E \ip{\lt(\bf x^{(1)}\rt)^\top x^{(2)}}_\lambda$
\end{hint}
\item Using \eqref{eq:free-energy-mmse}, deduce I-MMSE relation
$$
\frac{\partial}{\partial \lambda}I(\bf{X}; \bf{Y}) = \frac{1}{2}\operatorname{MMSE}(\lambda).
$$
\end{enumerate}

\renewcommand{\sectionletter}{$\bm \pi$} 
\subsection{Group \grouppi}
We will consider the Gaussian prior $P_0 = \mathcal{N}(0,1)$ and $n=1$ (i.e. the scalar Gaussian channel).

\begin{enumerate}[(a)]
    \item Show that 
    $$
    \bb E[X | Y] = \frac{\sqrt{\lambda}}{1 + \lambda }Y.
    $$
    You may use the following property of scalar Gaussian variables without proof. Let $G_1 \sim \mathcal{N}(0, \sigma_1^2)$ and $G_2 \sim \mathcal{N}(0, \sigma_2^2)$. Then conditioned on $G_1+G_2 = z$, $G_1$ is distributed normally with mean $z \frac{\sigma_1^2}{\sigma_1^2+\sigma^2_2}$ and variance $\frac{\sigma_1^2\sigma_2^2}{\sigma_1^2 + \sigma_2^2}$. 
    \item Deduce that $\operatorname{MMSE} = \frac{1}{1 + \lambda}$. Draw a sketch of MMSE depending on $\lambda$.
 
\end{enumerate}
We will now prove that the Gaussian prior is the worst-case prior for the scalar channel in \eqref{eq:scalar-channel}.
\begin{enumerate}[(a)]
\setcounter{enumi}{2}
    \item Let $P_X$ be a probability distribution on $\bb R$ with unit second moment $\bb E_{P_X} X^2 = 1$. Prove that $\operatorname{MMSE}_{P_X}(\lambda) \le \frac{1}{1+\lambda}$. Conclude that
    $$
    \sup_{P_X} \operatorname{MMSE}_{P_X}(\lambda) = \frac{1}{1+\lambda}.
    $$

    \begin{hint}
        Consider estimator $\hat{x}(Y) = \frac{\sqrt{\lambda}}{1 + \lambda }Y$.
    \end{hint}
\end{enumerate}

\renewcommand{\sectionletter}{\faPuzzlePiece} 
\subsection{Jigsaw}

\begin{enumerate}[(a)]
\item Share results of your group and discuss them with your peers.
    \item Using your results, prove that $F(\lambda)$ is non-decreasing and $\frac 1 2 \bb E \|X\|^2$-Lipschitz function over $\bb R_{\ge 0}$. 
    \item For the Gaussian scalar channel (see group $\bm \pi$), find the mutual information $I(X;Y)$ and the free energy $F(\lambda)$.
\end{enumerate}

\printbibliography[segment=\therefsegment] 
\section{Free energy derivative and needle in a haystack}
\renewcommand{\sectionletter}{1}

In this session, we will show how to reduce the computation of the minimal mean square error to the derivative of the free energy and illustrate application of these theoretical tools on a toy inference problem. Main reference of this session is \cite{miolane2018phase}.

\subsection{Background}
We consider the same inference problem as before:
\begin{equation}\label{eq:scalar-channel}
\mathbf{Y}=\sqrt{\lambda} \mathbf{X}+\mathbf{Z}
\end{equation}
where the signal $\mathbf{X} \sim P_X$, where $P_X$ is a probability distribution over $\mathbb{R}^n$, and the noise $\mathbf{Z}=\left(Z_1, \ldots, Z_n\right) \stackrel{\text { i.i.d. }}{\sim} \mathcal{N}(0,1)$ is independent from $\mathbf{X}$. The parameter $\lambda \geq 0$ is a signal-to-noise ratio. We assume that $\mathbb{E}\|\mathbf{X}\|^2<\infty$.

\begin{definition}
    $F(\lambda)=\mathbb{E} \log \mathcal{Z}(\lambda, \mathbf{Y})$ is called the \emph{free energy}, where 
$$
\mathcal{Z}(\lambda, \mathbf{Y})=\int d P_X(\mathbf{x}) e^{H_{\lambda, \mathbf{Y}}(\mathbf{x})}
$$
is the normalizing constant of the Hamiltonian $H_{\lambda, \mathbf{Y}}(\mathbf{x})=\sqrt{\lambda} \mathbf{x}^{\top} \mathbf{Y}-\frac{\lambda}{2}\|\mathbf{x}\|^2=\sqrt{\lambda} \mathbf{x}^{\top} \mathbf{Z}+\lambda \mathbf{x}^{\top} \mathbf{X}-\frac{\lambda}{2}\|\mathbf{x}\|^2$.
\end{definition}

Recall from the last session that we denote by Gibbs brackets $\langle \cdot \rangle$ the expectation with respect to the posterior $\mathbb{P}(\mathbf{X}=\cdot \mid \mathbf{Y})$. The usual notation $\bb E$ is the expectation with respect to $(\mathbf{X}, \mathbf{Y})$.

\begin{proposition}[Nishimori identity]
Let $k \geq 1$ and let $\mathbf{x}^{(1)}, \ldots, \mathbf{x}^{(k)}$ be $k$ be sampled from the distribution $\mathbb{P}(\mathbf{X}=\cdot \mid \mathbf{Y})$ independently given $\bf Y$. Then, for all continuous bounded functions $f$ it holds
$$
\mathbb{E}\left\langle f\left(\mathbf{Y}, \mathbf{x}^{(1)}, \ldots, \mathbf{x}^{(k)}\right)\right\rangle=\mathbb{E}\left\langle f\left(\mathbf{Y}, \mathbf{x}^{(1)}, \ldots, \mathbf{x}^{(k-1)}, \mathbf{X}\right)\right\rangle
$$
\end{proposition}

\renewcommand{\sectionletter}{$\bm \lambda$} 
\subsection{Group \grouplambda}
The goal of the exercises is to prove that we can reduce the computation of the MMSE to the computation of the free energy, the formula we saw in the last session: 
\begin{equation}\label{eq:free-energy-mmse}
F^{\prime}(\lambda)=\frac{1}{2}\left(\mathbb{E}\|\mathbf{X}\|^2-\operatorname{MMSE}(\lambda)\right).
\end{equation}
This will be useful because the free energy $F$ is much easier to handle than the MMSE. 

\begin{enumerate}[(a)]
    \item For $\lambda > 0$, derive $\frac{\partial}{\partial \lambda} \log \mathcal Z(\lambda, \bf Y)$
    
    \begin{hint}
        The final expression should be  $\frac{\partial}{\partial \lambda} \log \mathcal Z(\lambda, \bf Y) = \ip{\frac{1}{2 \sqrt{\lambda}} \mathbf{x}^{\top} \mathbf{Z}+\mathbf{x}^{\top} \mathbf{X}-\frac{1}{2}\|\mathbf{x}\|^2}_\lambda$
    \end{hint}
    \item 
    Show that 
    $$
F\left(\lambda_2\right)-F\left(\lambda_1\right)=\int_{\lambda_1}^{\lambda_2} \mathbb{E}\left\langle\frac{1}{2 \sqrt{\lambda}} \mathbf{x}^{\top} \mathbf{Z}+\mathbf{x}^{\top} \mathbf{X}-\frac{1}{2}\|\mathbf{x}\|^2\right\rangle_\lambda d \lambda .
$$
You may assume that the right hand side of the expression of $\frac{\partial}{\partial\lambda} \log Z(\lambda, \bm Y)$ is integrable (it is so, because $\bb E \|X\|^2 < \infty$).
 \item  You may use the following fact without proof: $\bb E \bf \ip{x^\top \bf Z}_\lambda  = \sqrt{\lambda} \bb E [ \ip{\|\bf x\|^2}_\lambda - \ip{\bf x^\top X}_\lambda ]$. Conclude that for all $\lambda \geq 0$,
\begin{equation}\label{eq:free-energy-diff}
F(\lambda)-F(0)=\frac{1}{2} \int_0^\lambda \mathbb{E}\left\langle\mathbf{x}^{\top} \mathbf{X}\right\rangle_\gamma d \gamma.
\end{equation}
    
    You may also use the fact that $F$ is continuous at $\lambda = 0$ without proof.
 \item To derive \eqref{eq:free-energy-mmse}, use the following statement, that we proved in the last session:
    $$
\operatorname{MMSE}(\lambda)=\mathbb{E}\|\mathbf{X}\|^2-\mathbb{E}\left\langle\mathbf{x}^{\top} \mathbf{X}\right\rangle_\lambda.
$$

\end{enumerate}

\renewcommand{\sectionletter}{$\bm \mu$} 
\subsection{Group \groupmu}

\begin{enumerate}[(a)]
    \item  To compute the integral, prove that for each $i \in [n]$ and $\lambda > 0$,
    $$\bb E Z_i \ip{x_i}_\lambda = \sqrt{\lambda} \mathbb{E}\left[\left\langle x_i^2\right\rangle_\lambda-\left\langle x_i X_i\right\rangle_\lambda\right].$$ 
    
     \begin{hint}
        1. Gaussian integration by parts states that for a normally distributed normal variable $Z$, it holds $\mathbb{E}[Z g(Z)] = \mathbb{E}[g'(Z)]$.

        2. Apply Nishimori identity to $\ip{x_i}_\lambda^2$.
    \end{hint}

    \item Conclude that $\bb E \bf \ip{x^\top \bf Z}_\lambda  = \sqrt{\lambda} \bb E [ \ip{\|\bf x\|^2}_\lambda - \ip{\bf x^\top X}_\lambda ]$.
\end{enumerate}

\renewcommand{\sectionletter}{$\bm \pi$} 
\subsection{Group \grouppi}

In order to illustrate the results from your groups, we consider now a very simple inference model, which we will refer to as 'needle in a haystack' problem. Let $\left(e_1, \ldots, e_{2^n}\right)$ be the canonical basis of $\mathbb{R}^{2^n}$. Let $\sigma_0 \sim \operatorname{Unif}\left(\left\{1, \ldots, 2^n\right\}\right)$ and define $\mathbf{X}=e_{\sigma_0}$ (i.e. $\mathbf{X}$ is chosen uniformly over the canonical basis of $\mathbb{R}^{2^n}$ ). Suppose here that we observe:

$$
\mathbf{Y}=\sqrt{\lambda n} \mathbf{X}+\mathbf{Z}
$$

where $\mathbf{Z}=\left(Z_1, \ldots, Z_{2^n}\right) \stackrel{\text { Li.d. }}{\sim} \mathcal{N}(0,1)$, independently from $\sigma_0$. The goal here is to estimate $\mathbf{X}$ or equivalently to find $\sigma_0$. 

\begin{enumerate}[(a)]
    \item Write the posterior distribution and the partition function for this model.
     \item We will be interested in computing the free energy $F_n(\lambda)=\frac{1}{n} \mathbb{E} \log \mathcal{Z}_n(\lambda)$ in order to deduce then the minimal mean squared error using the I-MMSE relation. Use Jensen's inequality to show that
    $$
    F_n(\lambda) \leq \frac{1}{n} \mathbb{E} \log \left(1-\frac{1}{2^n}+e^{\sqrt{\lambda n} Z_{\sigma_0}+\frac{\lambda n}{2}-\log (2) n}\right)
    $$
    \item Denote $\mu = \lambda n / 2 - n \log 2$. Show that 
    $$
    \bb E\lt[ \log(1 + e^{\mu + \sqrt{\lambda n}Z_{\sigma_0}}  ) \rt]\le \log(1 + e^\mu) + \bb E \abs{\sqrt{\lambda n}Z_{\sigma_0}}
    $$
    \begin{hint}
        Show that $f(x) = \log(1 + e^x)$ is 1-Lipschitz continuous. 
    \end{hint} 
    \item You may use the fact that for a gaussian variable $g$, $\bb E |g| = \sqrt{2/\pi}$ without proof. Show that 
    $$
    F_n(\lambda) \le \frac{1}{n}\log\lt(1 + e^{\frac{\lambda n}{2}-n\log 2 }\rt) + \sqrt{\frac{\lambda}{n}}.
    $$
    \item Show that 
    $$
    \frac{1}{n}\log\lt(1 + e^{\frac{\lambda n}{2}-n\log 2 }\rt) + \sqrt{\frac{\lambda}{n}} \to_{n\to\infty} \begin{cases}
        0  \quad &\text{if } \lambda \le 2 \log (2)\\
        \lambda / 2 - \log(2) \quad &\text{otherwise }
    \end{cases}
    $$
\end{enumerate}

\renewcommand{\sectionletter}{$\bm \gamma$} 
\subsection{Group \groupgamma}
We consider the same model as in group $\bm \pi$. 
\begin{enumerate}[(a)]
\item Write the posterior distribution and the partition function for this model.
\item Show that $F_n(0) = 0$ and $F_n$ is non-decreasing. Conclude that $F_n$ is therefore non-negative. 
\item Using the result of group $\bm \pi$ that 
$$
    F_n(\lambda) \le \frac{1}{n}\log\lt(1 + e^{\frac{\lambda n}{2}-n\log 2 }\rt) + \sqrt{\frac{\lambda}{n}}.
    $$
and that $\frac{1}{n}\log\lt(1 + e^{\frac{\lambda n}{2}-n\log 2 }\rt) + \sqrt{\frac{\lambda}{n}} \to 0$ when $\lambda \le 2 \log 2$ as $n \to \infty$, conclude that $\lim_{n\to\infty} F_n(\lambda) = 0$ for all $\lambda \in [0, 2 \log(2)]$.
    \item Show that 
    $$
    F_n(\lambda) \geq \frac{1}{n} \mathbb{E} \log \left(\frac{e^{\sqrt{\lambda n} Z_{\sigma_0}+\frac{\lambda n}{2}}}{2^n}\right)
    $$
    and deduce that $\lim_{n\to\infty} F_n(\lambda) = \lambda / 2 - \log(2)$ for $\lambda \ge 2\log(2)$.

\end{enumerate}

\renewcommand{\sectionletter}{\faPuzzlePiece} 
\subsection{Jigsaw}

\begin{enumerate}[(a)] 
\item Groups $\bm \lambda$ and $\bm \mu$, combine your results to finish the proof of the identity \eqref{eq:free-energy-mmse} and explain it to groups $\bm \pi$ and $\bm \gamma$.
\item Groups $\bm \pi$ and $\bm \gamma$, introduce the 'needle in a haystack' problem to your peers and combine your results to show the following fact. 

    For all $\lambda>0$,
    \begin{itemize}
        \item if $\lambda<2 \log (2)$, then $\operatorname{MMSE}_n(\lambda) \underset{n \to \infty}{\longrightarrow} 1$ : one can not recover $\mathbf{X}$ better than a random guess.
        \item if $\lambda>2 \log (2)$, then $\operatorname{MMSE}_n(\lambda) \underset{n \to \infty}{\longrightarrow} 0$ : one can recover $\mathbf{X}$ exactly with high probability. In particular, in this regime, one can use the maximum likelihood estimator to recover the signal. 
    \end{itemize} 

     To this end, you may use the following fact without proof. If $f_n(t)$ is a sequence of convex functions of $t$, its derivative converges to the derivative of its limit at each $t$ at which the limit is differentiable. 
\end{enumerate}

\begin{remark}
We can obtain the same result by analyzing the maximum likelihood estimator

$$
\widehat{\sigma}(\mathbf{Y})=\underset{1 \leq \sigma \leq 2^n}{\arg } \max Y_\sigma
$$
We have $\max _\sigma Z_\sigma \simeq \sqrt{2 \log (2) n}$ with high probability so that the maximum likelihood estimator recovers perfectly the signal for $\lambda>2 \log (2)$ with high probability.
\end{remark}

\printbibliography[segment=\therefsegment] 
\section{Free energy for rank-one matrix estimation}
\renewcommand{\sectionletter}{1}

For more sophisticated statistical inference problems, the computation of the free energy may become significantly more complicated. However, in many cases it is possible to reduce the computations to the free energy of a related simpler problem. In this session, we explore one such scenario in which the computation of the free energy for the rank-one matrix estimation problem can be reduced to that of a scalar Gaussian channel. The main references are \cite{dominguez2024Statistical,miolane2018phase}.

\subsection{Background}

We consider the spiked Wigner model. The observations $\bm Y \in \bb R^{n\times n}$ are as follows:

$$
Y_{i, j}=\sqrt{\frac{\lambda}{2n}} X_i X_j+Z_{i, j}\quad 1 \le i, j \le n,
$$

where $X_i {\sim} P_0$ and $Z_{i, j} {\sim} \mathcal{N}(0,1)$
are independent random variables. 

Assume that $\bb E_{P_0} X^2 < \infty$. Equivalently, we can write $\bm Y =  \sqrt{\frac{\lambda}{2n}} \bm X \bm X^\top + \bm Z$.

Our main quantity of interest is the Minimum Mean Squared Error (MMSE) defined as:
$$
\begin{aligned} \operatorname{MMSE}_n(\lambda) & =\min _{\widehat{\theta}} \frac{1}{n^2} \sum_{1 \leq i,j \leq n} \mathbb{E}\left[\left(X_i X_j-\widehat{\theta}_{i, j}(\mathbf{Y})\right)^2\right] \\ & =\frac{1}{n^2} \sum_{1 \leq i,j \leq n} \mathbb{E}\left[\left(X_i X_j-\mathbb{E}\left[X_i X_j \mid \mathbf{Y}\right]\right)^2\right].\end{aligned}$$

Similarly as before, we can write the posterior distribution of $\bf X$ given $\bf Y$ as 
$$
d P(\mathbf{x} \mid \mathbf{Y})=\frac{1}{\mathcal{Z}_n(\lambda)} d P_0^{\otimes n}(\mathbf{x}) e^{H_n(\mathbf{x})},
$$
where 
\begin{align*}
H_n(\mathbf{x})&=\sum_{i,j} \sqrt{\frac{\lambda}{n}} x_i x_j Z_{i, j}+\frac{\lambda}{n} X_i X_j x_i x_j-\frac{\lambda}{2 n} x_i^2 x_j^2 \\
&= \sqrt{\frac{\lambda}{n}} \bm X^\top \bm Z \bm X + \frac{\lambda}{n} (\bm X^\top \bm x)^2 - \frac{\lambda}{2 n}\|\bm X\|^4
\end{align*}
is the random Hamiltonian and $\mathcal{Z}_n(\lambda)$ is the normalizing constant.

The free energy is defined as
$$
F_n(\lambda)=\frac{1}{n} \mathbb{E}\left[\log \int d P_0^{\otimes n}(\mathbf{x}) e^{H_n(\mathbf{x})}\right]=\frac{1}{n} \mathbb{E} \log \mathcal{Z}_n(\lambda) .
$$

Our goal is to find the limit of MMSE by finding the limit of the free energy and using the I-MMSE relation. We express the limit of $F_n$ using the following function
\begin{equation}\label{eq:free-energy-limit-F}
\mathcal{F}:(\lambda, q) \mapsto \mathbb{E} \log \left(\int d P_0(x) \exp \left(\sqrt{\lambda q} Z x+\lambda q x X-\frac{\lambda}{2} q x^2\right)\right)-\frac{\lambda}{4} q^2 =: \psi_{P_0}(\lambda q)-\frac{\lambda}{4} q^2,
\end{equation}
 where $\psi_{P_0}(\gamma)$ is the free energy of the scalar Gaussian channel $Y = \sqrt{\gamma}X + Z$, where $X, Z\in \bb R$ and $X \sim P_0, Z\sim \mathcal{N}(0,1) $ are independent random variables.

\renewcommand{\sectionletter}{$\bm \lambda$} 
\subsection{Group \grouplambda}

\paragraph{BBP transition}
Let $\mathbf {Y}^\text{asym} = \sqrt{\frac{\lambda}{2n}}\bf X\bf X^\top + Z$, where $\bf Z $ is a matrix with independent standard Gaussian entries.

Denote the top eigenvector of the matrix $\bm Y^\text{asym}+(\bm Y^\text{asym})^\top$ by $v \in \mathbb{R}^N$. With probability one, the vector $v$ is well-defined up to multiplication by a scalar. One can show (see \cite{BBP}) the following convergence in probability:

$$
\frac{|\bm X^\top v|^2}{|\bm{X}|^2|v|^2} \xrightarrow[n \rightarrow \infty]{\text { (prob.) }} \begin{cases}
    0 \quad &\text{ if } \lambda \le 1 \\
    1 - \frac{1}{\lambda}\quad &\text{ if } \lambda > 1.
\end{cases}
$$
In the following exercises, you may use this statement as given.

\begin{enumerate}[(a)]
\item Prove an upper bound on MMSE, 
    $$
    \operatorname{MMSE}_n \le \frac{1}{n^2}\bb E\|\bm X\|^4 - \lt (\frac{(\bb E X^2)^2}{n} + \frac{n-1}{n}(\bb E X)^4\rt).
    $$

    Observe that in the limit, 
    $$
    \lim_{n\to\infty} \operatorname{MMSE}_n \le \left(\mathbb{E} X^2\right)^2 - (\bb E X)^4  =: \operatorname{DMSE}
    $$
    To see that, you may use the following fact without proof (try to prove it if you have extra time!):
    \begin{equation}\label{eq:asymptotic_var}
\frac{1}{n^2} \mathbb{E}\|\bm X\|^4=\left(\mathbb{E} X^2\right)^2+o(1) .
\end{equation}
    \item Assume that $\bb E_{P_0} X^2 = 1$. Show that PCA estimator (top eigenvector of $\bm Y + \bm Y^\top$) achieves 
    $$
    \operatorname{MSE}_{\operatorname{PCA} }(\lambda) \to \begin{cases}
        1 \quad &\text{ if } \lambda \le 1 \\
        \frac{1}{\lambda}(2 - \frac{1}{\lambda})\quad &\text{ otherwise}
    \end{cases}
    $$
    Here, you may use \eqref{eq:asymptotic_var} as well. 
\end{enumerate}

\renewcommand{\sectionletter}{$\bm \mu$} 
\subsection{Group \groupmu}
\begin{enumerate}[(a)]
  \item Check that $\psi_{P_0}(\gamma)$ is the free energy of the scalar channel
    $$
    Y = \sqrt{\gamma} X + Z,
    $$
    where $X, Z \in \bb R$ and $X \sim P_0, Z \sim \mathcal{N}(0, 1)$ are independent. 
    
    You may use the results from the previous session. 
\item 

Fix $\lambda > 0$ and denote $q^\star$ a maximizer of $\mathcal F(\lambda, \cdot)$. Show that 
\begin{equation*}
\frac{\partial}{\partial \lambda} \mathcal{F}\left(\lambda, q^*\right)=\frac{\left(q^*\right)^2}{4} .
\end{equation*}

\begin{hint}
    Write $\frac{\partial}{\partial q} \mathcal{F}$ and note that the optimality condition is $q^* = 2\psi_{P_0}^\prime (\lambda q^*)$.
\end{hint}

\item Denote
\begin{equation}
D=\left\{\lambda>0 \mid \mathcal{F}(\lambda, \cdot) \text { has a unique maximizer } q^*(\lambda)\right\} .
\end{equation}

By the I-MMSE relation (adapted to our model) 
$$
F_n^{\prime}(\lambda)=\frac{1}{4}\left(\frac{1}{n} \bb E X^4 + \frac{n-1}{n}(\mathbb{E}X^2)^2- {\operatorname{MMSE}}_n (\lambda)\right). 
$$ 
For now, you may use without proof that $F'_n(\lambda) \to \frac{(q^*)^2}{4}$ for all $\lambda \in D$ (in order to compute the limit, we need to swap limit, derivative and supremum. Group $\bm \pi$ is showing the details).

With these facts, deduce that for all  $\lambda \in D$,
$$
\operatorname{MMSE}_n(\lambda) \underset{n \rightarrow \infty}{\longrightarrow}\left(\mathbb{E}_{P_0} X^2\right)^2-q^*(\lambda)^2.
$$
\end{enumerate}

\renewcommand{\sectionletter}{$\bm \pi$} 
\subsection{Group \grouppi}
We will use the following result: 
\begin{theorem}[Replica-Symmetric formula for the spiked Wigner model]
    For all $\lambda>0$,

$$
F_n(\lambda) \xrightarrow[n \rightarrow \infty]{} \sup _{q \geq 0} \mathcal{F}(\lambda, q) .
$$
\end{theorem}

In order to understand the asymptotic behavior of $\text{MMSE}_n(\lambda)$, we take the limit as $n \to \infty$. Therefore, we need to evaluate $\lim_{n \to \infty} F'_n(\lambda)$. We know that $F_n(\lambda) \to \sup_{q \geq 0} \mathcal{F}(\lambda, q)$, and group $\bm \mu $ is computing $\frac{\partial}{\partial \lambda} \mathcal{F}(\lambda, q^*)$. Our goal is to swap the limit, supremum, and derivative. We will use the following two results.

\begin{proposition}\label{prop:derivative-supremum}
    Let $I \subset \mathbb{R}$ be an interval, and let $\left(f_n\right)_{n \geq 0}$ be a sequence of convex functions on $I$ that converges pointwise to a function $f$. Then for all $t \in I$ for which these inequalities have a sense, it holds
$$
f^{\prime}\left(t^{-}\right) \leq \liminf _{n \rightarrow \infty} f_n^{\prime}\left(t^{-}\right) \leq \limsup _{n \rightarrow \infty} f_n^{\prime}\left(t^{+}\right) \leq f^{\prime}\left(t^{+}\right) .
$$
\end{proposition}

\begin{theorem}[Envelope]\label{thm:envelope} Let $g \in C\left(\mathbb{R}^d \times \mathbb{R}^d ; \mathbb{R}\right)$ be a continuous function that is continuously differentiable in its first variable, let $f: \mathbb{R}^d \rightarrow \mathbb{R}$ be its envelope $f(x):=\sup _{y \in \mathbb{R}^d} g(x, y)$, and let $\mathcal{O}_x$ be defined as
\begin{equation}\label{eq:optimizer}
\mathcal{O}_x:=\left\{y \in \mathbb{R}^d \mid f(x)=g(x, y)\right\} .
\end{equation}
We fix $x \in \mathbb{R}^d$, and assume that there exists a compact set $K$ which contains $\mathcal{O}_{x^{\prime}} \neq \varnothing$ for every $x^{\prime}$ in a small enough neighbourhood of $x$. 
Then the envelope function $f: \mathbb{R}^d \rightarrow \mathbb{R}$ is differentiable at $x \in \mathbb{R}^d$ if and only if the set

$$
\mathcal{D}_x:=\left\{\nabla_x g(x, y) \mid y \in \mathcal{O}_x\right\}
$$

is a singleton. In this case, for any $y \in \mathcal{O}_x$, we have $\nabla f(x)=\nabla_x g(x, y)$.

\end{theorem}

In words, this theorem states that the function $f(x)=\sup _{y \in \mathbb{R}^d} g(x, y)$ is differentiable a point $x \in \mathbb{R}^d$ if and only if the function $y \mapsto \nabla_x g(x, y)$ is constant on the set of optimizers $\mathcal{O}_x$.
\begin{enumerate}[(a)]
    \item Recall that $\operatorname{MMSE}$ is a non-increasing function (see exercise $\bm \lambda$(a) from the last session).

    Show that $F(\lambda)$ is convex.

    \begin{hint}
        A univariate differentiable function is convex on an interval if and only if its derivative is monotonically non-decreasing on that interval.
    \end{hint}
    \item Using \Cref{prop:derivative-supremum}, show that $F^\prime_n(\lambda)\to f^\prime(\lambda)$.
    \item Denote
\begin{equation}
D=\left\{\lambda>0 \mid \mathcal{F}(\lambda, \cdot) \text { has a unique maximizer } q^*(\lambda)\right\} .
\end{equation}

Using \Cref{thm:envelope}, deduce that $D$ coincides with the set of points $\lambda > 0$ at which $f$ is differentiable, and that $f^\prime(\lambda) = q^*(\lambda)^2 / 4$, where $q^*(\lambda)$ maximizes $\mathcal{F}(\lambda, \cdot)$. 
\item Conclude that $F_n^\prime(\lambda) \to q^*(\lambda)^2 / 4$.
\end{enumerate}

\renewcommand{\sectionletter}{$\bm \gamma$} 
\subsection{Group \groupgamma}
You may use the following theorem as given for now. At the end of the session, we will prove this theorem using other groups' results.
\begin{theorem}\label{thm:MMSE-q}
    At every point of differentiability $\lambda \in D$ of the limit free energy, the limit of the free energy is given by 
$$
\operatorname{MMSE}_n(\lambda) \underset{n \rightarrow \infty}{\longrightarrow}\left(\mathbb{E}_{P_0} X^2\right)^2-q^*(\lambda)^2 ,
$$
where $q^*(\lambda)$ is any maximizer of function $\mathcal  F(\lambda, \cdot)$.
\end{theorem}

We will focus on the case when $P_0 =\mathcal{N}(0, 1)$. Recall from the last class that the MMSE for the scalar Gaussian channel $Y = \sqrt{\gamma}X + Z$ is $\operatorname{MMSE}(\lambda) = \frac{1}{1+\gamma}$ and $\psi(\lambda) = \frac{1}{2}(\lambda - \log( 1 + \lambda))$ by I-MMSE relation. 

\begin{enumerate}[(a)]
    \item Find $q^*(\lambda)$. 

    \begin{hint}
    $q^*(\lambda) = \max\{0, 1 - 1/ \lambda\}$.
    \end{hint}
    \item Using \Cref{thm:MMSE-q}, deduce that 
    $$
\lim _{n \rightarrow \infty} \operatorname{MMSE}_n(\lambda)= \begin{cases}0 & \text { if } \lambda \leq 1 \\ \frac{1}{\lambda}\left(2-\frac{1}{\lambda}\right) & \text { if } \lambda \geq 1\end{cases}
$$
\end{enumerate}
\paragraph{Optional*} 
In the last session, we had a vector channel $Y = \sqrt{\lambda} \bm \xi + \bm Z$, where $\bm \xi \in \bb R^N$. We showed the I-MMSE relation: 
$$
 F_{\bm \xi}^{\prime}(\lambda)=\frac{1}{2}\left(\mathbb{E}\|\bm{\xi}\|^2- {\operatorname{MMSE}}_{\bm \xi}(\lambda)\right) ,
$$
where $F(\lambda)_\xi = \bb E \log \mathcal Z (\lambda, \bm Y) = \bb E \log \int d P_{\xi} (\bm \xi)e^{H_{\lambda, \bm \xi}(\bm \xi)}$ and ${\operatorname{MMSE}}_{\bm \xi} = \bb E \norm{[\bm \xi - \bb E [\bm \xi | \bm Y]}^2$.

Observe that we can reduce one-rank matrix estimation to the vector channel if we flatten the matrix $\frac{1}{\sqrt{2n}}\bm X \bm X^\top$. 

Deduce that in our case the I-MMSE relation is as follows: 
$$
 F_n^{\prime}(\lambda)=\frac{1}{4}\left(\frac{1}{n} \bb E X^4 + \frac{n-1}{n}(\mathbb{E}X^2)^2- {\operatorname{MMSE}}_n (\lambda)\right) .
$$

Note that in our case the definition of MMSE and the free energy include scaling $\frac{1}{n^2}$ and $\frac{1}{n}$ respectively.

\renewcommand{\sectionletter}{\faPuzzlePiece} 
\subsection{Jigsaw}

\begin{enumerate}[(a)]
    \item Combine your results to deduce the main result: 
\begin{theorem}
    At every point of differentiability $\lambda \in D$ of the limit free energy, the limit of the free energy is given by 
$$
\operatorname{MMSE}_n(\lambda) \underset{n \rightarrow \infty}{\longrightarrow}\left(\mathbb{E}_{P_0} X^2\right)^2-q^*(\lambda)^2 ,
$$
where $q^*(\lambda)$ is any maximizer of function $\mathcal  F(\lambda, \cdot)$.

\end{theorem}
\item Define the information-theoretic threshold

$$
\lambda_c=\inf \left\{\lambda \in D \mid q^*(\lambda)>\left(\mathbb{E}_{P_0} X^2\right)^2\right\} .
$$

If the above set is empty, we define $\lambda_c=0$. 

 Denote by DMSE the MSE achieved by the trivial estimator: 
    $$
   \operatorname{DMSE} := \left(\mathbb{E} X^2\right)^2 - (\bb E X)^4  .
    $$

Conclude that
\begin{itemize}

    \item
    
    if $\lambda>\lambda_c$, then $\lim _{n \rightarrow \infty} \mathrm{MMSE}_n<\mathrm{DMSE}$: one can estimate the signal better than a random guess.
    \item if $\lambda<\lambda_c$, then $\lim _{n \rightarrow \infty} \mathrm{MMSE}_n=$ DMSE: one can not estimate the signal better than a random guess.
\end{itemize}
\item Compare MSE achieved by PCA and MMSE for the Gaussian prior. What can you say about performance of PCA for this problem? 
\item[(d*)] \textbf{Optional} Define the prior be $P_0=p \delta_{\sqrt{\frac{1-p}{p}}}+(1-p) \delta_{-\sqrt{\frac{p}{1-p}}}$. This prior is related to the stochastic block model (to see that, take $p = 1/2$). \Cref{fig:potential} contain the plots of the function $q \to -\mathcal{F}(\lambda, q)$ for different values of $\lambda$. What can you tell about asymptotic behavior of MMSE in different regimes of $\lambda$?

\begin{figure}[h]
    \centering
    \includegraphics[width=0.9\linewidth]{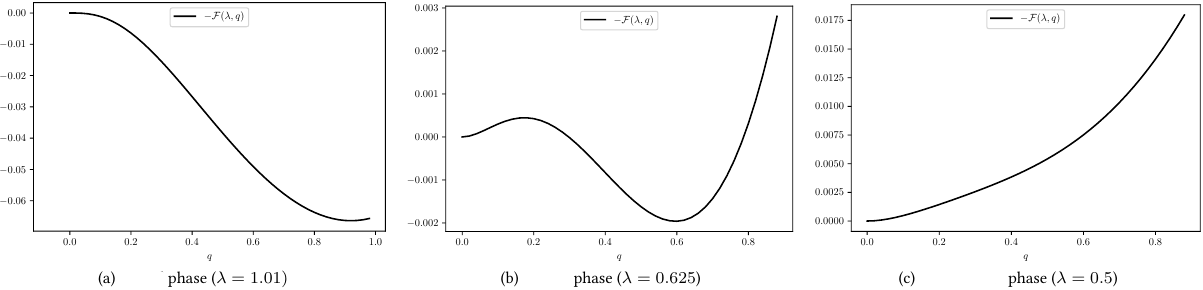}
    \caption{Plots of $q \mapsto-\mathcal{F}(\lambda, q)$ for different values of $\lambda$ and $P_0$ with $p=0.05$. Figure from \cite{miolane2018phase}}
    \label{fig:potential}
\end{figure}
\end{enumerate}

\printbibliography[segment=\therefsegment] 
\section{Franz-Parisi criterion} \label{sec:franz-parisi-I}
\renewcommand{\sectionletter}{1}
The goal of this session is to link low-degree hardness with concepts from statistical physics by introducing the Franz–Parisi criterion. We will demonstrate this connection using the class of Gaussian additive models and illustrate the limitations of the method on a specific boolean problem.

The main reference for the session is \cite{bandeira2022franz}.
\subsection{Background}

Given positive integers $n, N$, a distribution $\mu$ on $\bb R^n$ and a distribution $\bb P_u $ for each $u \in \operatorname{supp}(\mu)$ on $\bb R^N$, we consider hypothesis testing between 
\begin{align*}
&\mathbf{H}_0: \quad Y \sim \mathbb{Q} \quad \text{(Null model)},\\
&\mathbf{H}_1: \quad Y \sim \mathbb{P}=\underset{u \sim \mu}{\mathbb{E}} \mathbb{P}_u \quad \text{(Planted model)}.
\end{align*}

As before, we define the degree-$D$ likelihood ratio as
$$
\operatorname{LD}(D):=\left\|L^{\leq D}\right\|_Q^2=\left\|\left(\underset{u \sim \mu}{\mathbb{E}} L_u\right)^{\leq D}\right\|_Q^2=\underset{u, v \sim \mu}{\mathbb{E}}\left[\left\langle L_{{u}}^{\leq D}, L_{{v}}^{\leq D}\right\rangle_Q\right].
$$

\begin{definition}
    We define the \emph{low-overlap likelihood norm} at overlap $\delta \ge 0$ as
    $$
\mathrm{LO}(\delta):=\underset{u, v \sim \mu}{\mathbb{E}}\left[\mathds{1}_{|\ip{u, v}| \leq \delta} \cdot\left\langle L_u, L_v\right\rangle_Q\right],
$$
where $u, v \sim \mu$ are independent. 
\end{definition}

\begin{definition} We define the \emph{Franz-Parisi Criterion} at $D$ deviations to be the quantity
 $$ \operatorname{FP}(D):=\mathrm{LO}(\delta), \quad \text{for } \delta=\delta(D):=\sup \left\{\varepsilon \geq 0\right. \text{s.t.} \left.\bb{P}(|\langle u, v\rangle| \geq \varepsilon) \geq e^{-D}\right\}.$$
\end{definition}
\begin{remark}\label{rmk:tail_bound}
    The two basic properties of $\delta$ which you may use without proof are that $\bb P (|\langle u, v\rangle| \geq \delta) \geq e^{-D}$ and $\bb P (|\langle u, v\rangle| > \delta) \le e^{-D}$. This follows from continuity of measure (you can find the proof in Section 5.1 of \cite{bandeira2022franz}.)
\end{remark}

In particular, we will consider the Gaussian additive model. Suppose that all moments of distribution $\mu$ are finite. Given $\lambda \ge 0$, we perform hypothesis test between
$$
\begin{aligned}
&\begin{array}{llr}
\mathbf{H}_0 & \mathbb{Q}: Y=Z, & Z \sim \mathcal{N}(0, I), \\
\mathbf{H}_1 & \mathbb{P}: Y=\lambda u+Z, & Z \sim \mathcal{N}(0, I), u \sim \mu .
\end{array}
\end{aligned}
$$
Recall that for Gaussian additive model 
$$
\left\langle L^{\le D}_u, L^{\le D}_v\right\rangle_{\bb Q}=\exp^{\le D} \left(\lambda^2\langle u, v\rangle\right).
$$
and $LD(D) = \bb E_{u, v}\exp^{\le D} \left(\lambda^2\langle u, v\rangle\right)$. The same holds for $L_u$ without restriction on the degree:
$$
\left\langle L_u, L_v\right\rangle_{\bb Q}=\exp\left(\lambda^2\langle u, v\rangle\right).
$$

\renewcommand{\sectionletter}{$\bm \lambda$} 
\subsection{Group \grouplambda}
Our goal is to prove that for Gaussian additive models Franz-Parisi hardness implies low-degree hardness.
\begin{theorem}
    Suppose $\|u \|^2 \le M$ for all $u \in \operatorname{supp}(\mu)$ for some $M > 0$. Then for any $\lambda \ge 0$ and any odd integer $D \ge 1$, 
    $$
\mathrm{LD}(D, \lambda) \leq \mathrm{FP}(\tilde{D}, \lambda)+e^{-D}
$$
where $\tilde{D}:=D \cdot\left(2+\log \left(1+\lambda^2 M\right)\right).$

\end{theorem}
\begin{enumerate}[(a)]
    \item Show that for odd $D$, in Gaussian additive model, 
    $$\left\langle L_u^{\leq D}, L_v^{\leq D}\right\rangle_{\bb{Q}} \leq\left\langle L_u, L_v\right\rangle_{\bb{Q}}$$
    for all $u, v\in \operatorname{supp}(\mu)$.

    You may use the fact that $\exp^{\le D} (x) \le \exp(x)$ for all $x\in\R$ when $D$ is odd without proof. 
    \item Our next step is to show the following crude bound on $\| L_u^{\le D}\|_{\bb Q}$: 
    $$
\left\|L_u^{\leq D}\right\|_Q^2 \leq(D+1)\left(1+\lambda^2 M\right)^D .
$$
\item Decompose LD into low- and high-overlap terms: 
$$
\operatorname{LD}(D, \lambda)=\underset{u, v}{\mathbb{E}}\left[\left\langle L_u^{\leq D}, L_v^{\leq D}\right\rangle_{\bb Q}\right]=\underset{u, v}{\mathbb{E}}\left[\mathds{1}_{|(u, v)| \leq \delta} \cdot\left\langle L_u^{\leq D}, L_v^{\leq D}\right\rangle_{\bb Q}\right]+\underset{u, v}{\mathbb{E}}\left[\mathds{1}_{|(u, v\rangle|>\delta} \cdot\left\langle L_u^{\leq D}, L_v^{\leq D}\right\rangle_{\bb Q}\right] .
$$
Using the definition of $\delta$ in Franz-Parisi criterion, show that 
$$
\underset{u, v}{\mathbb{E}}\left[\mathds{1}_{|(u, v)| \leq \delta} \cdot\left\langle L_u^{\leq D}, L_v^{\leq D}\right\rangle_{\bb Q}\right] \le  \operatorname{FP}(\tilde D, \lambda).
$$
\item For the high-overlap term, show that 
$$
{\mathbb{E}}\left[\mathds{1}_{|(u, v\rangle|>\delta} \cdot\left\langle L_u^{\leq D}, L_v^{\leq D}\right\rangle_{\bb Q}\right]  \le \exp(-D).
$$

\begin{hint}
    Use Cauchy-Schwarz and (b) to show that $\ip{L_u^{\le D}, L_v^{\le D}} \le (D+1) (1 + \lambda^2M)^D$. Use the tail bound from \Cref{rmk:tail_bound} and invoke the definition of $\tilde D$ in the last step. 
\end{hint}
\end{enumerate}

\renewcommand{\sectionletter}{$\bm \mu$} 
\subsection{Group \groupmu}

Our goal is to prove that for Gaussian additive models low-degree hardness implies Franz-Parisi hardness. 
\begin{theorem}\label{thm:ld-hard-implies-fp-hard}
    For every $\varepsilon \in(0,1)$ there exists $D_0=D_0(\varepsilon)>0$ such that for any $\lambda \geq 0$ and any even integer $D \geq D_0$, if

$$
\operatorname{LD}(D,(1+\varepsilon) \lambda) \leq \frac{1}{e D}(1+\varepsilon)^D
$$

then

$$
\mathrm{FP}(D, \lambda) \leq \mathrm{LD}(D,(1+\varepsilon) \lambda)+\varepsilon
$$

\end{theorem}
Define overlap random variable $s = \ip{u, v}$ and denote $C :=\operatorname{LD}(D, \hat{\lambda})$, where $\hat \lambda = (1 + \varepsilon) \lambda$. 

The overall proof idea is to show the following sequence of inequalities:
\begin{equation}\label{eq:proof-idea}
\operatorname{FP}(D, \lambda)=\underset{s}{\mathbb{E}}\left[1_{|s| \leq \delta} \exp \left(\lambda^2 s\right)\right] \leq \underset{s}{\mathbb{E}}\left[\exp ^{\leq D}\left(\tilde{\lambda}^2 s\right)\right]+\varepsilon=\operatorname{LD}(D, \tilde{\lambda})+\varepsilon \leq \operatorname{LD}(D, \hat{\lambda})+\varepsilon,
\end{equation}
where $\tilde \lambda = (1 + \varepsilon^2/ 4) \lambda $ and the first inequality holds when $D \ge \delta \tilde\lambda^2$. 
\begin{enumerate}[(a)]
    \item To show the first inequality in \eqref{eq:proof-idea}, we want to show that $\delta \le D / \tilde\lambda^2$. 
    
    Since $\mu$ has all finite moments, one can show that $\bb E s^d \ge 0$ for every integer $d\ge 0$. You may use this fact without proof to show that $C \ge \bb E_s \frac{1}{D!}(\hat \lambda^2 s)^D$.
    \item Using the definition of $\delta$ and Markov's inequality, show that 
    $$
    \underset{s}{\mathbb{E}} \frac{1}{D!}\left(\hat{\lambda}^2 s\right)^D \geq e^{-D} \frac{1}{D!}\left(\hat{\lambda}^2 \delta\right)^D .
    $$
    \item Combining (a) and (b), we get that $\left(\hat{\lambda}^2 \delta\right)^D\le Ce^D D!$. Apply the following bound on the factorial:
    $$
    k! \le \frac{k^{k+1}}{e^{k-1}} \quad \text{for any integer } k\ge 1, 
    $$
    to show that $\delta \le D \hat \lambda^{-2} (C e D)^{1/D}$.
    \item Use the theorem assumption to show that $\delta \le D / \tilde\lambda^2$.
    \item To finish the proof, you may use the following facts without proof. Firstly, if $\delta \le D / \tilde\lambda^2$, then for all $s \in [-\delta, \delta]$, $$
    \exp \left(\lambda^2 s\right) \leq \exp ^{\leq D}\left(\tilde{\lambda}^2 s\right)+\varepsilon ,
    $$
    and secondly, $\operatorname{LD}(D, \lambda)$ is monotone increasing in $\lambda$.
\end{enumerate}

\renewcommand{\sectionletter}{$\bm \pi$} 
\subsection{Group \grouppi}

Our goal is to show limitation of the criterion by analyzing a boolean problem for which the Franz-Parisi criterion makes an incorrect hardness prediction.

Let $H_0 = \operatorname{Rad}(1/2)^{\otimes n}$, where $\operatorname{Rad}$ is Rademacher distribution (i.e. a distribution of a random variable that takes values $\pm 1$ with probability 1/2). We take $H_1$ to be a mixture of biased distributions $H_1 = \bb E_{u \sim \mu} H_u$, where $\mu$ is distribution over $[-1, 1]^n$, and we sample $x \sim H_u$ by independently sampling
$$
x_i= \begin{cases}1 & \text { with probability } \frac{1}{2}+\frac{u_i}{2} \\ -1 & \text { with probability } \frac{1}{2}-\frac{u_i}{2}\end{cases}.
$$

\begin{enumerate}[(a)]
\item Show that $\ip {L_u, L_v} = \prod_{i=1}^n (1 + u_i v_i)$.

\begin{hint}
    Show first that by definition $L_u(x)  = \prod_{i=1}^n (1 + u_i x_i)$ and then use the independence to compute the inner product.
\end{hint}
\item Show that
$$
\mathrm{LD}(D)=\sum_{\substack{S \subset[n] \\|S| \leq D}} \mathbb{E}_{u, v \sim \mu}\left[\prod_{i \in S} u_i v_i\right].
$$

To show it, recall the Fourier characters that we introduced in the third class: $\chi_S(x) = \prod_{i\in S} x_i$ with $|S| \le D$ which form orthonormal basis for polynomials of degree at most $D$ under $\bb Q$. We showed that $L^{\le D}_u(x) = \sum_{S \subseteq n, |S| \le D } \hat L_u(S)\chi_S(x)$. You may use these results without proving them again. 

\item Consider $\mu$ to be distribution over $\{\pm 1\}^n $ (instead of the whole interval $[-1, 1]^n$). Show that in this case if $u$ and $v$ disagree on at least one coordinate, then $\ip{L_u, L_v} = 0$ and thus for any  $\delta$ small enough to exclude case $u = v$, 
$$
\mathrm{LO}(\delta) \leq \mathbb{E}_{u, v \sim \mu}\left[1_{u \neq v} \cdot\left\langle L_u, L_v\right\rangle\right]=0 .
$$
\item Conclude that Franz-Parisi criterion will predict that the problem is hard for any $D = n^{\Omega(1)}$ even if $H_0$ and $H_1$ are easy to distinguish. The latter can happen for example if $H_1$ chooses vector $u$ uniformly from the set $\{u \in {\pm 1} : \sum_{i=1}^n u_i = 0.9n\}$.
\end{enumerate}

\renewcommand{\sectionletter}{\faPuzzlePiece} 
\subsection{Jigsaw}

\begin{enumerate}[(a)]
\item  Share and discuss your results with your peers from other groups.  
\item Show the following proposition that follows from Theorem $\bm \mu$.1.1:
\begin{proposition}
Fix any constant $\varepsilon^{\prime}>0$ and suppose $D=D_n, \lambda=\lambda_n, N=N_n$, and $\mu=\mu_n$ are such that $D$ is an even integer, $D=\omega(1)$, and $\operatorname{LD}\left(D,\left(1+\varepsilon^{\prime}\right) \lambda\right)=O(1)$. Then
$$
\operatorname{FP}(D, \lambda) \leq \mathrm{LD}\left(D,\left(1+\varepsilon^{\prime}\right) \lambda\right)+o(1)
$$
\end{proposition}
For the proof, use the fact that $\rm{LD}(D, \lambda)$ is monotone increasing in $\lambda$ (try to prove it if you have extra time!)
\item 
Show that $\rm{LD}(D, \lambda)$ and $\rm{FP}(D, \lambda)$ are monotone increasing in $D$.
\end{enumerate}

\printbibliography[segment=\therefsegment] 
\section{Markov Chain Monte Carlo and free energy wells}\label{sec:MCMC}
\renewcommand{\sectionletter}{1}
In this session, we introduce Markov Chain Monte Carlo (MCMC), a class of algorithms originally developed for drawing samples from a probability distribution. In the context of statistical inference, when the posterior has a Gibbs form, sampling from the posterior distribution is closely related to finding the low-energy states of the associated Hamiltonian. Additionally, we will show that the existence of free energy barriers obstructs the mixing of the Markov chain, suggesting computational hardness for this class of algorithms.

The main references for this session are \cite{levin2017markov,arous2023free}.
\subsection{Background}
Recall first the basic definitions related to Markov chains. 
\begin{definition}
    A sequence of random variables $\left(X_0, X_1, \ldots\right)$ is a \emph{Markov chain} with state space $\mathcal{X}$ and transition matrix $P$ if for all $x, y \in \mathcal{X}$, all $t \geq 1$, and all events $H_{t-1}=$ $\bigcap_{s=0}^{t-1}\left\{X_s=x_s\right\}$ satisfying $\mathbb{P}\left(H_{t-1} \cap\left\{X_t=x\right\}\right)>0$, we have

$$
\mathbb{P}\left\{X_{t+1}=y \mid H_{t-1} \cap\left\{X_t=x\right\}\right\}=\mathbb{P}\left\{X_{t+1}=y \mid X_t=x\right\}=P(x, y) .
$$
\end{definition}

\begin{definition}
    We say that the probability distribution $\mu$ on $\mathcal{X}$ is a \emph{stationary distribution} if it satisfies $\mu = \mu P$. 
\end{definition}

\begin{proposition}\label{prop:stationary}
 Let $P$ be the transition matrix of a Markov chain with state space $\mathcal{X}$. Any distribution $\pi$ satisfying \emph{the detailed balance equations}:
$$
\pi(x) P(x, y)=\pi(y) P(y, x) \quad \text { for all } x, y \in \mathcal{X} .
$$
is stationary for $P$.
\end{proposition}

\paragraph{Model}
We consider the following variant of sparse PCA in the spiked Wigner model (also called principal submatrix recovery). Let $W$ be a $\operatorname{GOE}(n)$ matrix, i.e., $n \times n$ symmetric with off-diagonal entries $\mathcal{N}(0,1 / n)$ and diagonal entries $\mathcal{N}(0,2 / n)$, all independent aside from the symmetry constraint $W_{i j}=W_{j i}$. Let $x$ be an unknown $k$-sparse vector in $\{0,1\}^n$ . We are interested in recovering $x$ from the observation

$$
Y=\frac{\lambda}{k} x x^{\top}+W,
$$

where $\lambda>0$ is the signal-to-noise ratio. 

For this model, we define the Hamiltonian $H(v)=-v^{\top} Y v$. In the following, we define the free energy well, which as we will see creates an obstruction for local MCMC.
\begin{definition}
Consider the Gibbs distribution $\mu_\beta(v) \propto \exp (-\beta H(v))$ on the space of $k^{\prime}$ sparse vectors $S_{k^{\prime}}=\left\{v \in\{0,1\}^n:\|v\|_0=k^{\prime}\right\}$, where $\beta \geq 0$ is the inverse temperature. For some $\ell>0$, let $A=\left\{v \in S_{k^{\prime}}: 0 \leq\langle v, x\rangle<\ell\right\}$ and $B=\left\{v \in S_{k^{\prime}}: \ell \leq\langle v, x\rangle \leq 2 \ell\right\}$ (where $x$ is the planted signal). We say that the depth of the free energy well at correlation $\ell$ is
$$
D_{\beta, \ell}:=\log \mu_\beta(A)-\log \mu_\beta(B).
$$
If $D_{\beta, \ell} \leq 0$ then there is no free energy well at correlation $\ell$.
\end{definition}

\renewcommand{\sectionletter}{$\bm \lambda$} 
\subsection{Group \grouplambda}
\paragraph{Metropolis chain}
Suppose we are given a Markov chain with transition matrix $\Psi$, and our goal is to modify the transitions so that the new chain has stationary distribution $\pi$. A possible example of such setting can be to sample uniformly from a vertex set $V$ of a graph, without access to either $V$ or its edge set. The only operation allowed is performing a simple random walk (this scenario may arise in computer networks). If the graph is non-regular, it is easy to verify that the stationary distribution of the random walk is not uniform. However, using the algorithm described below, we can achieve uniform sampling.

Assume that $\Psi$ is symmetric and consider the following chain: when at state $x$, a candidate move is generated from the distribution $\Psi(x, \cdot)$. If the proposed new state is $y$, then the move is censored with probability $1-a(x, y)$. That is, with probability $a(x, y)$, the state $y$ is "accepted" so that the next state of the chain is $y$, and with the remaining probability $1-a(x, y)$, the chain remains at $x$. 
\begin{enumerate}[(a)]
    \item Write the transition matrix $P$ of the described chain.

    \begin{hint}
        You should get the following expression: 
        $$
P(x, y)= \begin{cases}\Psi(x, y) a(x, y) & \text { if } y \neq x \\ 1-\sum_{z: z \neq x} \Psi(x, z) a(x, z) & \text { if } y=x .\end{cases}
$$
    \end{hint}
    \item We would like to choose the acceptance probabilities so that the stationary distribution of $P$ is $\pi(x)$ and we have as little rejections as possible. Using \Cref{prop:stationary}, write the sufficient condition for $P$ to have the stationary distribution $\pi(x)$. 
    \item Show that we should choose $a(x, y) = (\pi(y) / \pi(x)) \wedge 1 $ to minimize the number of rejections and achieve the desired stationary distribution. 
\end{enumerate}
This way we just defined the \textit{Metropolis chain}: 
$$
P(x, y)= \begin{cases}\Psi(x, y)\left[1 \wedge \frac{\pi(y)}{\pi(x)}\right] & \text { if } y \neq x \\ 1-\sum_{z: z \neq x} \Psi(x, z)\left[1 \wedge \frac{\pi(z)}{\pi(x)}\right] & \text { if } y=x\end{cases}
$$

\begin{remark}
  For a general irreducible transition matrix $\Psi$ (not necessarily symmetric), we can define the Metropolis chain as follows: 
$$
P(x, y)= \begin{cases}\Psi(x, y)\left[\frac{\pi(y) \Psi(y, x)}{\pi(x) \Psi(x, y)} \wedge 1\right] & \text { if } y \neq x \\ 1-\sum_{z: z \neq x} \Psi(x, z)\left[\frac{\pi(z) \Psi(z, x)}{\pi(x) \Psi(x, z)} \wedge 1\right] & \text { if } y=x .\end{cases}
$$
\end{remark}

\renewcommand{\sectionletter}{$\bm \mu$} 
\subsection{Group \groupmu}

Our goal is prove that the presence of the free energy well presents an obstruction to the Metropolis chain. 

Assume the model as described in the background section. Consider the undirected graph $\mathcal{X}$ with $\binom{n}{k^{\prime}}$ vertices, where each vertex corresponds to a unique $k^{\prime}$-sparse binary vector. We connect two vertices if the Hamming distance between their associated vectors is exactly 2 (the minimal nonzero distance). Additionally, each vertex has a self-loop.

Let $ X_0 \sim \mu_\beta(\cdot \mid A) $, and let $ X_0, X_1, X_2, \ldots $ be any Markov chain on the vertices of $\mathcal{X}$  (with transitions allowed only on the edges of $\mathcal{G}$) whose stationary distribution is $\mu_\beta$. A canonical choice is the Metropolis chain with the following update step:

If the current state is $ v $, choose a random neighbor $ u $, and 
    \begin{itemize}
        \item Move to $ u $ with probability $ \min \{ 1, \mu_\beta(u) / \mu_\beta(v) \} $.
        \item Otherwise, remain in the state $ v $.
    \end{itemize}
One can check that its stationary distribution is $\mu_\beta$.

Define the hitting time,
$$
\tau_\beta:=\inf \left\{t \in \mathbb{N}: X_t \in B\right\}.
$$

Our goal is to show the following result.
\begin{proposition}[Proposition~2.2 in \cite{arous2023free}]\label{prop:hitting-time-lower-bound}
    Consider a fixed $Y$ for which $\mu_\beta$ has a free energy well of depth $D_{\beta, \ell}$ at correlation $\ell$. As above, let $X_0, X_1, \ldots$ be any Markov chain on $\mathcal{X}$ with stationary distribution
$\mu_\beta$, initialized from $X_0 \sim \mu_\beta(\cdot \mid A)$. Then for any $t \geq 1$,

$$
\operatorname{Pr}\left\{\tau_\beta \leq t\right\} \leq t \exp \left(-D_{\beta, \ell}\right) .
$$
\end{proposition}
\begin{enumerate}[(a)]
    \item Suppose that $X_0 \sim \mu_\beta$. Deduce that $X_1, X_2, \dots$ are also distributed according to $\mu_\beta$.
    \item Show that 
    $$ 
    \bb P(\tau_\beta \le t) \le \frac{\bb{P}\left\{\exists i \in\{1, \ldots, t\}: X_i \in B\right\}}{\bb{P}\left\{X_0 \in A\right\}}.
    $$

    \begin{hint}
        You can define $\tau_\beta(x_0)$ as the hitting time when $X_0 = x_0$, $\tau_\beta(x_0) = \inf \{t \in \bb N: X_t \in B, X_0 = x_0\}$. Recall that $X_0 \sim \mu_\beta(\cdot \mid A)$ and use the definition of conditional probability. 
    \end{hint}

    \item Deduce that 
    $$ 
    \bb P(\tau_\beta \le t) \le \frac{t \mu_\beta (B)}{\mu_\beta(A)}.
    $$
    \begin{hint}
        Use union bound.
    \end{hint}
    \item Conclude the proof using the definition of the free energy well.

\end{enumerate}

\renewcommand{\sectionletter}{$\bm \pi$} 
\subsection{Group \grouppi}
\paragraph{Hill climber algorithm}

Let $f$ be a real-valued function defined on the vertex set $\mathcal{X}$ of a graph. In many applications it is desirable to find a vertex $x$ where $f(x)$ is maximal. If the domain $\mathcal{X}$ is very large, then an exhaustive search may be too expensive.

\begin{figure}
    \centering
    \includegraphics[width=0.5\linewidth]{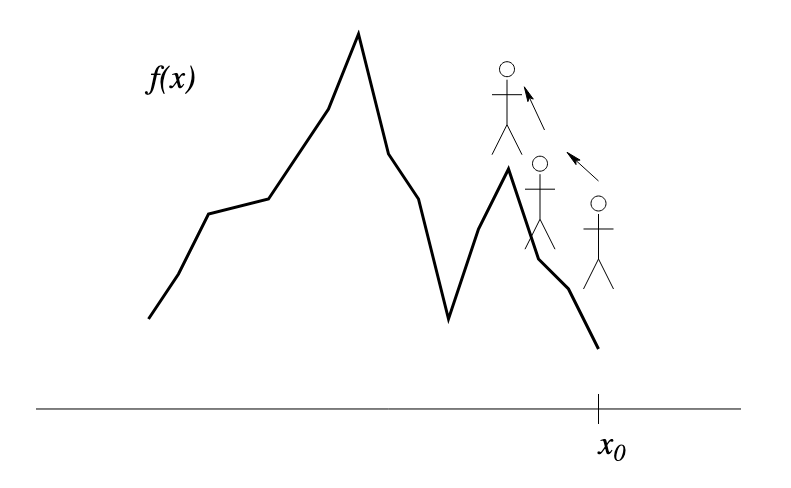}
    \caption{A hill climb algorithm may become trapped at a local maximum. Figure from \cite{levin2017markov}}
    \label{fig:hill-climb}
\end{figure}

A hill climb is an algorithm which attempts to find the maximum values of $f$ as follows: if there is at least one neighbor $y$ of the current position $x$ satisfying $f (y) > f (x)$, move to a neighbor with the largest value of $f$. A downside of such approach is that the climber may become stranded at local maxima, see \Cref{fig:hill-climb}.

One solution is to randomize moves so that instead of always remaining at a local maximum, with some probability the climber moves to lower states. Suppose for simplicity that $\mathcal{X}$ is a regular graph, so that simple random walk on $\mathcal{X}$ has a symmetric transition matrix. Define 
$$
\mu_\beta(x) = \frac{\exp(\beta f(x))}{Z(\beta)},
$$
where $\beta \ge 0$ and $Z(\beta):= \sum_{x \in \mathcal{X}} \exp(\beta f(x))$ is the normalizing constant. Consider now a Metropolis chain with the stationary distribution $\mu_\beta(x)$ as defined in group $\bm \mu$. Note that the Metropolis chain does not require computation of $Z(\beta)$. Since $\mu_\beta(x)$ is increasing in $f(x)$, the measure $\mu_\beta(x)$ favors vertices $x$ for which $f (x)$ is large.

Define
$$
\mathcal{X}^{\star}:=\left\{x \in \mathcal{X}: f(x)=f^{\star}:=\max _{y \in \mathcal{X}} f(y)\right\} .
$$

\begin{enumerate}[(a)]
    \item Show that 
    $$
\lim _{\beta \rightarrow \infty} \mu_\beta(x)=\frac{\mathbf{1}_{\left\{x \in \mathcal{X}^{\star}\right\}}}{\left|\mathcal{X}^{\star}\right|}.
$$
That is, as $\beta \to \infty$, the stationary distribution $\mu_\beta$ of this Metropolis chain converges to the uniform distribution over the global maxima of $f$.
\item Define the Hamiltonian for the sparse PCA as $H(v) = -v^\top Y v$, where 
$$
v \in S_{k^\prime} := \{v \in \{0, 1\}^n: \|v\|_0 = k^\prime\},
$$
and $\mu_\beta = \frac{1}{Z_\beta}\exp(-\beta H(v))$. 

     Show that when $\beta \to \infty$, the metropolis chain reduces to the deterministic hill climb for $f(v) = - H(v)$ (which is equivalent to "hill descent" for $H(v)$).
\end{enumerate}

\renewcommand{\sectionletter}{$\bm \gamma$} 
\subsection{Group \groupgamma}
We will define the Hamiltonian for the sparse PCA as $H(v) = -v^\top Y v$, where 
$$
v \in S_{k^\prime} := \{v \in \{0, 1\}^n: \|v\|_0 = k^\prime\},
$$
and $\mu_\beta = \frac{1}{Z_\beta}\exp(-\beta H(v))$.

In the following exercises we will prove that when $\beta = 0$, the chain will visit all states within time $\exp(\tilde{O}(\frac{k^2 }{ \lambda^2 n}))$ with high probability. This statement is formalized in the following proposition. 

\begin{proposition}
    Let $\mathcal{G}$ be the graph with vertex set $\left\{v \in\{0,1\}^n:\|v\|_0=k^{\prime}\right\}$, with an edge between each pair of vertices whose associated vectors differ in exactly 2 coordinates. Let $X_0, X_1, X_2, \ldots$ be the Markov chain on the vertices of $\mathcal{G}$ where $X_0$ is a uniformly random initialization and $X_{i+1}$ is a uniformly random neighbor of $X_i$. Fix a vertex $x$ and let $\tau$ be the hitting time $\tau=\inf \left\{t \in \mathbb{N}: X_\tau=x\right\}$. Then for any $t \geq 0$ we have $\operatorname{Pr}\{\tau \geq t\} \leq k^{\prime} n^{2 k^{\prime}} / t$.
\end{proposition} 

\begin{enumerate}[(a)]
    \item Show that when $\beta = 0$, the Metropolis chain reduces to a random walk (ignoring the data $Y$).
    \item Show that $\mathcal{G}$ is a $d$-regular graph with $d = k^\prime (n - k^\prime) \le n^2$. 
    \item Denote by $y$ the current state and by $\ell$ the path length from $y$ to $x$ (note $\ell \le k^\prime$). Show that the probability of reaching $x$ from $y$ over the next $\ell $ steps is at least $d^{-k^\prime}$.
    \item Denote by $N$ number of such trials, each consisting of at most $k^\prime$ steps before $x$ is reached. Show that $\bb E N \le d^{k^\prime}$.

    \begin{hint}
        Consider $N$ as a random variable distributed according to geometric distribution (number of trials before the first success). 
    \end{hint}

    \item Deduce that $\tau \le k^{\prime} N$ and $\bb E \tau \le k^\prime n^{2k^\prime}$. Use Markov's inequality to finish the proof. 
\end{enumerate}
\begin{remark}
    We have just proven that the Metropolis chain with $\beta = 0$ effectively performs an exhaustive search over all vectors of sparsity $k^\prime$. We can then output the configuration with the highest Hamiltonian value. It was shown in \cite{ding2023subexponential} that such an algorithm, with $k^\prime \approx \frac{k^2}{\lambda^2 n}$ followed by a certain boosting procedure, achieves exact recovery.
\end{remark}

\renewcommand{\sectionletter}{\faPuzzlePiece} 
\subsection{Jigsaw}
Read through the following results from \cite{arous2023free}. 
We will call $k^\prime$ informative, if
$$
\frac{k^2 \log n}{\lambda^2 n} \leq k^{\prime} \leq \frac{n \lambda^2}{\log n}.
$$
Additionally, we call $\ell$ informative, if 
$$
\max \left\{1, \frac{k k^{\prime}}{n}\right\} \ll \ell \leq \frac{k}{2 \lambda} \sqrt{\frac{k^{\prime}}{n} \log n} .
$$

\begin{theorem}[informal]\label{thm:main}
    Suppose $\lambda$ is in the "hard" regime $\sqrt{k / n} \ll \lambda \ll$ $\min \{1, k / \sqrt{n}\}, k^{\prime}$ is informative, and that additionally, either (i) $k^{\prime} \leq k$ or (ii) $\lambda \ll(k / n)^{1 / 4}$. For any $\beta \geq 0$ (possibly depending on $n$), there exists an informative $\ell$ such that $D_{\beta, \ell} \geq \tilde{\Omega}\left(\frac{k^2}{\lambda^2 n}\right)$ with high probability.
\end{theorem}

\begin{enumerate}[(a)] 
    \item Discuss your results and share your findings with your peers. 
    \item What is the motivation to use the Metropolis chain with the described Hamiltonian? 
    \item What does \Cref{thm:main} imply about the performance of MCMC?
\end{enumerate}
\renewcommand{\sectionletter}{*} 
\subsection{Optional} If you have extra time and would like to further explore this problem as homework, you may try proving the existence of the free energy well in a certain parameter regime.

\begin{theorem}[Theorem~3.8 in \cite{arous2023free}] Fix a constant $\delta>0$. For any $n$ exceeding some $n_0=n_0(\delta)$, for any $k$, for any $k^{\prime} \leq n^{1-\delta}$, for any $k$-sparse signal $x \in\{0,1\}^n$, for any $\beta \geq 0$, and for any $\ell \geq 2 e k\left(k^{\prime} / n\right)^{1-\delta}$ (with $\left.1 \leq 2 \ell \leq \min \left\{k, k^{\prime}\right\}\right)$, with probability at least $1-2^{-(\ell-2) / 2}$ over $W$, the free energy well at correlation $\ell$ has depth bounded by

$$
D_{\beta, \ell} \geq-\frac{4 \beta \lambda}{k} \ell^2+\frac{\log 2}{2} \ell-\log 2 .
$$

\end{theorem}

Recall that $Y = \frac{\lambda}{k}x x^\top + W$ and $H(v) = -v^\top Y v$, where $x, v \in \{0, 1\}^n$ with $\|x\|_0 = k$ and $\|v\|_0 = k^\prime$. 

\begin{enumerate}[(a)]
    \item For $S \subseteq \{0, 1\}^n$, set 
    $$
Z_\beta(S)=\sum_{v \in S} e^{-\beta H(v)}.
$$

Show that $D_{\beta, \ell} = \log Z_\beta(A) - \log Z_\beta(B)$.
\item Define the pure noise Hamiltonian $\tilde{H}(v)=-v^{\top} W v$ and the corresponding $\tilde{Z}_\beta(S)$ as follows:

$$
\tilde{Z}_\beta(S)=\sum_{v \in S} e^{-\beta \tilde{H}(V)} .
$$

Show that 
$$
\log Z_\beta(A) - \log Z_\beta(B) \ge -\frac{4 \beta \lambda \ell^2}{k}+\log \tilde{Z}_\beta(A)-\log \tilde{Z}_\beta(B).
$$

\begin{hint}
    Use the fact that $H(v)=-\frac{\lambda}{k}\langle v, x\rangle^2+\tilde{H}(v)$. In the second step, use that for all $v\in B$, $\ip{v, x} \le 2 \ell$ and for all $v \in A$ $\ip{v,x} \ge 0$.
\end{hint}
\item Define the complement set as $A^c=\left\{v \in\{0,1\}^n:\|v\|_0=k^{\prime}\right\} \backslash A$ and the pure noise Gibbs measure as $\tilde \mu_\beta(v) \propto e^{-\tilde H(v)}$. Show that
$$
\frac{\tilde{Z}_\beta(A)}{\tilde{Z}_\beta(B)}=\frac{\tilde{\mu}_\beta(A)}{\tilde{\mu}_\beta(B)} \geq \frac{\tilde{\mu}_\beta(A)}{\tilde{\mu}_\beta\left(A^c\right)}=\frac{1-\tilde{\mu}_\beta\left(A^c\right)}{\tilde{\mu}_\beta\left(A^c\right)}.
$$
\item You may use the following fact without proof. Fix $\gamma > 0$. With probability $1 - \gamma$, 
$$
\tilde \mu_\beta(A^c) \le \frac{2^{1 - \ell}}{\gamma},
$$
provided that $\ell \ge 2ek(k^\prime/n)^{1 - \delta}$.

Derive that for $\gamma=2^{-(\ell-2) / 2}$, it holds
$$
D_{\beta, \ell} \geq-\frac{4 \beta \lambda}{k} \ell^2+\frac{\log 2}{2} \ell-\log 2 
$$
with probability at least $1 - 2^{-(\ell-2) / 2}$ over $W$.

\begin{hint}
    First, note that for $\gamma \ge 2^{-(\ell-2)}$, $\tilde \mu_\beta(A^c) < 1/2$. 
    
    In the intermediate step you should arrive at the following expression: 
$$D_{\beta, \ell} \geq-\frac{4 \beta \lambda \ell^2}{k}+(\ell-2) \log 2+\log \gamma.
$$
\end{hint}

\end{enumerate}

\printbibliography[segment=\therefsegment] 
\section{Franz-Parisi criterion and MCMC}\label{s13:franz-parisi-II}
\renewcommand{\sectionletter}{1}
Recall that in Session~\ref{sec:MCMC}, we showed that a free energy barrier implies a lower bound on the hitting time of the Markov chain. In this session, we will explore a technique for proving the existence of such a free energy barrier using the Franz–Parisi criterion, introduced in Session~\ref{sec:franz-parisi-I}. As a consequence of this analysis, we also establish a connection between low-degree hardness and lower bounds on the hitting time of Markov chains.

The main reference for this session is \cite{bandeira2022franz}.
\subsection{Background}

We consider Gaussian additive model, $Y = \lambda u + Z$, where $u\sim\mu$. We make a few assumptions on the prior $\mu$: we assume that $\mu$ is uniform on some finite set $S\subset \bb R^N$ with transitive symmetry which defined as follows.

\begin{definition}
We say $S \subseteq \mathbb{R}^N$ has transitive symmetry if for any $u, v \in S$ there exists an orthogonal matrix $R \in O(N)$ such that $R u=v$ and $R S=S$.
\end{definition}

An example of such measure is the uniform measure on the unit ball in $\bb R^N$.

Consider the associated Gibbs measure $\nu_{\beta}$ defined on $S$ as $$
\nu_{\beta}(v) = \frac{1}{\mathcal Z_{\beta}}\exp(-\beta H(v)),
$$
where $\beta\ge 0$ is the inverse temperature parameter, $H(v)=-\langle v, Y \rangle$ is the Hamiltonian, and $\mathcal{Z_{\beta}}$ is the normalizing constant, $\mathcal{Z_{\beta}} = \sum_{v\in S}\exp(-\beta H(v))$.

We will consider a Markov chain with stationary distribution $\nu_{\beta}$ (for example, Metropolis chain, but not necessarily).
Recall the definition of the Franz-Parisi criterion.

\begin{definition}\label{def:franz-parisi} We define the \emph{Franz-Parisi Criterion} at $D$ deviations to be the quantity
 $$ \operatorname{FP}(D):=\mathrm{LO}(\delta), \quad \text{for } \delta=\delta(D):=\sup \left\{\varepsilon \geq 0\right. \text{s.t.} \left.\bb{P}(|\langle u, v\rangle| \geq \varepsilon) \geq e^{-D}\right\},$$
 where the \emph{low-overlap likelihood norm} at overlap $\delta \ge 0$ is defined as
    $$
\mathrm{LO}(\delta):=\underset{u, v \sim \mu}{\mathbb{E}}\left[\mathds{1}_{|\ip{u, v}| \leq \delta} \cdot\left\langle L_u, L_v\right\rangle_Q\right],
$$
where $u, v \sim \mu$ are independent. 
\end{definition}

\renewcommand{\sectionletter}{$\bm \lambda$} 
\subsection{Group \grouplambda}

\begin{theorem}[Theorem~5.3 in \cite{bandeira2022franz}, FP-Hard Implies Free Energy Barrier]\label{thm:fp-hard-implies-free-energy-barrier}
    Let $\mu$ be the uniform measure on $S$, where $S \subseteq \mathbb{S}^{N-1}$ is a finite, transitive-symmetric set. The following holds for any $\varepsilon \in(0,1 / 2), D \geq 2, \lambda \geq 0$, and $\beta \geq 0$. Fix a ground-truth signal $u \in S$ and let $Y=\lambda u+Z$ with $Z \sim \mathcal{N}\left(0, I_N\right)$. Define $\delta=\delta(D)$ as in \Cref{def:franz-parisi}. Let

$$
A=\{v \in S:|\langle u, v\rangle| \leq \delta\} \quad \text { and } \quad B=\{v \in S:\langle u, v\rangle \in(\delta,(1+\varepsilon) \delta]\} .
$$

With probability at least $1-e^{-\varepsilon D}$ over $Z$, the Gibbs measure associated to $Y$ satisfies

$$
\frac{\nu_\beta(B)}{\nu_\beta(A)} \leq 2(2 \cdot \mathrm{FP}(D+\log 2, \tilde{\lambda}))^{1-2 \varepsilon} e^{-\varepsilon D}
$$

where

$$
\tilde{\lambda}:=\sqrt{\beta \lambda \cdot \frac{2+\varepsilon}{1-2 \varepsilon}} .
$$

\end{theorem}

Your goal is to prove a part of this theorem, while group $\bm \mu$ will show the rest. 

\begin{enumerate}[(a)]
    \item Define the pure noise Hamiltonian $\tilde{H}(v)=-\ip{v, Z}$ and the corresponding $\tilde{\nu}_\beta \propto \exp(-\beta \tilde H(v))$. Note that $H(v) = -\lambda \ip{u, v} + \tilde H(v)$ and show that 
    $$
    \frac{\nu_\beta(B)}{\nu_\beta(A)}\le\exp(\beta \lambda \delta(2+ \varepsilon)) \frac{\tilde \nu_\beta(B)}{\tilde \nu_\beta(A)}.
    $$
    \begin{hint}
        Use the fact that $\ip{u, v} \le (1 + \varepsilon)\delta$ for all $v \in B$ and $\ip{u, v} \ge -\delta$ for $v\in A$. 
    \end{hint}

    \item Denote $A^c = S \backslash A$ and show that 
    $$
    \frac{\tilde{\nu}_\beta(B)}{\tilde{\nu}_\beta(A)} \leq \frac{\tilde{\nu}_\beta\left(A^c\right)}{1-\tilde{\nu}_\beta\left(A^c\right)},
    $$
    so now it remains only to bound $\tilde \nu_\beta (A^c)$. Group $\bm \mu$ is showing that 
    $$
    \frac{\tilde{\nu}_\beta\left(A^c\right)}{1-\tilde{\nu}_\beta\left(A^c\right)} \leq 2 e^{-(1-\varepsilon) D}.
    $$
    and you may use this fact without proof. 

    \item Define $\tilde D = D + \log 2$ and $\tilde \delta = \delta(\tilde D)$. Show $\bb{P}_{v, v^{\prime} \sim \mu}\left(\left|\left\langle v, v^{\prime}\right\rangle\right|>\tilde{\delta}\right) \leq \frac{1}{2} e^{-D} $ and conclude that 
    $$
    \bb{P}_{v, v^{\prime} \sim \mu}\left(\left|\left\langle v, v^{\prime}\right\rangle\right| \in[\delta, \tilde{\delta}]\right) \geq \frac{1}{2} e^{-D}.
    $$
    \item Deduce from (c) that $$
\operatorname{FP}(\tilde{D}, \tilde{\lambda})=\underset{v, v^{\prime} \sim \mu}{\mathbb{E}}\left[\mathds{1}_{\left|\left\langle v, v^{\prime}\right\rangle\right| \leq \tilde{\delta}} \cdot \exp \left(\tilde{\lambda}^2\left\langle v, v^{\prime}\right\rangle\right)\right] \geq \frac{1}{2} e^{-D} \cdot \exp \left(\tilde{\lambda}^2 \delta\right) .
$$
\item Now we are ready to finish the proof by using the choice $\tilde \lambda^2 = \beta \lambda (2 + \varepsilon) / (1 - 2 \varepsilon )$: 
$$
\frac{\nu_\beta(B)}{\nu_\beta(A)}   \leq 2(2 \cdot \operatorname{FP}(\tilde{D}, \tilde{\lambda}))^{1-2 \varepsilon} e^{-\varepsilon D}.
$$
\item (Optional) Show that at the "Bayesian temperature" $\beta=\lambda$, $\nu_\lambda$ is the posterior distribution for the signal $u$ given the observation $Y=\lambda u + Z$.
\end{enumerate}

\renewcommand{\sectionletter}{$\bm \mu$} 
\subsection{Group \groupmu}
Our goal is to prove \Cref{thm:fp-hard-implies-free-energy-barrier} with group $\bm{\lambda}$. We will focus on a part of the proof, while group $\bm{\lambda}$ will show the rest.

We define the pure noise Hamiltonian as $\tilde{H}(v)=-\ip{v, Z}$ and the corresponding Gibbs measure as $\tilde{\nu}_\beta \propto \exp(-\beta \tilde H(v))$. Group $\bm \lambda$ is showing
    $$
    \frac{\nu_\beta(B)}{\nu_\beta(A)}=\exp(\beta \lambda \delta(2+ \varepsilon)) \frac{\tilde \nu_\beta(A^c)}{1 - \tilde \nu_\beta(A^c)},
    $$
    where $A^c = S \backslash A$. Our goal is to bound $\nu_\beta(A^c)$ and consequently the quantity on the right hand side of the above expression.

\begin{enumerate}[(a)]
    \item We first claim that $\underset{Z}{\mathbb{E}}\left[\tilde{\nu}_\beta(v)\right]=\underset{Z}{\mathbb{E}}\left[\tilde{\nu}_\beta\left(v^{\prime}\right)\right]$ for all $v, v^{\prime} \in S$. To show that, let $R \in O(N)$ be the orthogonal matrix from the definition of the transitive symmetry, i.e., such that $R v = v^\prime$ and $RS = S$. Show that 
    $$
    \tilde{\nu}_\beta\left(v^{\prime}\right)=\frac{\exp \left(\beta\left\langle v, R^{\top} Z\right\rangle\right)}{\sum_{w \in S} \exp \left(\beta\left\langle w, R^{\top} Z\right\rangle\right)} 
    $$
    which equals to $ \tilde{\nu}_\beta\left(v\right)$ in distribution. 
    \item Conclude that $\bb E_Z [\tilde \nu_\beta (v) ] = 1/ |S|$ for every $v \in S$ and consequently, $\underset{Z}{\mathbb{E}}\left[\tilde{\nu}_\beta\left(A^c\right)\right]=\frac{\left|A^c\right|}{|S|}$.
    \item Observe that $\bb P_{v\sim \mu}(|\ip{u, v}| > \delta) = \frac{|A^c|}{S}$. By transitivity, it also holds that $\bb P_{v, v^\prime\sim \mu}(|\ip{v, v^\prime}| > \delta) = \frac{|A^c|}{S}$.

    \begin{hint}
        For the latter claim, observe that the volume of set $A = A_u = \{ v \in S : |\ip{u, v}| \le \delta \}$ does not depend on $u$. Due to transitive symmetry of $S$, for every $v^\prime$ there is $R \in O(N)$ such that $v^\prime = R u$ and $RS = S$ and thus the volume of the set $A_{v^\prime}$ remains the same as $|A_u|$.
    \end{hint}
    \item Use Markov's inequality and \Cref{rmk:tail_bound}, to show that
    $$
    \operatorname{Pr}_Z\left(\tilde{\nu}_\beta\left(A^c\right) \geq e^{-(1-\varepsilon) D}\right) \leq e^{-\varepsilon D} .
    $$
    \item Show that with probability at least $1 - e^{-\varepsilon D}$
    $$
    \frac{\tilde\nu_\beta(B)}{\tilde\nu_\beta(A)}   \leq 2 e^{-(1-\varepsilon) D}.
    $$

    \begin{hint}
        You may use that with high probability $\tilde{\nu}_\beta\left(A^c\right) \leq e^{-(1-\epsilon) D} \le 1/2$ if $D > 1$ and $\varepsilon < 1/2$.
    \end{hint}
\end{enumerate}

\renewcommand{\sectionletter}{$\bm \pi$} 
\subsection{Group \grouppi}
\begin{definition}
     We say a Markov chain on a finite state space $S \subseteq \mathbb{S}^{N-1}$ is $\Delta$-local if for every possible transition $v \rightarrow v^{\prime}$ we have $\left\|v-v^{\prime}\right\|_2 \leq \Delta$.
\end{definition}

Your goal is to derive a lower bound on the hitting time of MCMC. Recall the result from \Cref{sec:MCMC}, specifically \Cref{prop:hitting-time-lower-bound},  which we reformulate here for convenience.

\begin{proposition}\label{prop:barrier-implies-hitting-time-lb} Suppose $X_0, X_1, \ldots$ is a Markov chain on a finite state space $S$, with some stationary distribution $\nu$. Let $A$ and $B$ be two disjoint subsets of $S$ and define the hitting time $\tau_B:=\inf \left\{t \in \mathbb{N}: X_t \in B\right\}$. If the initial state $X_0 \sim \nu \mid A$, then for any $t \in \mathbb{N}$, we have $\operatorname{Pr}\left(\tau_B \leq t\right) \leq t \cdot \frac{\nu(B)}{\nu(A)}$. 

In particular, for any $t \in \mathbb{N}$ there exists a state $v \in A$ such that if $X_0=v$ deterministically, then $\operatorname{Pr}\left(\tau_B \leq t\right) \leq t \cdot \frac{\nu(B)}{\nu(A)}$.
\end{proposition}

\begin{corollary}[Corollary~5.4 in \cite{bandeira2022franz}, FP-Hard Implies Hitting Time Lower Bound] \label{cor:fp-hard-mcmc-hard} Assume the setting of \Cref{thm:fp-hard-implies-free-energy-barrier} with some $\delta = \delta(D)$. Let $\varepsilon > 0$ and define the hitting time as $\tau:=\inf \left\{t \in \mathbb{N}:\left\langle u, X_t\right\rangle>\delta\right\}$. Suppose $X_0, X_1, \ldots$ is a $\Delta$-local Markov chain with state space $S$ and stationary distribution $\nu_\beta$, for some $\Delta \leq \varepsilon \delta$. With probability at least $1-e^{-\varepsilon D}$ over $Z$, there exists a state $v \in A$ such that for the initialization $X_0=v$, with probability at least $1-e^{-\varepsilon D / 2}$ over the Markov chain,

$$
\tau \geq \frac{e^{\varepsilon D / 2}}{2(2 \cdot \operatorname{FP}(D+\log 2, \tilde{\lambda}))^{1-2 \varepsilon}}
$$

\end{corollary}
\begin{enumerate}[(a)]
\item Using the definition of the Markov chain locality, \Cref{prop:barrier-implies-hitting-time-lb}, and \Cref{thm:fp-hard-implies-free-energy-barrier}, prove \Cref{cor:fp-hard-mcmc-hard}.

\item Using \Cref{thm:ld-hard-implies-fp-hard}, derive the following corollary:
    \begin{corollary}\label{cor:ld-implies-mcmc-hard}
        Suppose $D=D_n$ is a sequence with $D=\omega(1)$ such that $D+\log 2$ is an even integer. In the setting of \Cref{thm:fp-hard-implies-free-energy-barrier}, assume for some constant $B>0$ that

$$
\mathrm{LD}(D+\log 2,(1+\varepsilon) \tilde{\lambda}) \leq B
$$
Then there is a constant $C=C(B, \varepsilon)>0$ only depending on $B$ and $\varepsilon$ such that the following holds for all sufficiently large $n$. With probability at least $1-e^{-\varepsilon D}$ over $Z$, there exists a state $v \in A$ such that for the initialization $X_0=v$, with probability at least $1-e^{-\varepsilon D / 2}$ over the Markov chain,

$$
\tau \geq C(B, \varepsilon) e^{\varepsilon D / 2}
$$

    \end{corollary}
\end{enumerate}

\renewcommand{\sectionletter}{\faPuzzlePiece} 
\subsection{Jigsaw}

\begin{enumerate}[(a)]
\item Discuss your results and share your findings with your peers. Discuss connections between Franz-Parisi criterion, low-degree hardness and lower bounds on hitting time for MCMC.
\end{enumerate}

\printbibliography[segment=\therefsegment] 
\section{Overlap Gap Property in number partitioning problem}\label{s14:ogp-I}
\renewcommand{\sectionletter}{1}

In the last few sessions~\ref{sec:MCMC} and \ref{s13:franz-parisi-II}, we demonstrated that a free energy barrier creates an obstruction to Markov Chain Monte Carlo chains. In this session, we explore another property, namely, the overlap gap property, which implies the existence of such a barrier and, in turn, rules out certain local algorithms for optimization problems. We consider the standard notion of the OGP in this session and will analyze more intricate variants in the coming sessions.

The main references are \cite{gamarnik2023algorithmic,gamarnik2021overlap}.

\subsection{Background}

We consider the number partitioning problem: given $n$ items, each with an associated weight (where $n$ is a positive integer), divide them into two bins, $A$ and $B$, such that the total weights of the items in $A$ and $B$ are as close as possible. 

More formally, we can write this problem as follows. Given $X \in \bb R^n$, find an assignment vector $\sigma^\star \in \{\pm 1\}^n$ such that it minimizes $\mathcal H(\sigma, X)$, where
$$
\mathcal H(\sigma, X):= \frac{1}{\sqrt{n}}\left |\sum_{1 \le i \le n } \sigma_i X_i\right| =  \frac{1}{\sqrt{n}}|\ip{ \sigma, X}|.
$$
Define additionally the normalized overlap as $\mathcal{O}(\sigma, \tau) = \frac{1}{n}\abs{\ip{\sigma, \tau}}$ for any $\sigma, \tau \in \{\pm 1\}^n$.

\begin{definition}
    We say that the optimization problem $\min _\sigma \mathcal{H}(\sigma,X)$ exhibits the OGP with values $0 < \eta<\beta < 1$ at energy level $E_n$ if for every two solutions $\sigma, \tau$ which satisfy $\mathcal{H}(\sigma, X) \leq 2^{-E_n}, \mathcal{H}(\tau, X) \leq 2^{-E_n}$, it is the case that either $\mathcal O(\sigma, \tau) \leq \beta - \eta$ or $\mathcal O(\sigma, \tau) \geq \beta$. 
\end{definition}

In words, the above definition says that in case of OGP, any pair of two near-optimal solutions must be either very close or far away in the sense of the defined overlap. 

\begin{remark}
    The notation of the above definition is adjusted to the notation of \cite{gamarnik2023algorithmic}. In the notation of \cite{gamarnik2021overlap}, $\eta_1 = \beta - \nu$ and $\eta_2 = \beta$.
\end{remark}

\renewcommand{\sectionletter}{$\bm \lambda$} 
\subsection{Group \grouplambda}

The goal of your group and group $\bm \mu$ is to prove presence of the OGP at energy levels $E_n=\epsilon n$ where $\epsilon \in\left(\frac{1}{2}, 1\right]$.

\begin{theorem}[Theorem~2.2 in \cite{gamarnik2023algorithmic}]\label{thm:ogp-number-partitioning}
    Let $X \sim \mathcal{N}\left(0, I_n\right)$ and $\epsilon \in\left(\frac{1}{2}, 1\right]$ be arbitrary. Then, there exists a $\rho = \rho(\epsilon) \in$ $(0,1)$ such that with probability $1-\exp (-\Theta(n))$, there are no pairs $\left(\sigma, \tau\right) \in \{\pm 1\}^n \times \{\pm 1\}^n$ for which the following holds simultaneously: $\mathcal{O}\left(\sigma, \tau\right) \in\left[\rho, \frac{n-2}{n}\right] ; \frac{1}{\sqrt{n}}|\langle\sigma, X\rangle|=O\left(2^{-n \epsilon}\right)$, and $\frac{1}{\sqrt{n}}\left|\left\langle\tau, X\right\rangle\right|=O\left(2^{-n \epsilon}\right)$.
\end{theorem}
To prove the theorem, we will show that with high probability the number of configurations satisfying the theorem conditions is zero. More specifically, define
$$
N = \sum_{\left(\sigma, \tau\right) \in \mathcal{Z}(\rho)} \mathds{1}\left\{\frac{1}{\sqrt{n}}|\langle\sigma, X\rangle|, \frac{1}{\sqrt{n}}\left|\left\langle\tau, X\right\rangle\right|=O\left(2^{-n \epsilon}\right)\right\},
$$
where 
$$
\mathcal{Z}(\rho) := \left\{\left(\sigma, \tau\right) \in \{\pm 1\}^n \times \{\pm 1\}^n: \mathcal{O}\left(\sigma, \tau\right) \in\left[\rho, \frac{n-2}{n}\right]\right\} .
$$

We aim to prove that $N=0$ with high probability using the first-moment method.

Your task is to upper bound cardinality of the set of configurations with high overlap, and group $\bm \mu$ will upper bound a probability of an event inside the definition of $N$.

\begin{enumerate}[(a)]
    \item Show that if $k := d_{\rm H}(\sigma, \tau)$ then $\mathcal{O}(\sigma, \tau) = \abs{1 - 2\frac{k}{n}}$, where $d_{\rm H}$ is the Hamming distance. 
    \item Show that for a fixed $\sigma$, there are 
    $$
    2\cdot \sum_{1 \le k \le \ceil{n \frac{1-\rho}{2}}} \binom{n}{k}
    $$
    ways to choose $\tau$ so that $(\sigma, \tau) \in \mathcal{Z}(\rho)$.
    \item Define the binary entropy function as $h(x) = -x \log _2 x-(1-x) \log _2(1-x)$. Show that for $1 \le \ell \le n$,
    $$
    \log_2 \binom{n}{\ell} = n h(\ell /n )+O(\log_2 n).
    $$
    You may use Stirling's approximation without proof: 
    $$
\log _2 n!=n \log _2 n-n \log _2 e+O\left(\log _2 n\right)
$$
\item Combining everything together, show that
$$
|\mathcal{Z}(\rho)| \leq \exp _2\left(n+n h\left(\frac{1-\rho}{2}\right)+O\left(\log _2 n\right)\right) ,
$$
where notation $\exp_2(\gamma)$ means $2^\gamma$.
\end{enumerate}

\renewcommand{\sectionletter}{$\bm \mu$} 
\subsection{Group \groupmu}

 Together with group $\bm \lambda$, you will prove \Cref{thm:ogp-number-partitioning}. Define $N$ and $\mathcal Z(\rho)$ as in group $\lambda$. The proof is based on the first-moment method, and your task is to show that with high probability $N=0$, while group $\bm \lambda$ will upper bound cardinality of the $\mathcal Z(\rho)$
.
\begin{enumerate}[(a)]
    \item Set 
    $$
Y_\sigma = \frac{1}{\sqrt{n}}\langle\sigma, X\rangle \quad \text { and } \quad Y_{\tau}= \frac{1}{\sqrt{n}}\left\langle\tau, X\right\rangle .
$$

Show that $(Y_\sigma, Y_\tau) \sim \mathcal{N}(0, \Sigma)$, where $\Sigma = \begin{pmatrix}
    1 &\bar \rho \\
    \bar \rho & 1
\end{pmatrix}$, where $\bar \rho := \mathcal{O}(\sigma, \tau).$
\item Fix constant $C > 0$ and denote by $\mathcal{R}_C$ the region 
$$
\mathcal{R}_C =\left[-C 2^{-n \epsilon}, C 2^{-n \epsilon}\right] \times\left[-C 2^{-n \epsilon}, C 2^{-n \epsilon}\right].
$$
Show that 
$$
\mathbb{P}\left(\left(Y_\sigma, Y_{\tau}\right) \in \mathcal{R}_C\right) \le  C_1 \frac{1}{\sqrt{1-\left(1-\frac{2}{n}\right)^2}} 2^{-2 n \epsilon},
$$
where $C_1 = \frac{4C^2}{2\pi}$.

\begin{hint}
    The probability of $\left(Y_\sigma, Y_{\tau}\right) \in \mathcal{R}_C$ equals to $\int_{\mathcal R_C}f(x, y) dx dy$, where $f(x,y)$ is the joint density of $Y_\sigma, Y_{\tau}$. Use the fact that $(Y_\sigma, Y_\tau) \sim \mathcal{N}(0, \Sigma)$ and that $\bar \rho \le n-2/n$ by the definition of $\mathcal{Z}_\rho$. 
\end{hint}

\item Show that 
$$
\sqrt{1-\left(1-\frac{2}{n}\right)^2}=\frac{2}{\sqrt{n}}\left(1+o_n(1)\right) .
$$
\item Conclude that 
$$
\bb P\lt(\lt \{ \mathcal{H}(\sigma, X), \mathcal{H}(\tau, X)=O\left(2^{-n \epsilon}\right)\rt\}\rt) = {C_2}\left(1+o_n(1)\right) 2^{-2 n \epsilon} \sqrt{n},
$$
where $C_2 > 0$ is some absolute constant.
\end{enumerate}

\renewcommand{\sectionletter}{$\bm \pi$} 
\subsection{Group \grouppi}
Your task is to show that the presence of the OGP implies the free energy well, which in turn constitutes an obstruction to local algorithms. 

Groups $\bm \lambda$ and $\bm \mu$ show that the number partitioning problem with $X \sim \mathcal{N}(0, I_n)$ exhibits OGP, see \Cref{thm:ogp-number-partitioning}.

Denote by  $\sigma^\star$ the configuration achieving the minimum of $\mathcal{H}(\sigma, X)$. Using probabilistic methods, it can be shown that $\mathcal{H}(\sigma^\star, X) = \Theta(2^{-n})$ with high probability. You may use this result without proof.

Similarly as before, we define the Gibbs measure $\pi_\beta(\cdot)$ at temperature $\beta^{-1}$ as
$$
\pi_\beta(\sigma)=\frac{1}{Z_\beta} \exp (-\beta H(\sigma, X)) \quad \text { where } \quad Z_\beta \triangleq \sum_{\sigma \in \{\pm 1\}^n} \exp (-\beta H(\sigma, X)).
$$
Define the sets:
\begin{itemize}
    \item $I_1=\left\{\sigma \in \{\pm 1\}^n:-\rho \leq \frac{1}{n}\left\langle\sigma, \sigma^*\right\rangle \leq \rho\right\}$.
    \item $I_2 = \left\{\sigma \in \{\pm 1\}^n: \rho \leq \frac{1}{n}\left\langle\sigma, \sigma^*\right\rangle \leq \frac{n-2}{n}\right\}$, and $\overline{I_2}=\left\{-\sigma: \sigma \in I_2\right\}$.
    \item $I_3=\left\{\sigma^*\right\}$ and $\overline{I_3}=\left\{-\sigma^*\right\}$.
\end{itemize}
The following theorem establishes the presence of the free energy well. 
\begin{theorem}[Theorem~3.3 in \cite{gamarnik2023algorithmic}]\label{thm:number-partitioning-energy-barrier}
    Let $\epsilon \in\left(\frac{1}{2}, 1\right)$ be arbitrary and $\beta=\Omega\left(n 2^{n \epsilon}\right)$. Then
$$
\min \left\{\pi_\beta\left(I_1\right), \pi_\beta\left(I_3\right)\right\} \geq \exp \left(\Omega\left(\beta 2^{-n \epsilon}\right)\right) \pi_\beta\left(I_2\right)
$$
holds with high probability over the randomness of $X$, as $n \rightarrow \infty$.
\end{theorem}
We will show only a part of this theorem, specifically that 
\begin{equation}\label{eq:few3}
\pi_\beta\left(I_3\right) \geq \exp \left(\Omega(n)\right) \pi_\beta\left(I_2\right).
\end{equation}
\begin{enumerate}[(a)]
    \item Show that 
    $$
\pi_\beta\left(I_3\right)=\frac{1}{Z_\beta} \exp \left(-\beta 2^{-n}\right) .
$$
\item On the other hand, show that for any $\sigma \in I_2$, with high probability,
$$
\pi_\beta(\sigma) \leq \frac{1}{Z_\beta} \exp \left(-\beta 2^{-n \epsilon}\right) .
$$

\begin{hint}
    Use the result on OGP, specifically, \Cref{thm:ogp-number-partitioning}. 
\end{hint}
\item For this subtask, you may use the following upper bound without proof: 
$$
\left|I_2\right| \leq 2\cdot \sum_{1 \leq k \leq\left\lceil\frac{n(1-\rho)}{2}\right\rceil}\binom{n}{k}=\exp _2\left(n h\left(\frac{1-\rho}{2}\right)+O\left(\log _2 n\right)\right),
$$
(you can ask group $\bm \lambda$ for more details of this proof).
Using that derive that, with high probability,
$$
\pi_\beta\left(I_3\right) \geq \exp \left(-\beta 2^{-n}+\beta 2^{-n \epsilon}-n h\left(\frac{1-\rho}{2}\right)+O\left(\log _2 n\right)\right) \pi_\beta\left(I_2\right) .
$$
\item Conclude \eqref{eq:few3} using the assumption on $\beta$. 
\end{enumerate}

\renewcommand{\sectionletter}{$\bm \gamma$} 
\subsection{Group \groupgamma}
Define the sets $I_1, I_2, I_3$ as in group $\bm \pi$. Your goal is to show a part \Cref{thm:number-partitioning-energy-barrier} concerning $\pi_\beta(I_1)$, specifically, 
$$
\pi_\beta\left(I_1\right) \geq \exp \left(\Omega\left(\beta 2^{-n \epsilon}\right)\right) \pi_\beta\left(I_2\right),
$$
while group $\bm \pi$ will show the statement for $\pi_\beta(I_3)$. 

You may use the following fact without proof. 

Fix $\varepsilon^\prime \in (\varepsilon, 1)$. With probability $1 - O(1/n)$ there exists $\sigma^\prime \in \{ \pm 1 \}^n$ such that $H(\sigma^\prime) = \Theta(2^{-n\varepsilon^\prime}) $. 

\begin{enumerate}[(a)]
    \item Note that $\sigma^{\prime} \notin\left(\overline{I_2} \cup I_2\right) \cup\left(\bar{I}_3 \cup I_3\right)$.
    \item Deduce that then 
    $$
\pi_\beta\left(I_1\right) \geq \pi_\beta\left(\sigma^{\prime}\right)=\frac{1}{Z_\beta} \exp \left(-\beta 2^{-n \epsilon^{\prime}}\right).
$$
\item Repeat the same steps as before using that $\varepsilon^\prime > \varepsilon$ and conclude that $\pi_\beta\left(I_1\right) \geq \exp \left(\Omega\left(\beta 2^{-n \epsilon}\right)\right) \pi_\beta\left(I_2\right)$.
\end{enumerate}

\renewcommand{\sectionletter}{\faPuzzlePiece} 
\subsection{Jigsaw}

\begin{enumerate}[(a)]
\item Discuss your results. Combine results of group $\bm \lambda$ and group $\bm \mu$ to finish the proof of \Cref{thm:ogp-number-partitioning} and results of group $\bm \pi$ and group $\bm \gamma$ to finish the proof of \Cref{thm:number-partitioning-energy-barrier}.
\item Recall that a free energy barrier (implied by \Cref{thm:number-partitioning-energy-barrier}) constitutes the obstruction to local algorithms. Discuss the implications of the free energy well in the setting of the number partitioning problem.
\end{enumerate}

\printbibliography[segment=\therefsegment] 
\section{Ensemble overlap gap property and low-degree polynomials}\label{s15:e-OGP}

In this session, we explore a more intricate variant of the overlap gap property, namely, the ensemble overlap gap property. We will demonstrate its connection to the performance of low-degree polynomials by showing that they are stable in a certain sense. Our model problem is the $p$-spin glass model, which can be viewed as a special case of a Gaussian additive model where the signal forms a tensor. The main references are \cite{gamarnik2024hardness,gamarnik2021overlap}.

\renewcommand{\sectionletter}{1}
\subsection{Background}

Fix $p\ge 2$. We consider the optimization of the $p$-spin glass Hamiltonian, which is defined as follows. Let $Y \in (\bb R^n)^{\otimes p}$. For $x \in \bb R^n$, define the objective function as follows
\begin{equation}\label{eq:objective}
H_n(x ; Y)=\frac{1}{n^{(p+1) / 2}}\left\langle Y, x^{\otimes p}\right\rangle .
\end{equation}
We will focus on the case when $Y$ contains i.i.d. Gaussian entries $\mathcal{N}(0, 1)$ and $x \in \mathcal S_n = \{x\in \bb R^n: \|x\|_2 = \sqrt{n}\}$. 

Our goal is to understand the performance of randomized low-degree polynomials for optimizing $\max_{x \in \mathcal{S}_n} H_n(x; Y)$. To simplify the notation, we will refer to the space of $p$-tensors on $\bb R^n$ by $\bb R^m$, where $m = n^p$. 

We consider polynomial function $ f : \bb R^m \to \bb R^n$, where $f(Y) = (f_1(Y), \dots, f_n(Y))$ and each $f_i$ is a polynomial in entries of $Y$. Moreover, the coefficients of $f$ may be random, i.e., for some probability space $(\Omega, \bb P_\omega)$, we can define $f: \bb R^m \times \Omega \to \bb R^n$ so that $f(\cdot, \omega)$ is a polynomial for each $\omega \in \Omega$. 

We will round the polynomial input by normalizing it in the standard way, 
$$
g_f(Y, \omega)=\sqrt{n} \frac{f(Y, \omega)}{\|f(Y, \omega)\|_2},
$$
where $g_f(\cdot, \omega) : \bb R^m \times \Omega \to \mathcal S_n \cup \{\infty\}$ (we let $g_f(Y, \omega)=\infty$ if $f(Y, \omega)=0$).

\begin{definition}\label{def:random-polynom-optimizes}
    For parameters $\mu \in \mathbb{R}, \delta \in[0,1], \gamma \in[0,1]$, and a random polynomial $f: \mathbb{R}^m \rightarrow$ $\mathbb{R}^n$, we say that $f$ $(\mu, \delta, \gamma)$-optimizes the objective \eqref{eq:objective} on $\mathcal{S}_n$ if the following are satisfied:
\begin{itemize}
    \item $\underset{Y, \omega}{\mathbb{E}}\|f(Y, \omega)\|_2^2=n \quad$, where $Y\sim \bb P_Y$ and $\omega \sim \bb P_\omega$ independently (normalization).
    \item With probability at least $1-\delta$ over $Y$ and $\omega$, we have both $H_n\left(g_f(Y, \omega) ; Y\right) \geq \mu$ and $\|f(Y, \omega)\|_2 \geq$ $\gamma \sqrt{n}$.
\end{itemize} 
\end{definition}

\begin{definition}
    We say that a family of real-valued functions $\mathcal{F}$ with common domain $\mathcal{X} \subset \mathbb{R}^n$ satisfies the ensemble-overlap gap property (e-OGP) for an overlap $R: \mathcal{X} \times \mathcal{X} \rightarrow[0,1]$ with parameters $\mu \in \mathbb{R}$ and $0 \leq \nu_1<\nu_2 \leq 1$ if for every $f_1, f_2 \in \mathcal{F}$ and every $x_1, x_2 \in \mathcal{X}$ satisfying $f_k\left(x_k\right) \geq \mu$ for $k=1,2$, we have that $R(x, y) \in\left[0, \nu_1\right] \cup\left[\nu_2, 1\right]$.
\end{definition}

\begin{definition}
    We say that two real-valued functions $f, g$ with common domain $\mathcal{X}$ are $\nu$-separated above $\mu$ with respect to the overlap $R: \mathcal{X} \times \mathcal{X} \rightarrow[0,1]$ if for any $x, y \in \mathcal{X}$ with $f(x) \geq \mu$ and $g(y) \geq \mu$, we have that $R(x, y) \leq \nu$.
\end{definition}
In this session, we will focus only on the case when overlap $R(x,y) = \frac{1}{n}\ip{x,y}$.

\renewcommand{\sectionletter}{$\bm \lambda$} 
\subsection{Group \grouplambda}

Our goal is to prove the following theorem. As oftentimes before, the proof is split into two parts between your group and group $\bm \mu$.

Consider interpolation path $\left(Y_\tau\right)_{\tau \in[0, \pi / 2]}$ defined by

\begin{equation}\label{eq:Y_tau}
Y_\tau=\cos (\tau) Y+\sin (\tau) Y^{\prime}    
\end{equation}

and let $\mathcal{A}\left(Y, Y^{\prime}\right)=\left\{H_n\left(x ; Y_\tau\right): \tau \in[0, \pi / 2]\right\}$.

\begin{theorem}[Theorem~3.7 in \cite{gamarnik2024hardness}]\label{thm:eOGP-spin-model}
    For any $0 \leq \nu_1<\nu_2 \leq 1$, there exists a constant $\delta^*>0$ such that the following holds. Let $p, n, D \in \mathbb{N}$ and $\mu \in \mathbb{R}$. Suppose that $Y, Y^{\prime}$ are independent p-tensors with i.i.d. standard Gaussian entries and let $\mathcal{A}\left(Y, Y^{\prime}\right)$ be as above. Suppose further that with probability at least $3 / 4$ over $Y, Y^{\prime}$, we have that $\mathcal{A}\left(Y, Y^{\prime}\right)$ has the $\left(\mu, \nu_1, \nu_2\right)$-eOGP on domain $\mathcal{S}_n$, and that $H_n(\cdot, Y)$ and $H_n\left(\cdot, Y^{\prime}\right)$ are $\nu_1$ separated above $\mu$. Then for any $\delta \leq \min \left\{\delta^*, \frac{1}{4} \exp (-2 D)\right\}$ and any $\gamma \geq(2 / 3)^D$, there is no random degree-D polynomial that $(\mu, \delta, \gamma)$-optimizes \eqref{eq:objective} on $\mathcal{S}_n$.
\end{theorem}

In words, the above theorem states that if solutions of end points of the interpolation path $\mathcal A$ have small overlap, while $\mathcal A(Y, Y^\prime)$ has the $(\mu, \nu_1, \nu_2)$-eOGP, then it implies the failure of stable polynomials in the sense of \Cref{def:random-polynom-optimizes}.

For the sake of contradiction, we will assume that there exists a random polynomial $f$ of degree at most $D$ which $(\mu, \delta, \gamma)$-optimizes $H_n(\cdot, Y)$. 
\begin{enumerate}[(a)]
    \item First, we will reduce to the case when $f$ is deterministic. For now, you may use the following result as given, and you can prove it if you have extra time in the end. 

\begin{lemma}
    Let $A(Y, \omega)$ denote the "failure" event

$$
A(Y, \omega)=\left\{H_n\left(g_f(Y, \omega) ; Y\right)<\mu \vee\|f(Y, \omega)\|_2<\gamma \sqrt{n}\right\}.
$$
There exists $\omega^*$ such that $\mathbb{E}_Y\left\|f\left(Y, \omega^*\right)\right\|_2^2 \leq 3 n$ and $\mathbb{P}_Y\left\{A\left(Y, \omega^*\right)\right\} \leq 3 \delta$. 
\end{lemma}

We fix $\omega = \omega^*$ and consider deterministic function $f(\cdot) = f(\cdot, \omega^*)$ in what follows.

\begin{hint}
    Use the fact that $\mathbb{E}\|f(Y, \omega)\|_2^2=n$ and $\mathbb{P}(A(Y, \omega)) \leq \delta$ and then apply Markov's inequality to show that $\mathbb{P}_\omega\left\{\mathbb{E}_Y\|f(Y, \omega)\|_2^2 \geq 3 n\right\} \leq 1 / 3 \quad$ and $\quad \mathbb{P}_\omega\left(\mathbb{P}_Y(A(Y, \omega)) \geq 3 \delta\right) \leq 1 / 3$. Conclude that it implies the existence of $\omega^*$ so that the claim in (a) holds.
\end{hint}

\item For some $L$, divide the interval $[0, \pi/2]$ into $L$ subintervals with $0 = \tau_0 < \tau_1 < \dots < \tau_L = \pi/2$. Let $x_\ell = g_f(Y_{\tau_\ell})$, where $Y_\tau$ is defined by \eqref{eq:Y_tau}. We will show that under the assumptions of the theorem the following three events can not occur simultaneously (group $\bm \mu$ will show that these events occur simultaneously with positive probability, which leads to contradiction):
\begin{enumerate}[(i)]
    \item The family $\mathcal{A}\left(Y, Y^{\prime}\right)$ has the $\left(\mu, \nu_1, \nu_2\right)$-eOGP on $\mathcal{S}_n$ and $H_n(\cdot ; Y)$ and $H_n\left(\cdot ; Y^{\prime}\right)$ are $\nu_1$-separated above $\mu$.
\item For all $\ell \in\{0,1, \ldots, L\}, f$ succeeds on input $Y_{\tau_{\ell}}$, i.e., the event $A\left(Y_{\tau_{\ell}}, \omega^*\right)^c$ holds.
\item For all $\ell \in\{0,1, \ldots, L-1\},\left\|f\left(Y_{\tau_{\ell}}\right)-f\left(Y_{\tau_{\ell+1}}\right)\right\|_2^2<\gamma^2 c n$ for $c \leq\left(\nu_2-\nu_1\right)^2$.
\end{enumerate}
Show that under events (i) and (ii), there exists some $\ell$ such that
$$
\nu_2 - \nu_1 \le \frac{1}{\sqrt{n}} \| x_\ell - x_{\ell+1}\|.
$$

\begin{hint}
    Observe that $\left|\frac{1}{n}\left\langle x_0, x_L\right\rangle\right| \in\left[0, \nu_1\right]$ and $\left|\frac{1}{n}\left\langle x_0, x_0\right\rangle\right|=1$. Use the triangle inequality and Cauchy-Schwarz inequality. 
\end{hint}
\item Show that if $\|x\|_2=\|y\|_2=1$ and $a \geq \gamma, b \geq \gamma$ then $\|x-y\|_2 \leq \gamma^{-1}\|a x-b y\|_2$.
\item Using (c), show that under event (ii), $
\left\|x_{\ell}-x_{\ell+1}\right\|_2 \leq \frac{1}{\gamma }\left\|f\left(Y_{\tau_{\ell}}\right)-f\left(Y_{\tau_{\ell+1}}\right)\right\|_2.
$
\item Combine (b) and (d) to show that (iii) cannot hold under events (i) and (ii).
\end{enumerate}

\renewcommand{\sectionletter}{$\bm \mu$} 
\subsection{Group \groupmu}
Our goal is to prove \Cref{thm:eOGP-spin-model}. As oftentimes, the proof is split into two parts between your group and group $\bm \lambda$.

For the sake of contradiction, we will assume that there exists a random polynomial $f$ of degree at most $D$ which $(\mu, \delta, \gamma)$-optimizes $H_n(\cdot, Y)$. Let $A(Y, \omega)$ denote the "failure" event
$$
A(Y, \omega)=\left\{H_n\left(g_f(Y, \omega) ; Y\right)<\mu \vee\|f(Y, \omega)\|_2<\gamma \sqrt{n}\right\}.
$$

For this proof, you may assume that $f$ is deterministic and $\mathbb{E}_Y\left\|f\left(Y\right)\right\|_2^2 \leq 3 n$ and $\mathbb{P}_Y\left\{A\left(Y\right)\right\} \leq 3 \delta$. 
\begin{enumerate}[(a)]

\item Let events (i)-(iii) be defined as in $\bm \lambda$ (b). We will show that under the assumptions of \Cref{thm:eOGP-spin-model} the following three events occur simultaneously with positive probability (group $\bm \lambda$ will show that these events cannot occur simultaneously under the theorem assumptions, which leads to contradiction)

Show that (ii) does not hold with probability at most $1/3$ provided that $L \le \frac{1}{9\delta} - 1$. 
\item For event (iii), you may use \Cref{thm:ld-stable} as given. 
Show that the theorem implies that for some $\tilde{D} \ge D$ and a fixed $\ell$,
$$
\mathbb{P}\left(\left\|f\left(Y_{\tau_{\ell}}\right)-f\left(Y_{\tau_{\ell+1}}\right)\right\|_2^2 \geq 6 n(6 e)^{\tilde{D}}\left(1-\rho^{\tilde{D}}\right)\right) \leq \exp (-2 \tilde{D})
$$
where $\rho = \cos(\frac{\pi}{2L})$.
\item Show that $$
1-\rho^{\bar{D}}\leq \frac{\tilde{D}}{2}\left(\frac{\pi}{2 L}\right)^2
$$

\begin{hint}
    Use that $\cos(x) \ge 1 - x$ for any $x \in \bb R$.
\end{hint}
\item Show that when $L \geq \frac{\pi}{2 \gamma} \sqrt{\frac{3 \tilde{D}}{c}}(6 e)^{\tilde{D} / 2}$, 
$$
\mathbb{P}\left(\left\|f\left(Y_{\tau_{\ell}}\right)-f\left(Y_{\tau_{l+1}}\right)\right\|_2^2 \geq \gamma^2 c n\right) \leq \exp (-2 \tilde{D}).
$$
Consequently, the event (iii) fails with probability at most $1/3$ when $L \le 1/3 \exp(2\tilde D)$.
\item Using theorem assumption, (b), and (e), show that (i), (ii), (iii) occur simultaneously with positive probability.

\end{enumerate}
\begin{remark}
     To finish the proof, we will need to show that there exist integers $\tilde{D}$ and $L$ so that they satisfy all assumptions made throughout the proof, namely, 
     $$
\frac{\pi}{2 \gamma} \sqrt{\frac{3 \tilde{D}}{c}}(\sqrt{6 e})^{\tilde{D}} \leq L \leq \min \left\{\frac{1}{9 \delta}-1, \frac{1}{3}\left(e^2\right)^{\tilde{D}}\right\}.
$$
The proof of this claim is rather technical and therefore optional, but you may choose to prove it if you have extra time.
\end{remark}

\renewcommand{\sectionletter}{$\bm \pi$} 
\subsection{Group \grouppi}

Our goal is to prove that low-degree polynomials are stable. 
\begin{theorem}[Theorem~3.1 in \cite{gamarnik2024hardness}]\label{thm:ld-stable}
    Let $0 \leq \rho \leq 1$. Let $X, Y\sim \mathcal{N}(0, I_d)\in \bb R^d$ such that $Cov(X, Y) = \rho I$. Let $f: \mathbb{R}^d \rightarrow \mathbb{R}^k$ be a deterministic polynomial of degree at most $D$ with $\mathbb{E}\|f(X)\|_2^2=1$. For any $t \geq(6 e)^D$,
$$
\bb P\left(\|f(X)-f(Y)\|_2^2 \geq 2 t\left(1-\rho^D\right)\right) \leq \exp \left(-\frac{D}{3 e} t^{1 / D}\right)
$$
\end{theorem}

We will need several properties of degree-$D$ polynomials, which you may use without proof.
\begin{lemma}[Hypercontractivity for polynomials]\label{lm:hypercontr-multivar}
 Let $Y\sim \mathcal{N}(0, I_d)$. If $f: \mathbb{R}^d \rightarrow \mathbb{R}^k$ is a degree-$D$ polynomial and $q \in[2, \infty)$ then

$$
\mathbb{E}\left[\|f(Y)\|_2^{2 q}\right] \leq[3(q-1)]^{q D} \mathbb{E}\left[\|f(Y)\|_2^2\right]^q .
$$
\end{lemma}
\begin{lemma}\label{lm:gaussian-lip}
If $f: \mathbb{R}^d \rightarrow \mathbb{R}^k$ is a degree-D polynomial with $\mathbb{E}\|f(Y)\|_2^2=1$, then for any $\rho \in[0,1]$, if $X, Y\sim \mathcal N(0, I_d)$ such that $Cov(X, Y) = \rho I_d$,

$$
\mathbb{E}\|f(X)-f(Y)\|_2^2 \leq 2\left(1-\rho^D\right) .
$$

\end{lemma}
\begin{enumerate}[(a)]
\item First we will show that if $f: \mathbb{R}^d \rightarrow \mathbb{R}^k$ is a degree-D polynomial and $Y \sim \mathcal{N}(0, I_d)$, then for any $t \geq(6 e)^D$,

$$
\bb P\left(\|f(Y)\|_2^2 \geq t \mathbb{E}\left[\|f(Y)\|_2^2\right]\right) \leq \exp \left(-\frac{D}{3 e} t^{1 / D}\right).
$$
Use \Cref{lm:hypercontr-multivar} and Markov's inequality for some $q$. Pick $q$ to arrive at the conclusion. 

\begin{hint}
    $q=t^{1 / D} /(3 e) \geq 2$.
\end{hint}

\item Let $Y^\prime$ be an independent copy of $Y$ and let $\tilde{Y} = (Y, Y^\prime)$. Define new function 
$$
h(\tilde{Y})=f(Y)-f\left(\rho Y+\sqrt{1-\rho^2} Y^{\prime}\right).
$$
Note that $\tilde{Y}$ is standard Gaussian variable in $\bb R^{2d}$ and $h(\tilde{Y})$ is a polynomial in $\tilde{Y}$ of degree at most $D$. Show that applying \Cref{lm:gaussian-lip} gives 
$$
\mathbb{E}\|f(X)-f(Y)\|_2^2 \leq 2\left(1-\rho^D\right) 
$$
and conclude the proof of \Cref{thm:ld-stable} using (a).
\end{enumerate}

\renewcommand{\sectionletter}{\faPuzzlePiece} 
\subsection{Jigsaw}

\begin{enumerate}[(a)]
\item Discuss your results with your peers. 
\item It can be shown that the ensemble overlap gap property holds for $p$-spin glass Hamiltonian with respect to the overlap $R(x, y)=\frac{1}{n}|\langle x, y\rangle|$. This is formalized by the following theorem, and you may use it as given.
\begin{theorem}[Theorem~2.11 in \cite{gamarnik2024hardness}]
    Let $Y$ and $Y^{\prime}$ be independent p-tensors with i.i.d. $\mathcal{N}(0,1)$ entries. For every even $p \geq 4$, there exist $0<\mu<\lim_{n\to\infty} \max_{x \in\mathcal S_n} H(x; Y), 0 \leq \nu_1<\nu_2 \leq 1$, and $\gamma>0$ such that the following holds with probability at least $1-\exp (-\gamma n)$ for all large enough $n$ :
\begin{itemize}
    \item The family $\mathcal{F}=\mathcal{A}\left(Y, Y^{\prime}\right)$ satisfies the $\left(\mu, \nu_1, \nu_2\right)$-eOGP.
    \item $H_n(\cdot ; Y)$ and $H_n\left(\cdot ; Y^{\prime}\right)$ are $\nu_1$-separated above $\mu$.
\end{itemize}
\end{theorem}

Using this theorem and your results, deduce the following result. 
\begin{theorem}[Theorem~2.5 in \cite{gamarnik2024hardness}]
For any even integer $p \geq 4$ there exist constants $\mu<\lim_{n\to\infty} \max_{x \in\mathcal S_n} H(x; Y)$, $n^* \in \mathbb{N}$, and $\delta^*>0$ such that the following holds. For any $n \geq n^*$, any $D \in \mathbb{N}$, any $\delta \leq \min \left\{\delta^*, \frac{1}{4} \exp (-2 D)\right\}$, and any $\gamma \geq(2 / 3)^D$, there is no random degree-D polynomial that $(\mu, \delta, \gamma)$-optimizes $H_n(\cdot; Y)$ on $\mathcal{S}_n$.
\end{theorem}
\end{enumerate}

\printbibliography[segment=\therefsegment] 
\section{Sherrington-Kirkpatrick model and quiet planting}\label{s16:Sk-model}
In this session, we will primarily cover \cite{SKquietplanting-bandeira2020}. We will consider the problem of certifying an upper bound on the maximum value of a random optimization problem over search vectors in a constrained set.

In a special case, when the constrained set is a hypercube, this problem corresponds to certifying bounds on the Hamiltonian of the Sherrington-Kirkpatrick spin glass model from statistical physics.

We will try to understand algorithmic lower and upper bounds, i.e., bounds achievable by an efficient optimization algorithm.

For ease of exposition, we will state most of our results in expectation, although everything can also be done ``with high probability'' (as in~\cite{SKquietplanting-bandeira2020}).

\renewcommand{\sectionletter}{1}
\subsection{Background}

Throughout $W$ will be a symmetric $n\times n$ matrix drawn from the Gaussian Orthogonal Ensemble ($\GOE$), in other words $W$ is a symmetric matrix whose upper triangular (and diagonal) entries are independent and distributed as $W_{ij}\sim \NNN(0,1/n)$ if $i\neq j$, and $W_{ii}\sim\NNN(0,2/n)$. We will use without proof that
$\EE\,\lambda_{\max}(W) = 2+o(1)$.

In fact, (we will not need this) the distribution of the eigenvalues of $W$ converges (weakly) to the so-called semicircular law with support $[-2,2]$ given by $  d\mathrm{SC}(x) = \frac{1}{2\pi}\sqrt{4-x^2}1_{[-2,2]}(x)$.

All big-O notation will be in the sense of the dimension limit, $n\to\infty$. Expectations are taken with respect to the $\GOE$ matrix $W$.

Our goal will be understand the following random optimization problem
\begin{equation}\tag{\ding{100}}
\mathrm{SK}(W) := \frac{1}{n}\max_{x\in \{\pm1\}^n} x^\top Wx
\end{equation}
It is known (although difficult to prove, and a rather deep statement) that $\EE\,\mathrm{SK}(W) = 1.5264+o(1)$ (twice the so-called Parisi constant $P_\ast$). This week we will try to understand what algorithmic lower and upper bounds we can give.

\textbf{Optimization Algorithm} An efficient optimization algorithm is one that, given $W$, computes $x_{\srch}\in\{\pm1\}^\top$ such that $x_{\srch}^\top Wx_{\srch}$ is large. We can measure \emph{typical performance} of our algorithm by
$\mathrm{TP}_{\srch}=\EE[x_{\srch}^\top Wx_{\srch}]$.

\textbf{Certified Upper bound} The goal here is to have an efficient $\cert$ algorithm that, given $W$, computes $U_{\cert}(W)$ such that $U_{\cert}(W)\geq \SK(W)$ for all symmetric $n\times n$ matrices $W$. 

It is clear that
\begin{equation}\tag{\ding{51}}
\mathrm{TP}_{\srch} \leq \EE \mathrm{SK}(W) \leq \mathrm{TP}_{\cert} 
\end{equation}

The question we are interested in is understanding how small we can make the gap between these quantities, restricting ourselves to efficient algorithms.

\renewcommand{\sectionletter}{$\bm \lambda$} 
\subsection{Group \grouplambda}

\begin{enumerate}[(a)]

\item Show that $\mathrm{AbsSum}(W) = \sum_{i,j=1}^n|W_{ij}|$ is a certified upper bound.

\item Is $\EE \mathrm{AbsSum}(W) = O(1)$?

\item Find an efficient algorithm for which $\mathrm{TP}_{\cert} \leq 2024$.

\subsubsection*{Optional*}

\item Show that $\mathrm{SDP}(W)$ is a certified upper bound, where
\[
\mathrm{SDP}(W) = \sup_{X\succeq0,\  X_{ii}=1\,\forall_i } Tr(WX)
\]

\item Is this certified upper bound pointwise better, pointwise worse, or pointwise incomparable to the one you proposed above?

\item Compute $\mathrm{TP}_{\cert}$ for $\mathrm{SDP}(W)$.

\item (\emph{Rotation Invariance of $W$})  Show that $W$ is rotation invariant, meaning that for any orthogonal matrix $O$ we have that $OWO^\top\sim W$ (this explains the at-first seemingly bizarre choice of having different variance on the diagonal and off-diagonal entries)

\end{enumerate}

\renewcommand{\sectionletter}{$\bm \mu$} 
\subsection{Group \groupmu}

 You will compute a search algorithm for $\SK(W)$. Let $v_{\max}$ denote the leading eigenvector of $W$, a possible efficient algorithm is to take $x_{\srch}=\sign(v_{\max})$.

\begin{enumerate}[(a)]
    \item What is $\EE \langle \sign(v_{\max}),v_{\max}\rangle$ up to $(1+o(1))$ factors? (you can use that the eigenvectors of $W$ are uniformly distributed among the set of orthonormal basis -- this follows from one of the optional tasks; you can also just ``pretend'' $v_{\max}$ has independent Gaussian entries, this is not correct but it is almost true and one can show that the answer you get in the end is indeed correct.)

    \item Note that for this algorithm, $\mathrm{TP}_{\srch} = \EE\, \sign(v_{\max})^\top W \sign(v_{\max})$. Show that
    $$  \sign(v_{\max})^\top W \sign(v_{\max}) = \lambda_{\max} \langle \sign(v_{\max}),v_{\max}\rangle^2 + \sum_{i=1}^{n-1}\lambda_i(W) \langle \sign(v_{\max}),v_i\rangle^2 $$

    \item Can you compute $\mathrm{TP}_{\srch}$ for this algorithm, up to $(1-o(1))$ factors. It is okay if you are not fully rigorous in this particular computation.
    
\subsubsection*{Optional*}

\item (\emph{Rotation Invariance of $W$})  Show that $W$ is rotation invariant, meaning that for any orthogonal matrix $O$ we have that $OWO^\top\sim W$ (this explains the at-first seemingly bizarre choice of having different variance on the diagonal and off-diagonal entries)

\end{enumerate}

\renewcommand{\sectionletter}{$\bm \pi$} 
\subsection{Group \grouppi}

You will show a non-trivial upper bound on $\EE\SK(W)$, (lower than the best $\mathrm{TP}_{\cert}$ we know).

We will do it with Slepian's Comparison Lemma. Slepian's Comparison Lemma is a crucial tool to compare Gaussian processes. A Gaussian process is a family of Gaussian random variables indexed by some set $T$, more precisely, is a family of Gaussian random variables $\left\{X_t\right\}_{t\in T}$ (if $T$ is finite this is simply a Gaussian vector). Given a Gaussian process $X_t$, a particular quantity of interest is $\EE\left[\max_{t\in T}X_t\right]$. Intuitively, if we have two Gaussian processes $X_t$ and $Y_t$ with mean zero $\EE\left[X_t\right] = \EE\left[Y_t\right] = 0$, for all $t\in T$ and the same variances $\EE\left[X_t^2\right] = \EE\left[Y_t^2\right]$ then the process that has the ``least correlations'' should have a larger maximum (think of the maximum entry of a vector with i.i.d. Gaussian entries versus one always with the same Gaussian entry). A simple version of Slepian's Lemma makes this intuition precise:\footnote{Although intuitive in some sense, this is a delicate statement about Gaussian random variables, it turns out not to hold for other distributions.}

In the conditions above, if for all $t_1,t_2\in T$
$\EE\left[X_{t_1}X_{t_2}\right] \leq  \EE\left[Y_{t_1}Y_{t_2}\right]$, then $ \EE\left[\max_{t\in T}X_t\right] \geq \EE\left[\max_{t\in T}Y_t\right]$.

A slightly more general version of it asks that the two Gaussian processes $X_t$ and $Y_t$ have mean zero $\EE\left[X_t\right] = \EE\left[Y_t\right] = 0$, for all $t\in T$, but not necessarily the same variances. In that case, it says that:

\begin{lemma}[Slepian's Lemma]
    If $X_t$ and $Y_t$ are two mean zero Gaussian processes such that, for all $t_1,t_2\in T$
\begin{equation}
\EE\left[X_{t_1}-X_{t_2}\right]^2 \geq  \EE\left[Y_{t_1}-Y_{t_2}\right]^2,
\end{equation}
then
\[
 \EE\left[\max_{t\in T}X_t\right] \geq \EE\left[\max_{t\in T}Y_t\right].
\]
\end{lemma}

\begin{enumerate}[(a)]

 \item Note that if we take, for unit-norm $v$, $$Y_v \defeq  v^\top Wv$$ we have that
  \[
  \SK(W) = \EE\left[\max_{v\in\left\{\pm\frac1{\sqrt{n}}\right\}^n}Y_v\right].
 \]
 
 Use Slepian to compare $Y_v$ with $2X_v$ defined as
 \[
  X_v \defeq v^\top g, 
 \]
where $g\sim\NNN\left(0,\frac1nI_{n\times n}\right)$.

 \item Upper bound $\EE\left[\max_{v\in\left\{\pm\frac1{\sqrt{n}}\right\}^n}X_v\right]$.

 \item Conclude an upper bound on $\mathrm{TP}_{\cert}$ for this algorithm.

\subsubsection*{Optional*}

\item Can you use the machinery you developed to show $\EE\,\lambda_{\max}(W)\leq 2+o(1)$?

\item (\emph{Rotation Invariance of $W$})  Show that $W$ is rotation invariant, meaning that for any orthogonal matrix $O$ we have that $OWO^\top\sim W$ (this explains the at-first seemingly bizarre choice of having different variance on the diagonal and off-diagonal entries)

\end{enumerate}

\renewcommand{\sectionletter}{\faPuzzlePiece} 
\subsection{Jigsaw}
Here, you will try to show a computational lower bound on $\mathrm{TP}_{\cert}$, the goal being to show that, for any $\varepsilon>0$, no efficient algorithm has $\mathrm{TP}_{\cert}<2-\varepsilon$. 

One idea is to try to do what is referred to as \emph{quiet planting}: build a random variable $W'\sim \mathrm{\PPP}$ drawn from a distribution $\PPP$ such that:
\begin{enumerate}
\item $\EE_{W'\sim \mathrm{\PPP}} \SK(W')\geq c$
\item It is computationally hard (e.g. low-degree hard) to distinguish a sample $W'\sim\mathrm{\PPP}$ from a sample $W\sim\GOE$.
\end{enumerate}

\begin{enumerate}[(a)]
\item Discuss your results and share your findings with your peers. 

\item Argue that constructing a quiet planting for a constant $c$ implies that it is not possible to have an efficient algorithm with $\mathrm{TP}_{\cert}< c$.

\item We could try to construct a distribution $\PPP$ the following way: draw $x\in\{\pm1\}^n$ uniformly at random, then set $W' = \frac{c}{n}xx^\top+W$ where $W\sim\GOE$ (independent from $X$). Show that $\EE\SK(W')\geq c+o(1)$.

\item For which values of $c$ is the above planting actually quiet? (meaning that we can't efficiently distinguish a sample $W'\sim\mathrm{\PPP}$ from a sample $W\sim\GOE$.
       
\subsubsection*{Optional*}

\item What if we constructed the distribution $\PPP$ as follows: draw $x\in\{\pm1\}^n$ uniformly at random, then set $W'$ as a draw from $\GOE$ conditioned on $\frac1nx^\top Wx=c$ (or $\approx c$\ ; or $\geq c$)?

\item What if we constructed the distribution $\PPP$ as follows: draw $W$ from the $\GOE$ conditioned to $SK(W)=c$ (or $\approx c$\ ; or $\geq c$)?

\item Are the two above distributions the same?

\item Can you come up with a planting that is quiet up to $c=2-\varepsilon$?

\end{enumerate}

\subsection*{State of the art:} \cite{SKquietplanting-bandeira2020} provides a quiet planting for $c=2-\varepsilon$ by doing a reduction to a negatively Rademacher-spiked Wishart model for which we know a low-degree lower bound. The second conditional distribution suggested in the optional tasks above is a natural construction but, to this day, we do not understand it.

Also, incredibly, a recent breakthrough shows that there is no computational gap in $\mathrm{TP}_{srch}$ (see~\cite{Montanari_SK_OPT}).

\printbibliography[segment=\therefsegment] 

\end{document}